\documentclass[12pt,preprint]{aastex}
\usepackage{epsfig}
\usepackage{graphicx}
\begin{document}
\title{Triggered Star Formation and Dust around Mid-Infrared-Identified Bubbles}
\author{C.Watson\altaffilmark{1}, U.Hanspal\altaffilmark{2}, A.Mengistu\altaffilmark{1}}
\altaffiltext{1}{Department of Physics, Manchester College, 604 East College Avenue,
  North Manchester, IN 46962} 
\altaffiltext{2}{Ross University School of Medicine, 630 US Highway 1, North Brunswick, NJ 08902}
\begin{abstract}
We use Two Micron All Sky Survey, GLIMPSE, and MIPSGAL survey data to analyze the young stellar object (YSO) and warm dust distribution around several mid-infrared-identified bubbles. We identify YSOs using J-band to 8 $\mu$m photometry and correlate their distribution relative to the PDR (as traced by diffuse 8 $\mu$m emission) which we assume to be associated with and surround a H\,{\sc{ii}} region. We find that only 20\% of the sample H\,{\sc{ii}} regions appear to have a significant number of YSOs associated with their PDRs, implying that triggered star formation mechanisms acting on the boundary of the expanding H\,{\sc{ii}} region do not dominate in this sample. We also measure the temperature of dust inside 20 H\,{\sc{ii}} regions using 24 $\mu$m and 70 $\mu$m MIPSGAL images. In eight circularly symmetric sources we analyze the temperature distribution and find shallower temperature gradients than is predicted by an analytic model. Possible explanations of this shallow temperature gradient are a radially dependent grain-size distribution and/or non-equilibrium radiative processes.
\end{abstract}

\section{Introduction}
High-resolution infrared (IR) surveys, such as the Two Micron All Sky Survey (2MASS), GLIMPSE and MIPSGAL, have been recently used to study polycyclic aromatic hydrocarbon (PAH) and dust emission near ionized regions in a number of different ways. Churchwell et al. (2006, 2007) compiled a list of $\sim$600 objects with 8 $\mu$m emission in a circular morphology, which they termed bubbles. They posited that the bubbles were PAH emission in the photodissociation regions (PDRs) surrounding O and early B stars. Watson et al. (2008, 2009) analyzed 6 bubbles from this list in detail, identifying candidate ionizing stars and young stellar objects (YSOs). They also found a consistent morphological relationship between radio continuum, 24 $\mu$m emission and 8 $\mu$m emission. Specifically, the radio continuum (tracing ionized gas) and 24 $\mu$m emission (tracing warm dust) overlap and are surrounded by 8 $\mu$m emission (tracing PAH emission and the PDR). The consistent presence of warm dust in these objects indicates that it is either not destroyed or not blown out as easily as PAH molecules or is continuously replenished and may play an important role in the cooling of the H\,{\sc{ii}} region and absorbing the Lyman continuum (Jones et al. 1999; Everett \& Churchwell 2010). In contrast, PAH emission is not observed within the H\,{\sc{ii}} regions but is abundant in the PDR envelopes. Toward N49, Watson et al. (2008) identified the ionizing source as an O5 star and found the dust in a shell and suggested that the dust was being blown out by stellar winds. In several cases, Watson et al. (2008, 2009) found evidence suggestive of star formation being triggered by the H\,{\sc{ii}} region expanding into and compressing the surrounding interstellar medium (ISM). Zavagno et al. (2007) identified  a leaky structure in RCW 120 as gas ionized beyond the circular ionization front--PDR interface.

The effect of dust on H\,{\sc{ii}} regions has been studied both observationally and theoretically for several decades. Harper \& Low (1971) showed that there was an excess of IR from H\,{\sc{ii}} regions compared with what they would predict from radio continuum if dust were absent inside the H\,{\sc{ii}} region. Of course, one solution is to posit dust in the ionized gas which could absorb a significant amount of Lyman continuum photons. Several studies have also found that a dust cavity is necessary to explain the IR emission toward H\,{\sc{ii}} regions (Chini et al. 1986, 1987; Churchwell et al. 1990; Faison et al., 1998; Ghosh et al., 2000; Inoue, 2002). Predictions of the cavity radius range from $\sim$0.03 pc (Faison et al., 1998) to 0.3 pc (Inoue, 2002). However, these studies were limited by the lack of high-resolution broadband imaging at wavelengths of $\sim$10-100 $\mu$m where the dust inside the H\,{\sc{ii}} region should peak.

Several studies of PDRs have found continuum emission due to very small grains (VSGs) to dominate at $\lambda \gtrsim$10 $\mu$m in some regions (Cesarsky et al. 2000, Rapacioli et al. 2005, Berne et al. 2007). Based on Spitzer-IRS observations of PDRs, Berne et al. (2007) concluded that VSGs had a continuum component at $\lambda \sim$40 $\mu$m that dominated other emission. However, they also concluded that the VSGs were located further from the exciting star than the PAH molecules associated with the PDR. Partially based on the relative position of the PAH and VSG components, they concluded that the PAH molecules were a product of UV-radiation dissociating the VSGs. The mid-IR identified bubbles studied here were chosen, in part, based on the clear separation of 8 $\mu$m (PAH-dominated) emission surrounding 24 $\mu$m emission with the aim of identifying emission associated with a different grain population.  Based on ISO observations of M17, Jones et al. (1999) concluded that VSG survived inside the HII region. However, Crete et al. (1999) used ISO and CFHT observations of M17 to model the IR emission of several dust components, including big grains, VSG and PAHs. Based on their Figures 10(a) and 11(a), we conclude that big grains (BGs) appear to dominate emission inside the H\,{\sc{ii}} region at $\lambda \gtrsim$20 $\mu$m.

Several recent studies of gas and YSOs near H\,{\sc{ii}} regions have found evidence of triggered star formation. Deharveng and collaborators have studied several sources: Sh 217 (Deharveng et al., 2003), Sh 219 (Deharveng et al., 2003), Sh 104 (Deharveng et al., 2003), RCW 79 (Zavagno et al., 2006), Sh2-219 (Deharveng et al., 2006), RCW 120 (Zavagno et al. 2007, Deharveng et al. 2009), Sh2-212 (Deharveng et al., 2008) and RCW 82 (Pomar{\`e}s et al. 2009). They have identified 8 $\mu$m emission and molecular line emission consistently surrounding ionized gas (usually traced by H$\alpha$). In several sources (see Sh2-104 and RCW 79), they have identified an overdensity of YSOs coincident with dense gas, traced by molecular line emission (usually CO J=1-0). They interpreted several of these YSOs as being triggered by the H\,{\sc{ii}} region expanding into the surrounding ISM. In Sh 104 and RCW 79 molecular gas was observed elongated along the ionization front, as expected if the collect-and-collapse model is dominant. Kang et al. (2009) have observed CO emission surrounding an 8 $\mu$m double shell near W51A (N102 and N103 in the Churchwell catalog). They also identified YSOs coincident with the molecular emission and concluded that the YSOs may be triggered by the H\,{\sc{ii}} region expanding into the surrounding ISM.

Definitively identifying YSOs as being triggered by an expanding H\,{\sc{ii}} region, however, is difficult. Because of the short timescale of massive star formation and evolution, H\,{\sc{ii}} regions are frequently surrounded by young protostars that are likely coeval and not triggered (e.g., KR 140; Kerton et al. 2008). The studies cited above frequently invoke the collect-and-collapse model (Elmegreen \& Lada 1977), although radiation-driven implosion (RDI, Bertoldi 1989) is also a possible mechanism. Briefly, collect-and-collapse posits that stellar winds sweep up surrounding material (collect), which increases in density until some fragments become gravitationally unstable (collapse). Elmegreen \& Lada (1977) only propose this mechanism to explain the formation of OB stars. Most methods of identifying YSOs, however, are unable to distinguish massive and low-mass YSOs with precision. RDI posits that UV radiation, as it creates an H\,{\sc{ii}} region, can accelerate and, perhaps, compress already existent dense clumps, inducing star formation. As Bertoldi (1989) explains, however, it is possible for the dense clump to be in pressure equilibrium with the UV radiation and not collapse or for the dense clump to be gravitationally unstable before exposure to UV radiation, making the RDI somewhat incidental to the star formation process. The observational studies above typically confirm that the mechanism cited is physically reasonable, either by using timescale arguments or comparing the approximate density of the molecular clumps with that predicted by the proposed mechanism. Although these arguments may establish that it is physically reasonable for triggered star formation to have produced the identified YSOs, the possibility that the YSOs would have formed without an expanding H\,{\sc{ii}} region is untested. 

To establish evidence against this competing hypothesis, this study will analyze the YSO distribution relative to the PDR (as identified by 8 $\mu$m emission). If expanding H\,{\sc{ii}} regions commonly trigger star formation, then the YSO distribution should be correlated with the PDR. Such an analysis on a single object is difficult or impossible, however, because the number of identified YSOs is typically in the tens. Instead, we present a population study of several mid-IR identified bubbles. We also measure 24 $\mu$m and 70 $\mu$m emission from the MIPSGAL survey toward MIR-identified bubbles and calculate the dust temperature and temperature gradient. In $\S$2 we introduce the survey data analyzed and bubble selection criteria. In $\S$3 we describe how we identified YSOs toward each bubble and measured the MIR integrated flux density. In $\S$4 we analyze the YSO distribution around each bubble and the dust temperature distribution in the ionized gas. In $\S$5 we discuss the YSO distribution of three typical bubbles. In $\S$6 we summarize our conclusions.

\section{Data}
We have used data from the 2MASS All-Sky Point Source catalog (bands JHK), the GLIMPSE point-source catalog (3.6, 4.5, 5.8, and 8.0 $\mu$m; see the GLIMPSE Data Products Description\footnote{http://www.astro.wisc.edu/glimpse/glimpse1\_dataprod\_v2.0.pdf}) and the MIPSGAL (24 and 70 $\mu$m) survey. The GLIMPSE catalog includes 2MASS point-source fluxes band-merged with the IRAC fluxes. MIPSGAL is a survey of the Galactic plane ($|$l$| <$60$^\circ$,$|$b$| <$1$^\circ$) using the MIPS instrument aboard the Spitzer Space Telescope. We have used the second release of the MIPSGAL mosaics at 24 $\mu$m. At 70 $\mu$m, we have used the mosaics created by the Space Science Center. No additional corrections for artifacts were made and thus, all flux was conserved in the mosaicing process.

We chose 15 bubbles to perform temperature analysis and 46 bubbles to analyze the associated YSO population. The first group was chosen from the catalog of Churchwell et al. (2006) based on the criterion that there was detectable 24 and 70 $\mu$m emission from the MIPSGAL survey that was clearly associated with the interior of the bubble. Most of the bubble catalog sources either showed no emission at 70 $\mu$m, data were not currently released or emission (24 or 70 $\mu$m) was confused with nearby emission, making isolating the emission associated with the bubble interior impossible. The second group of bubbles was chosen from the Churchwell et al. (2006) catalog with the following criteria: the bubble radius was between $\sim$1' and $\sim$2', there was an associated object, usually a maser, with a velocity measurement available and the 8 $\mu$m emission appeared circular. Velocity measurements, citations and kinematic distances (using the rotation model of Brand \& Blitz, 1990) are given in Table \ref{ysobubbles}. The second sample was arbitrarily limited to 46 bubbles for tractability. Since both these samples were initially identified by their mid-IR emission morphology, this sample may not be statistically representative of compact H\,{\sc{ii}} regions. 

\section{Results}
\begin{figure}
\epsscale{0.8}
\plottwo{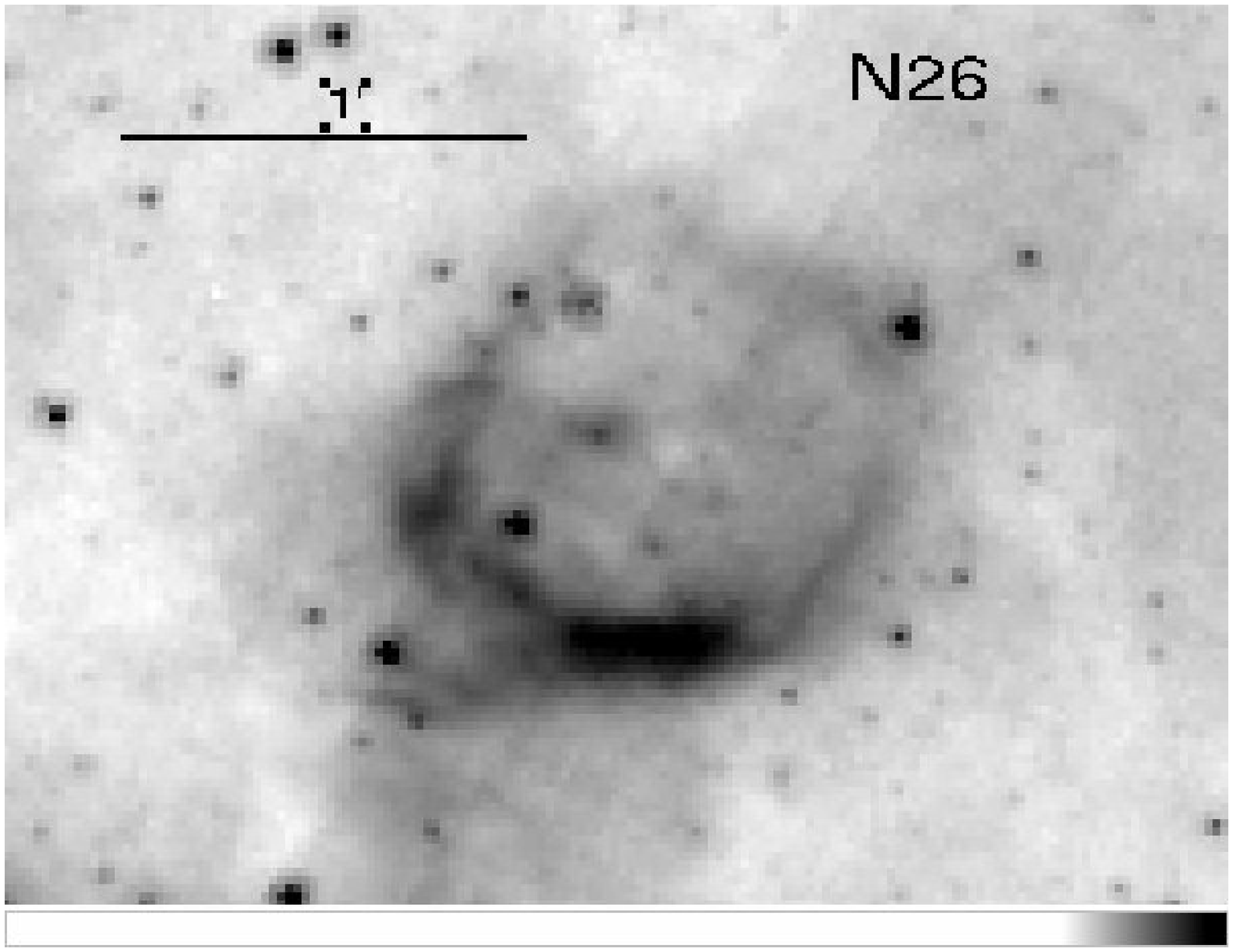}{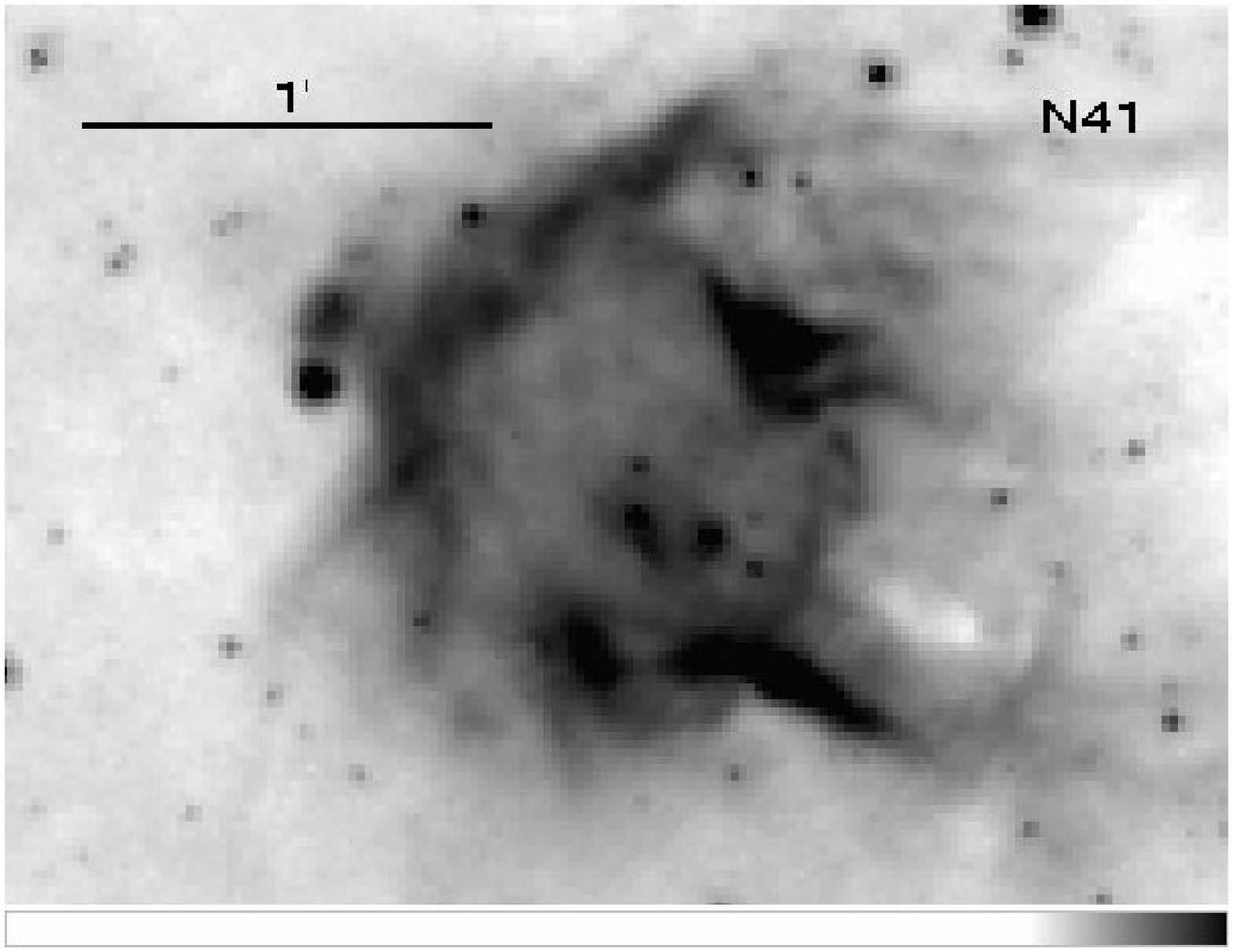}
\plottwo{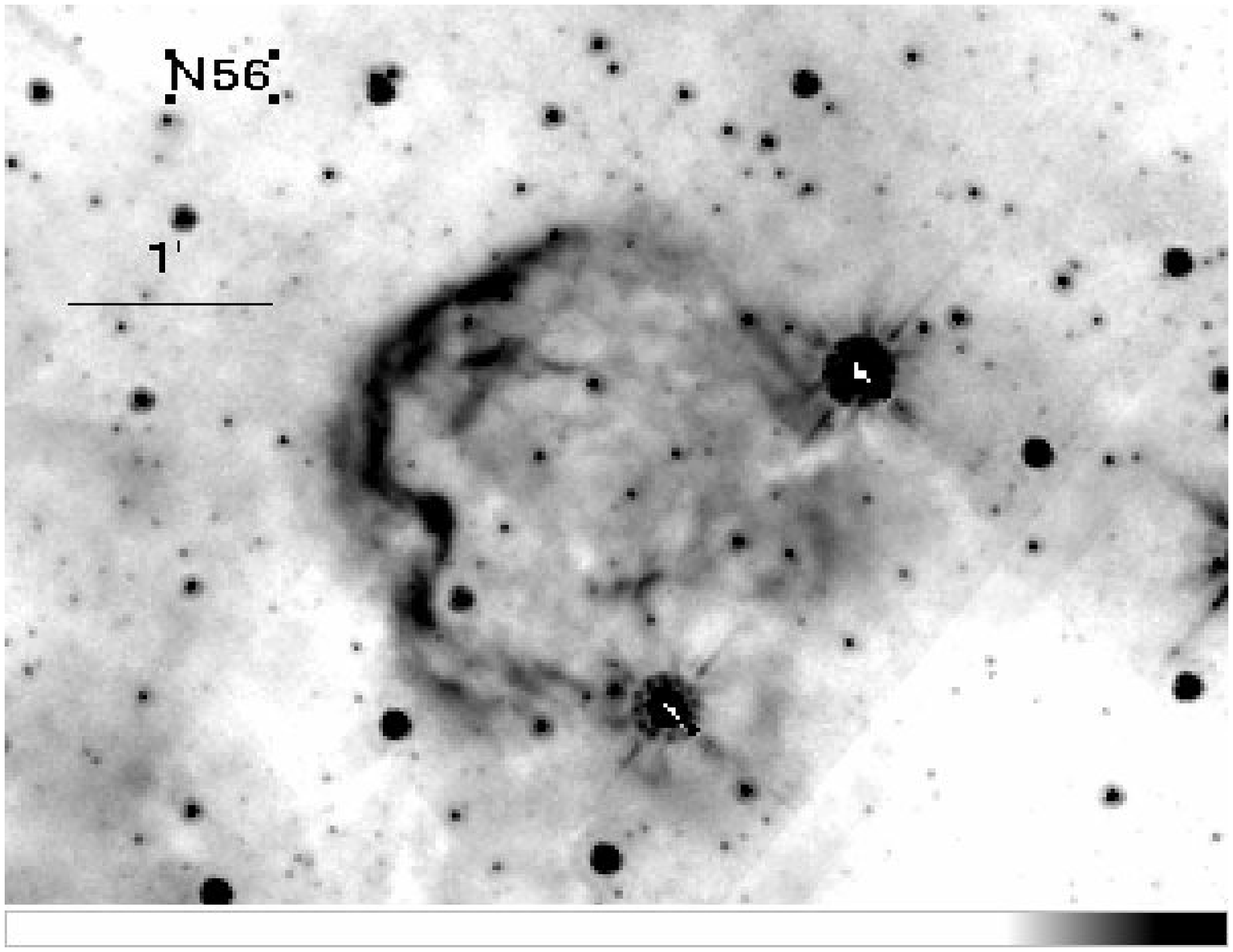}{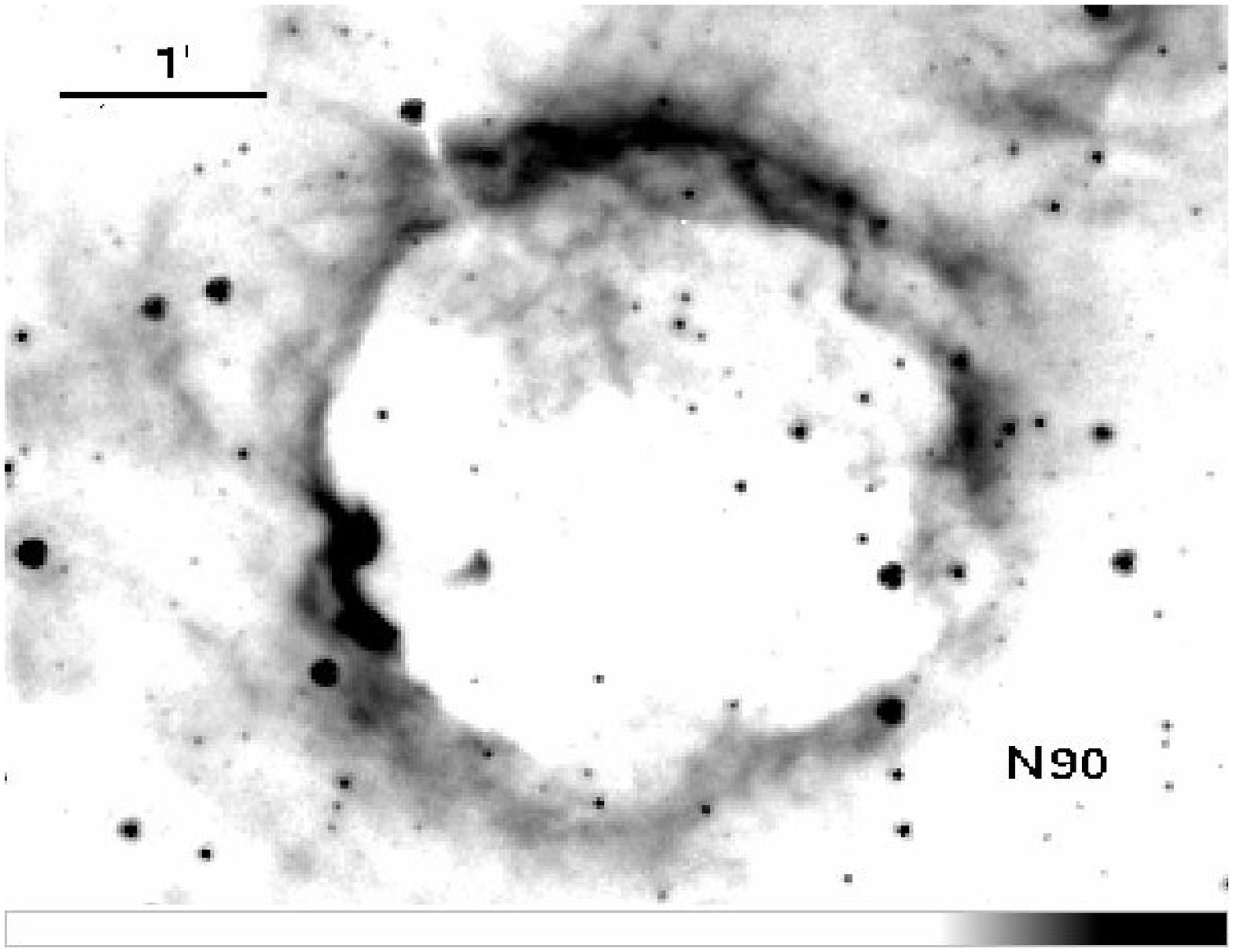}
\plottwo{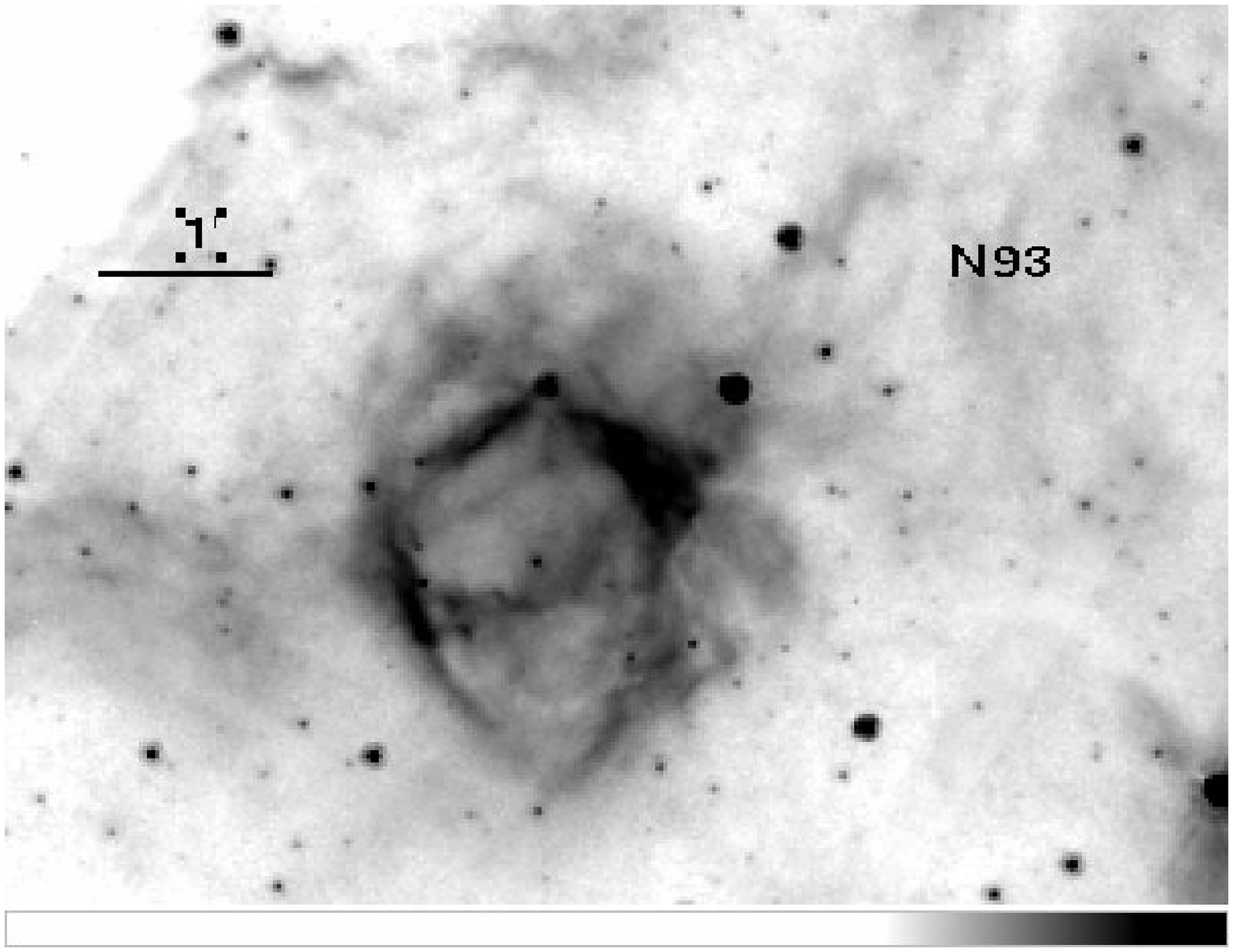}{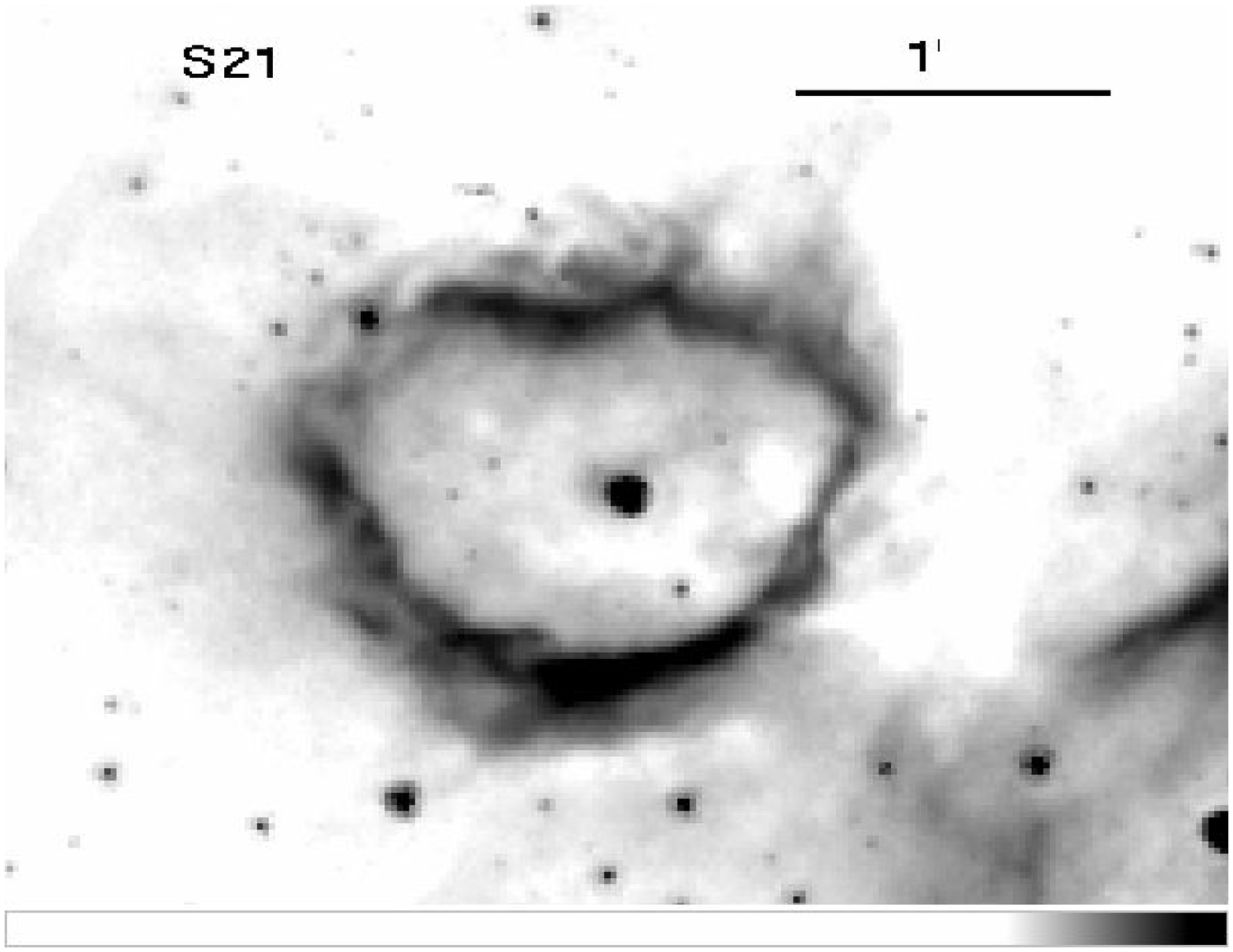}
\plottwo{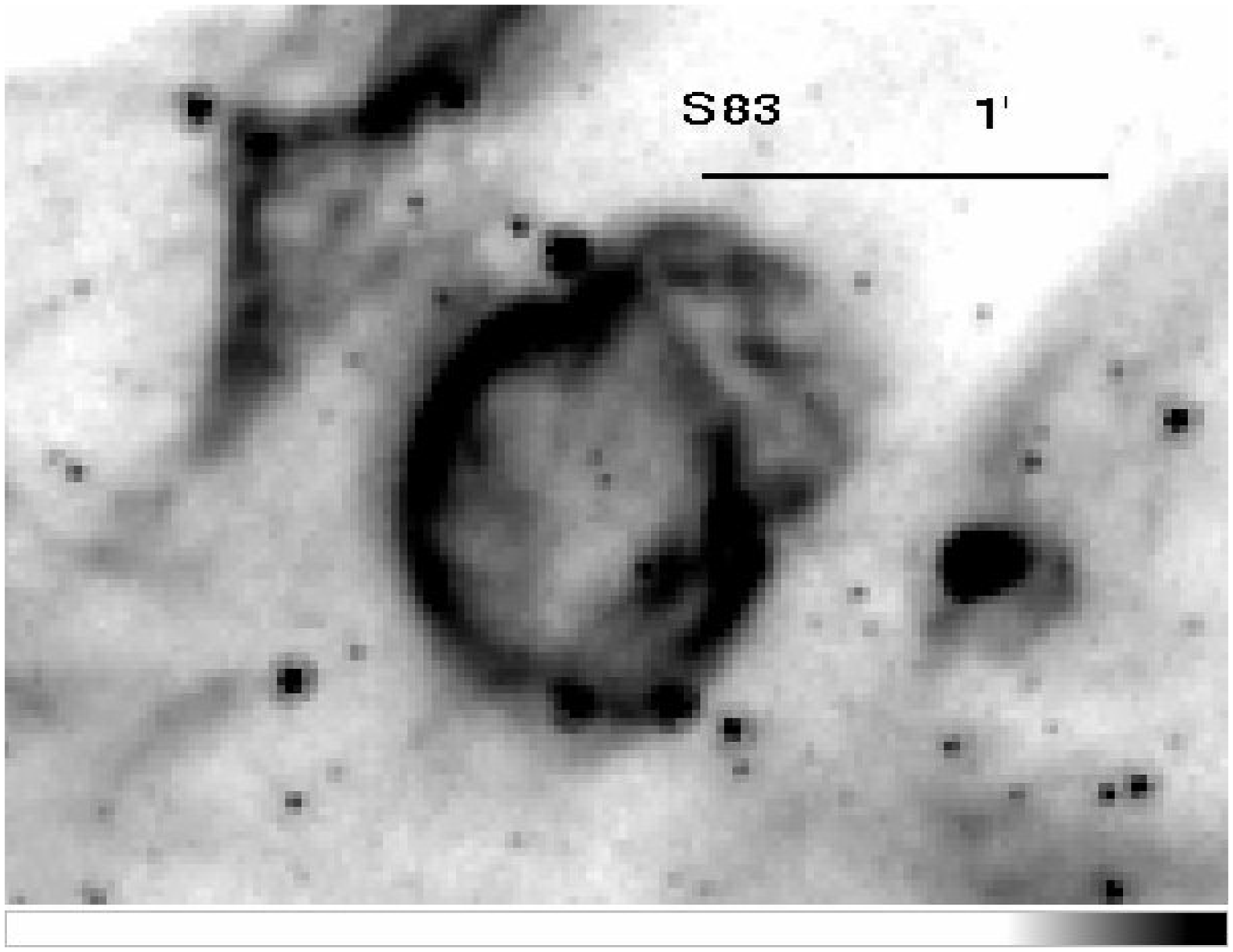}{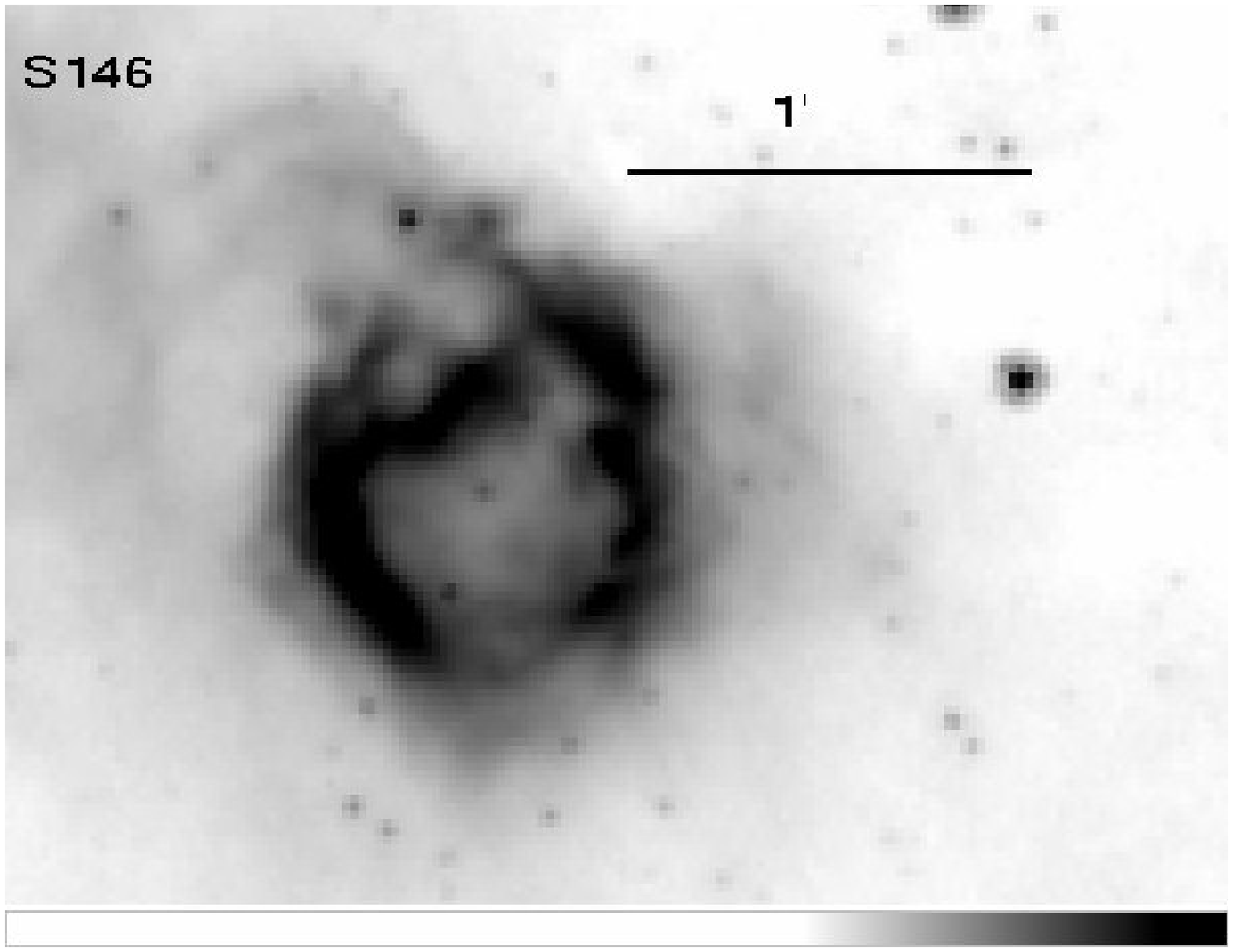}
\caption{8 $\mu$m emission from the GLIMPSE survey of 8 morphologically simple bubbles from the Churchwell et al. (2005) catalog. From upper-left by rows: N26, N41, N56, N90, N93, S21, S83, S146}
\label{bubbleim}
\end{figure}

\begin{figure}
\epsscale{1}
\plottwo{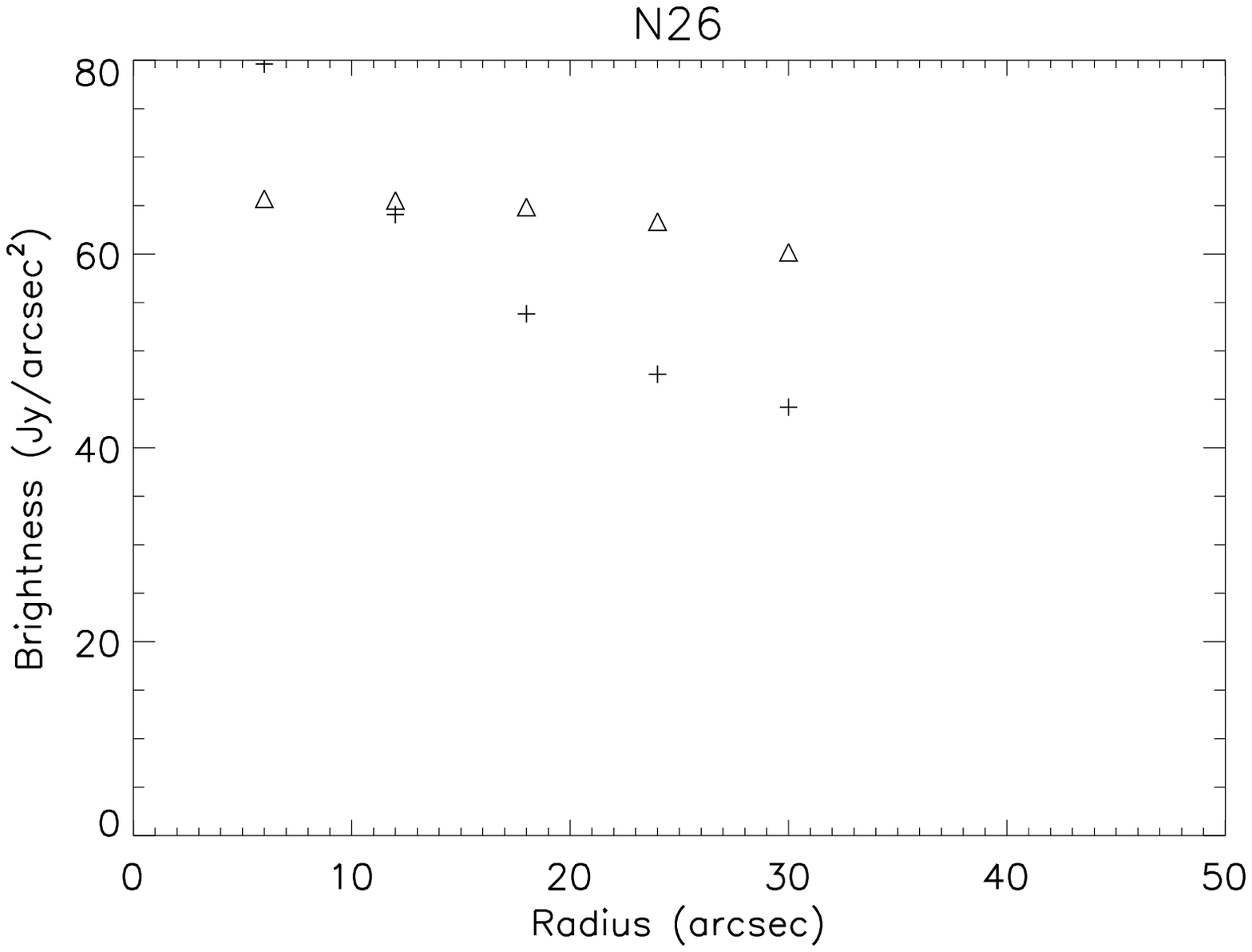}{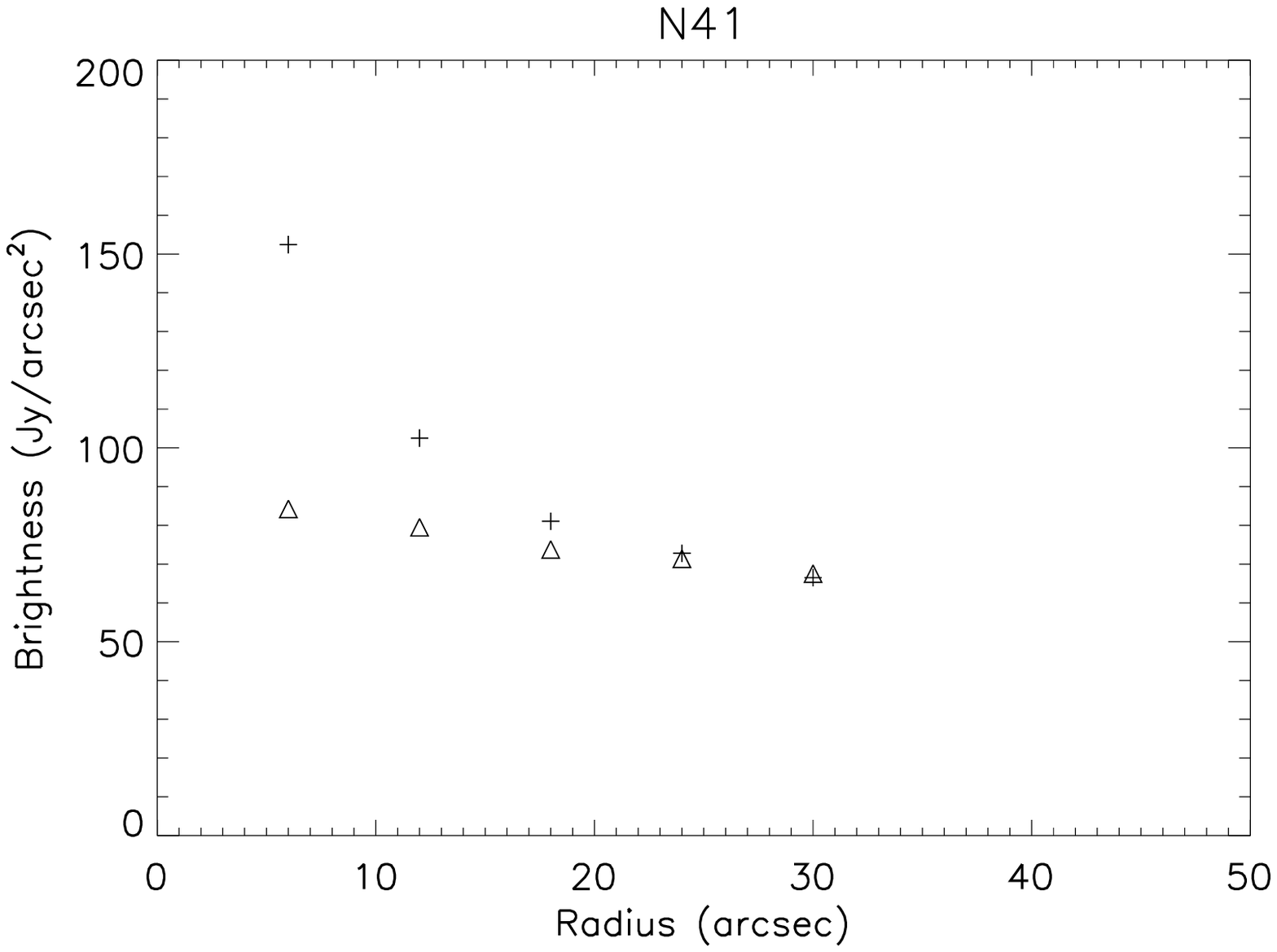}
\plottwo{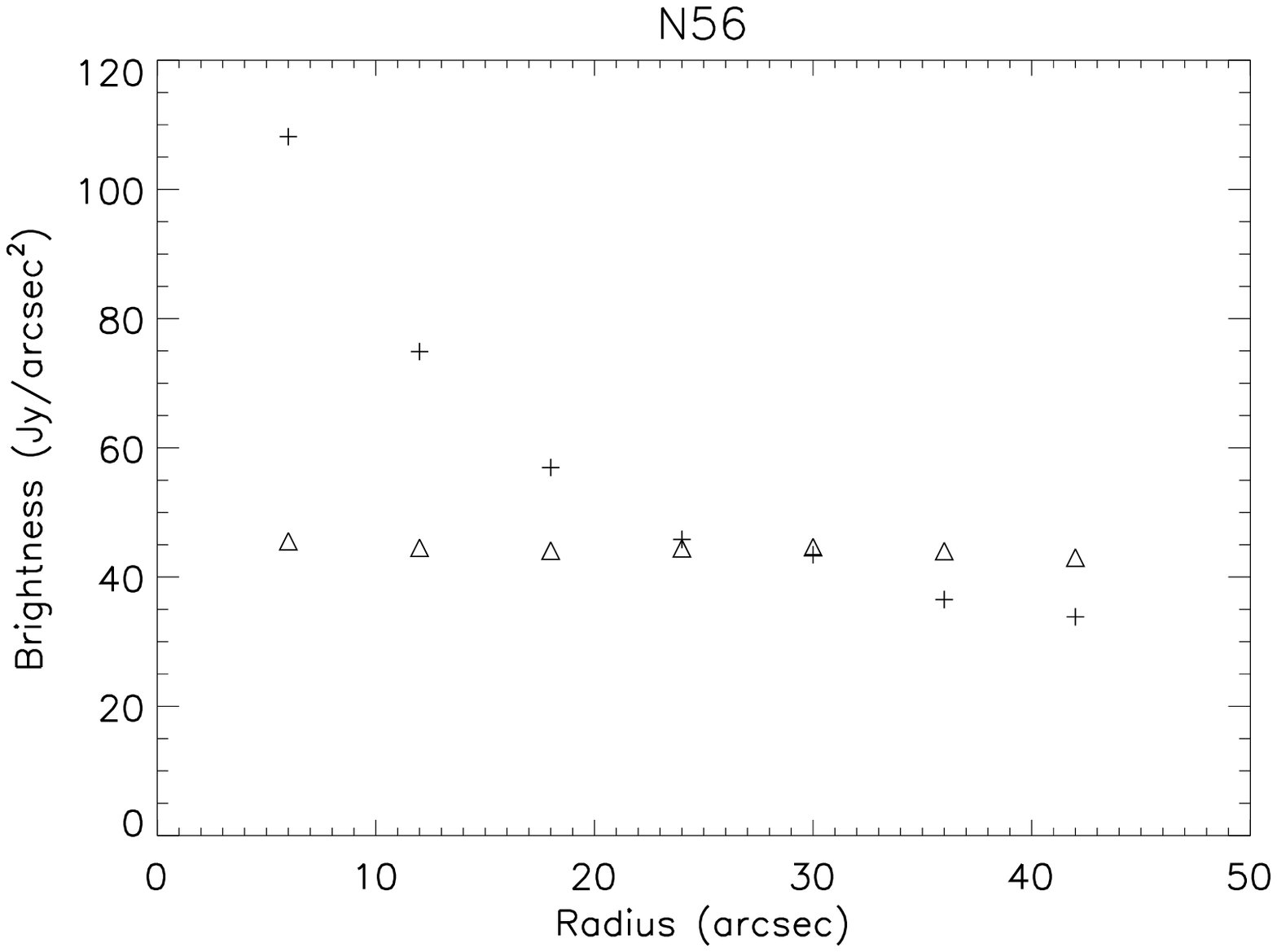}{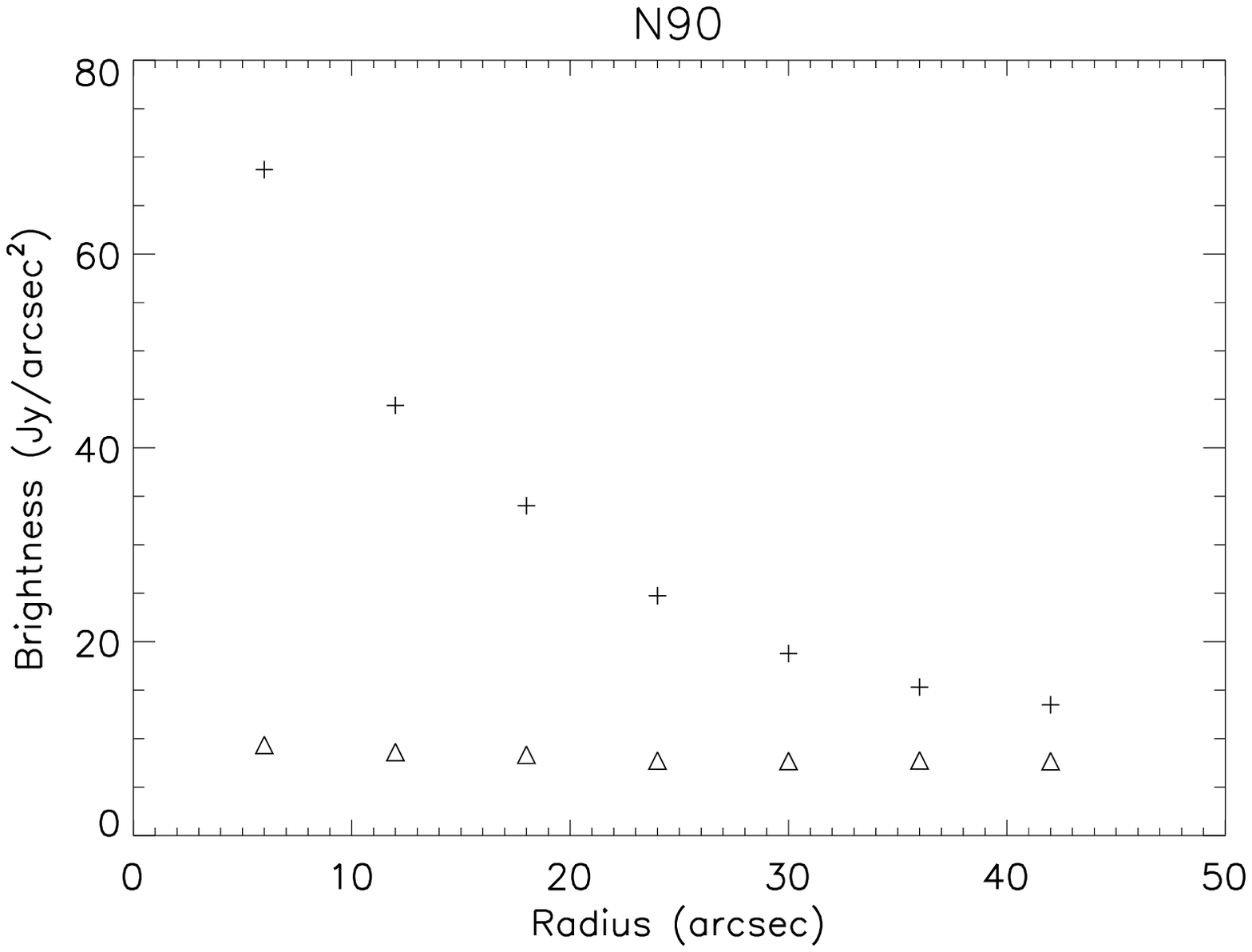}
\plottwo{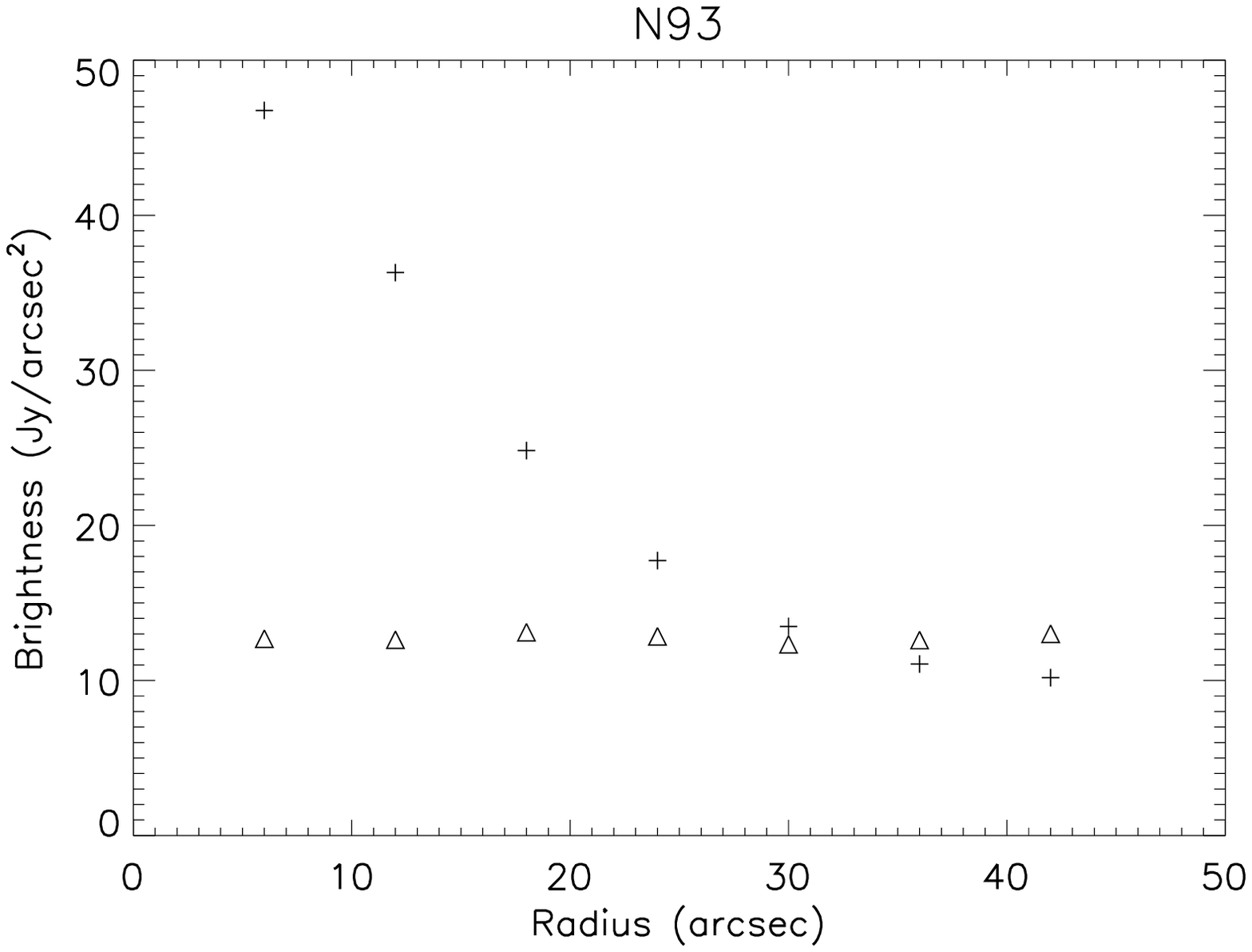}{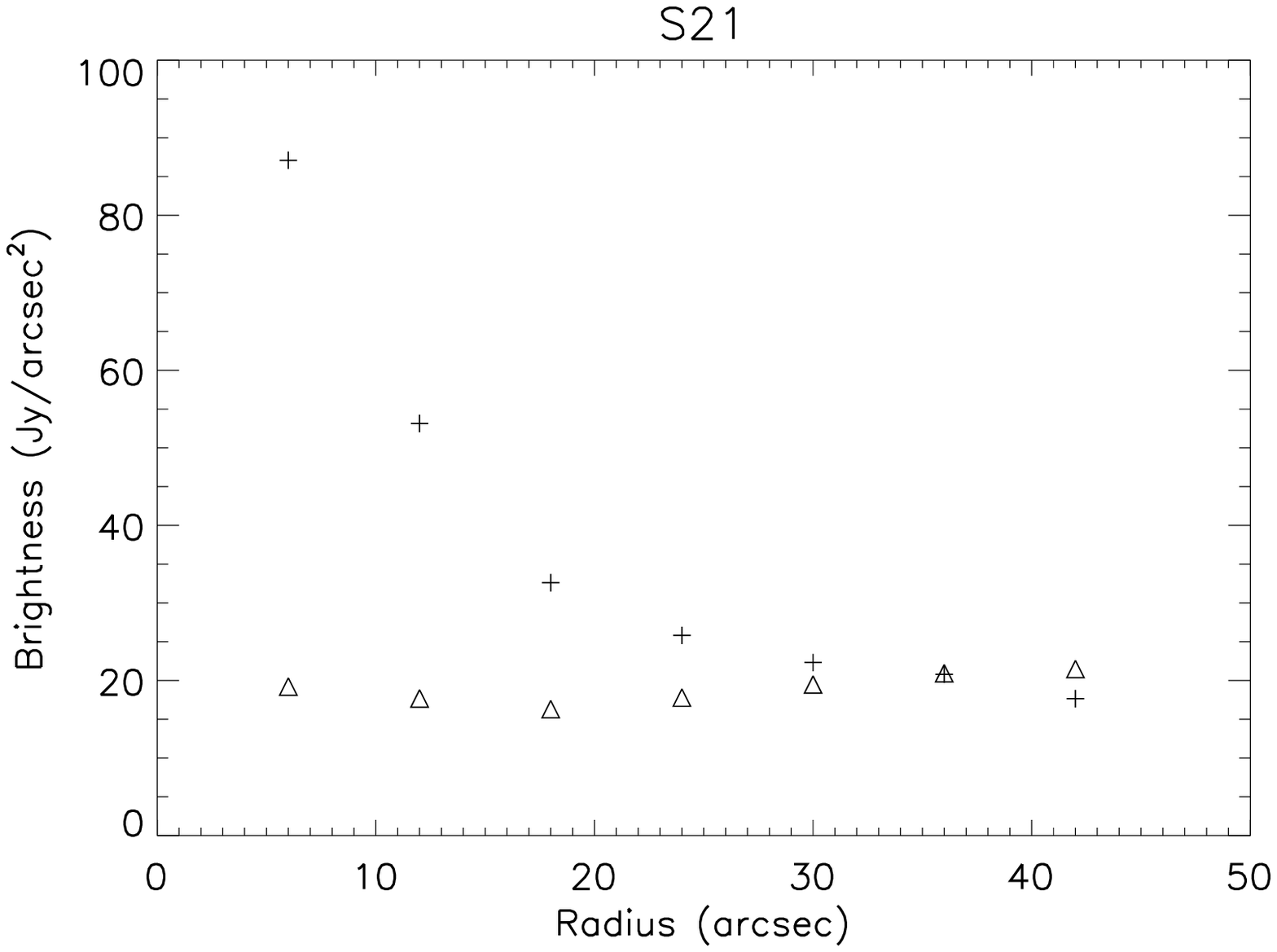}
\plottwo{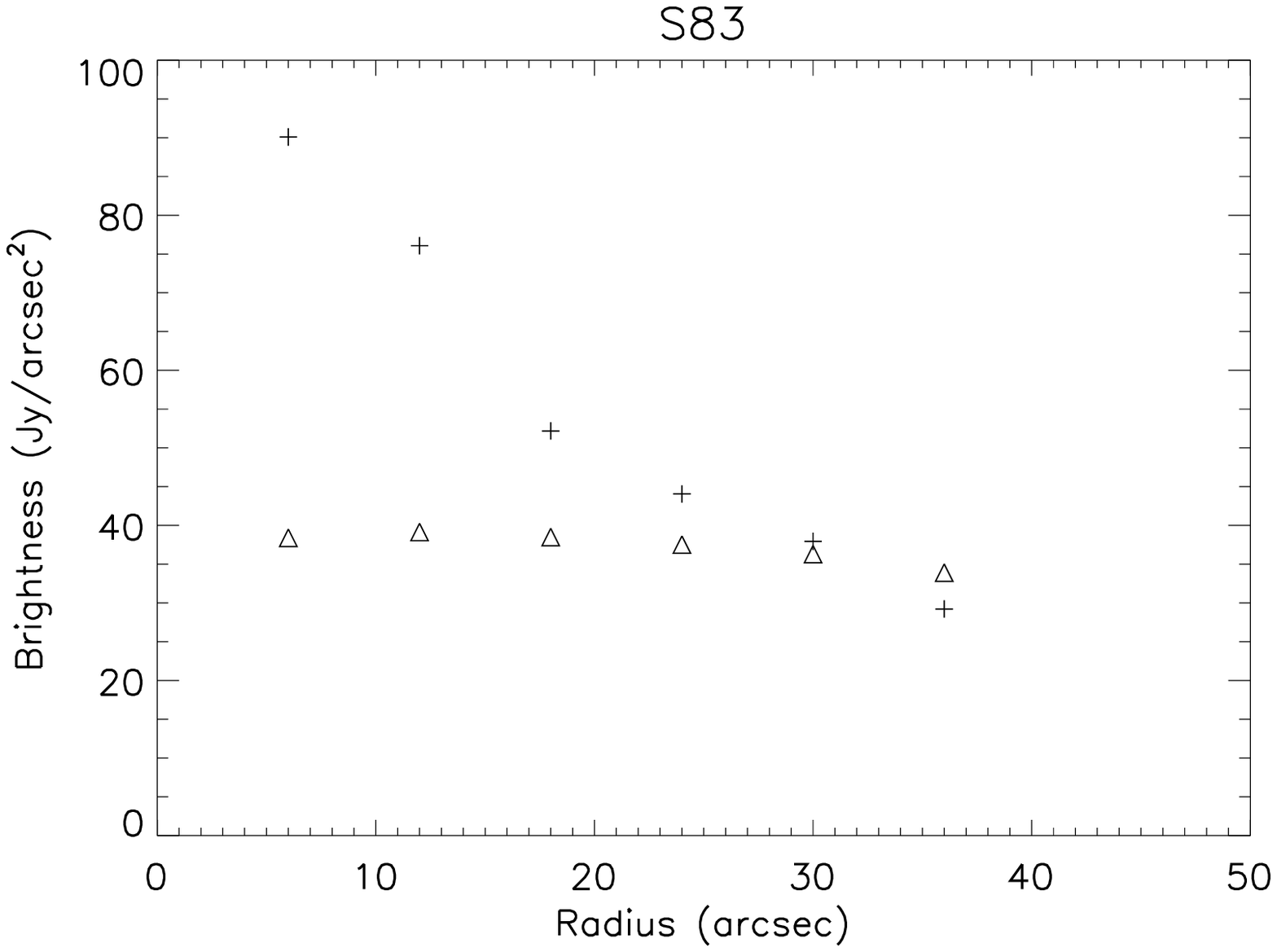}{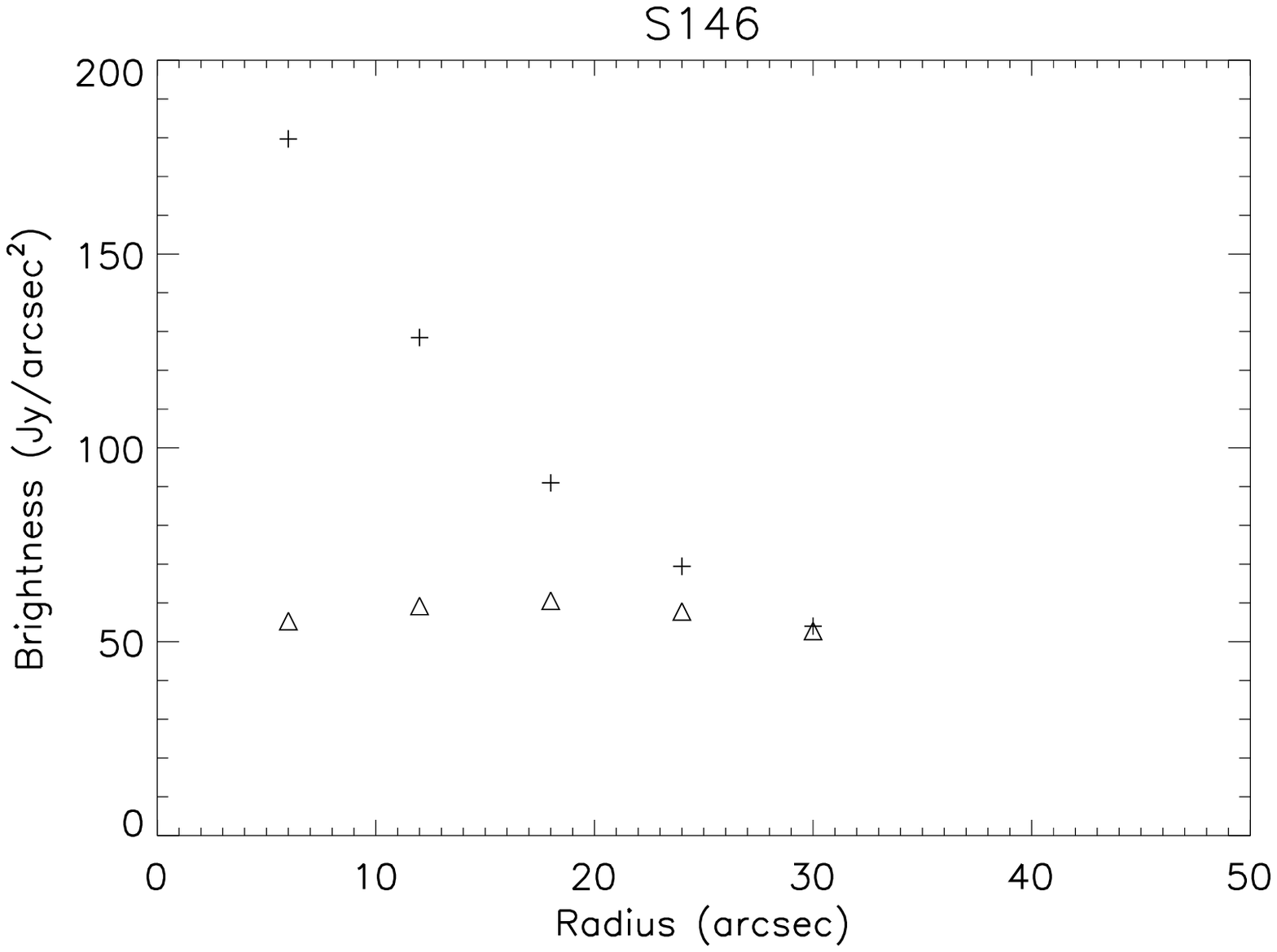}
\caption{Brightness of nearly symmetric bubbles averaged over concentric annuli in 24 (crosses) and 70 (triangles) $\mu$m emission.}
\label{fluxprofile}
\end{figure}

For each source in Table \ref{tempbubbles}, we used the ds9 software package to measure the integrated flux in a circle centered on each bubble with a radius equal to the inner-radius of the 8 $\mu$m shell from the Churchwell et al. (2006) catalog (typically between 1' and 2'). For all measurements, a background was subtracted from the flux measurement. 24 $\mu$m and 70$\mu$m background flux were measured in an equal-sized circle offset by 2' to 4' in the direction of the telescope slew. Multiple background measurements were made at different positions and found to vary at the 10\% level. In the 70 $\mu$m band from MIPSGAL, the background was clearly observation time-dependent, so subtracting a background in the telescope slew direction was critical to ensure a background measurement taken shortly before or after the source observation. Results of the flux measurements are given in Table \ref{tempbubbles}. Eight sources have a simple enough morphology that we were able to analyze the flux density in more detail (see Figure \ref{bubbleim}). For these sources, we measured the average brightness within concentric annuli with thickness of 6". The annuli thickness was chosen to match the resolution of the observations at 24 $\mu$m. Individual point sources within the diffuse emission were excluded from these measurements. Results for these eight sources are shown in Figure \ref{fluxprofile}. No sources show any emission decrease at small radii which would indicate a dust-free cavity. Although we do not have distances toward these sources, the absence of any cavity at a resolution of 6" and 15" is somewhat surprising if cavities of 0.3 pc ($\sim$ 20" at 3 kpc) are common. It may be that only early, hot stars are able to form dust cavities or that winds are not dominant in these sources (Marcolino et al. 2009).

For each source in Table \ref{ysobubbles}, we used the method of Robitaille et al. (2006, 2007) to identify all candidate YSOs within three bubble radii. This search area was choosen to include as many triggered YSOs as possible while limiting contamination by other sources along the line of sight. Briefly, Robitaille et al. (2007) simulated IR observations using a grid of Monte-Carlo radiative-transfer models of YSOs (Whitney et al., 2003) with a range of stellar masses, luminosities, disk masses, mass accretion rates and line-of-sight inclinations. Observations from J to 8 $\mu$m were fit using a $\chi^2$-minimization technique. First, the spectral energy distributions (SEDs) for all sources were fit with Kurucz models (Kurucz, 1993), as was done in Watson et al. (2008, 2009). Those sources that were poorly fit ($\chi^2 \ge$ 3) were then fit with the YSO model SEDs. The range of models that fit the observations ($\chi^2 \le$ 3) places limits on the possible YSO properties. Since we do not have 24 $\mu$m fluxes for these sources, however, we are unable to place strong limitations on mass or evolutionary class/stage. In Figure \ref{yso} we show three typical bubbles with the candidate YSOs marked. The average number of candidate YSOs identified surrounding each bubble is 35, but there is a wide range with a minimum of 4 (N89) and maximum of 161 (S11). It should be noted that there is likely significant contamination in this population of YSOs that are not associated with the region surrounding the bubble. Povich et al. (2008) studied M17 using a similar method and found $\sim$50\% of YSOs were either foreground or background YSOs.

\section{Analysis}
In the following subsections, we analyze the temperature structure and YSO distribution of several bubbles. In \S 4.1 we measure the average temperature of dust interior of the PDR radius for 15 bubbles and analyze the temperature structure of 8 circularly symmetric bubbles. In \S 4.2 we measure YSO distribution and discuss the results within the context of triggered star formation.

\subsection{Temperature Structure}
For each source in Table \ref{tempbubbles}, we fit a modified blackbody curve to the 24 and 70 $\mu$m fluxes to determine the average temperature of dust interior to the PDR (which is identified by 8 $\mu$m emission). We assumed the dust emission as follows:
\begin{eqnarray*}
S_\nu &=& A \left(\frac{\nu}{\nu_0}\right)^\beta B_\nu (T),
\end{eqnarray*}
where we assume that $\beta$ is 2 (Draine \& Li 2007). Although several studies have measured a variance in $\beta$ (Meny et al 2007, Dupac et al. 2002, Li et al. 2003), Netterfield et al. (2009) used simulations to predict that their temperature calculations would vary by 10\% if $\beta$=1.5 instead of $\beta$=2.0. Although a population of VSGs that deviates strongly from $\beta$=2 emission may exist, we anticipate that the large grains that are in radiative equilibrium dominate the emission inside the H\,{\sc{ii}} region at these wavelengths. Results are shown in Table \ref{tempresults}. The average dust temperature is 80 K and ranges between 49 K and 113 K. Eight sources are large enough and symmetrical enough to allow a more detailed analysis. We fit modified blackbody curves to the average brightness at 24 and 70 $\mu$m measured in concentric annuli for these sources. Results are shown in Figs \ref{n26}-\ref{s146}. For each source, the temperature decreases monotonically, as expected if the dust is heated by a central source.  The decrease, however, is very shallow, i.e. a weak dependence on radius. We fit these results with a power-law of the form:
\begin{eqnarray*}
T(r) &=& \frac{C}{r^\alpha}
\end{eqnarray*}
where T is the calculated temperature, r is the angular distance from the bubble center, and C and $\alpha$ are free parameters. Fit results are given in Table \ref{tempprofiles}. The average value of $\alpha$ is 0.12 and ranges between 0.04 and 0.19.

A simple model of the radial dependence of grain temperature is presented by Osterbrock \& Ferland (2006) which incorporates heating by radiation and collisions with gas particles as well as cooling by thermal radiation. They conclude that the grain temperature should go as:
\begin{eqnarray*}
T_D &\propto& \left(\frac{L}{4\pi r^2 a}\right)^\frac{1}{5}
\end{eqnarray*}
where L is the central stellar luminosity, r is the grain distance and a the grain radius. In the energy balance equation, the cooling has a T$^4$ dependence due to the Plank function and the collisional heating has a T dependence, resulting in the $\frac{1}{5}$ exponent in the above equation. The power-law fits in Table \ref{tempprofiles} are significantly shallower (lower $\alpha$) than this simple prediction ($\alpha$ = 0.4). There are several possible explanations for this difference. First, we have calculated temperature with projected distance from the central source. That is, the calculated temperature is based on emission from a range of physical distances from the central source. The effect of this projection would be to flatten the temperature dependence on projected distance, i.e. bring the theoretical prediction closer to agreement with the observations. Second, the grain size distribution could be complex. Grain size strongly affects the predicted grain-temperature in the model of Osterbrock \& Ferland (2006). Our measurement of dust temperature is an average over all grain sizes, weighted according to their radiation. If the grain size distribution is radially dependent, this average would not have the simple radial behavior predicted above. Third, one or more heating or cooling mechanisms could be out of thermal equilibrium at some of the distances measured here. A computer simulation is likely necessary to determine if these explanations or other physics, such as Ly$\alpha$ or stochastic heating, are necessary to explain both the 24 $\mu$m and 70 $\mu$m emission.

\subsection{YSO Population and Triggered star formation}
The presence of several YSOs distributed inside the bubble, in projection, and more than 2 bubble radii away ($\sim$300") from the H\,{\sc{ii}} region, as traced by the 8 $\mu$m emission (e.g., see S17 or S23), suggests that at least some of the YSOs are not directly triggered by the expanding H\,{\sc{ii}} region. These YSOs may have been formed by the same mechanism as the star responsible for ionizing the H\,{\sc{ii}} region or may have begun forming at another epoch by some other mechanism. Some of the YSOs, however, may be triggered by far-UV flux leaking from the ionized region. If triggered star formation, either by the collect-and-collapse mechanism or the RDI mechanism, is the dominant process then we expect a significant fraction of the YSOs to be associated with the bright 8 $\mu$m shell. 

Using the candidate YSOs identified near the bubbles in Table \ref{ysobubbles}, we have calculated the number of YSOs in equal area annuli centered  on each bubble center, as reported in Churchwell et al. (2006). Our results for all the sources are shown in Figs. \ref{profile1}-\ref{profile6}. Also shown in each figure is the diffuse 8 $\mu$m brightness averaged over concentric annuli centered on the same point. Note that the 8 $\mu$m emission is more finely sampled than the YSOs because there are more 8 $\mu$m data. Nine of the bubbles show a significantly increased YSO density at the 8 $\mu$m emission peak: N65, N77, N82, N90, N92, N101, N117, N128 and S23. Thus, we conclude that at most 20\% of these expanding H\,{\sc{ii}} regions have triggered an amount of star formation observationally distinguishable from the "background" star formation.

We have also identified four bubbles that have a peak YSO distribution not coincident with the 8 $\mu$m shell. In Table \ref{ysobubbleprop}, we list these bubbles along with their galactic coordinates, angular radius, kinematic distance, physical radius and the number of YSOs observed along the 8 $\mu$m shell. In order to estimate the completeness of our YSO catalog, we have estimated what percentage of a Taurus population of YSOs we would have detected. We start with the catalog of Taurus YSOs used by Robitaille et al. (2007) and observed using IRAC on Spitzer by Hartmann et al. (2005) and Luhman et al. (2006). For each bubble  we then adjust the expected flux for each YSO by the bubble's kinematic distance (assuming a Taurus distance of 140 pc). We also adjust the expected flux of each YSO by adding extinction equivalent to the average YSO as fit by the SED fitter. Since each YSO had multiple models fit the SED (each with a different interstellar extinction), we calculated the average interstellar extinction of the well-fit models ($\chi^2 \le$ 3), weighted according to $\frac{1}{\chi^2}$. We assumed that the interstellar extinction curve of Indebetouw et al. (2005). We assumed all sources with estimated flux densities greater than 0.6 mJy, 0.4 mJy, 2 mJy and 10 mJy in the 3.6 $\mu$m, 4.5 $\mu$m, 5.8 $\mu$m and 8 $\mu$m band respectively would be detected. We then simulated the GLIMPSE catalog criterion by requiring detection in two adjacent bands. In Table \ref{ysobubbleprop} we report the percentage of the 29 Taurus YSOs that we estimate would be detected. The average percentage is 71\% and range between 41\% and 100\%. Based on these estimates, we conclude that our YSO sample for each bubble is complete enough to support our conclusions on triggered star formation correlations. 

\section{Discussion}
\subsection{YSO Distribution}

Triggered star formation has been clearly observed in some environments similar to the mid-IR identified bubbles studied here. However, it is not clear how common this process is around massive stars and, when present, what the spatial distribution is. We have grouped the sample studied here into four categories according to their  8 $\mu$m and YSO distribution. The first group, with 11 members, does not show an easily distinguished 8 $\mu$m emission peak and will not be discussed further. The second group, with nine members, has a YSO distribution peak that is coincident or nearly coincident with an 8 $\mu$m emission peak. The third group, with 22 members, has an 8 $\mu$m peak but no clear YSO distribution peak. The fourth group, with four members, has an 8 $\mu$m emission peak and a YSO distribution peak, but they appear clearly separated. 

\begin{figure}
\includegraphics[width=3in]{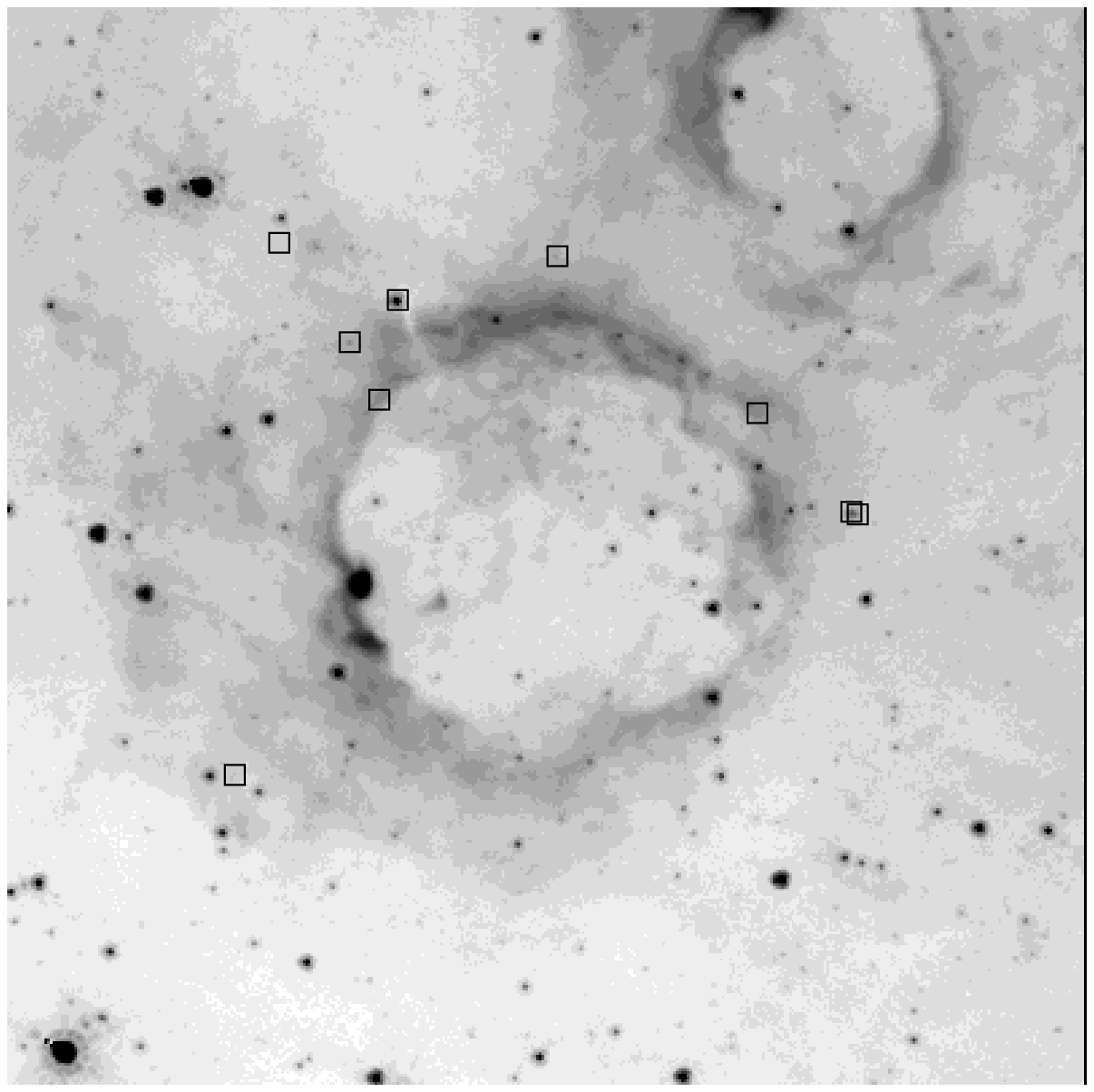}
\includegraphics[width=3in]{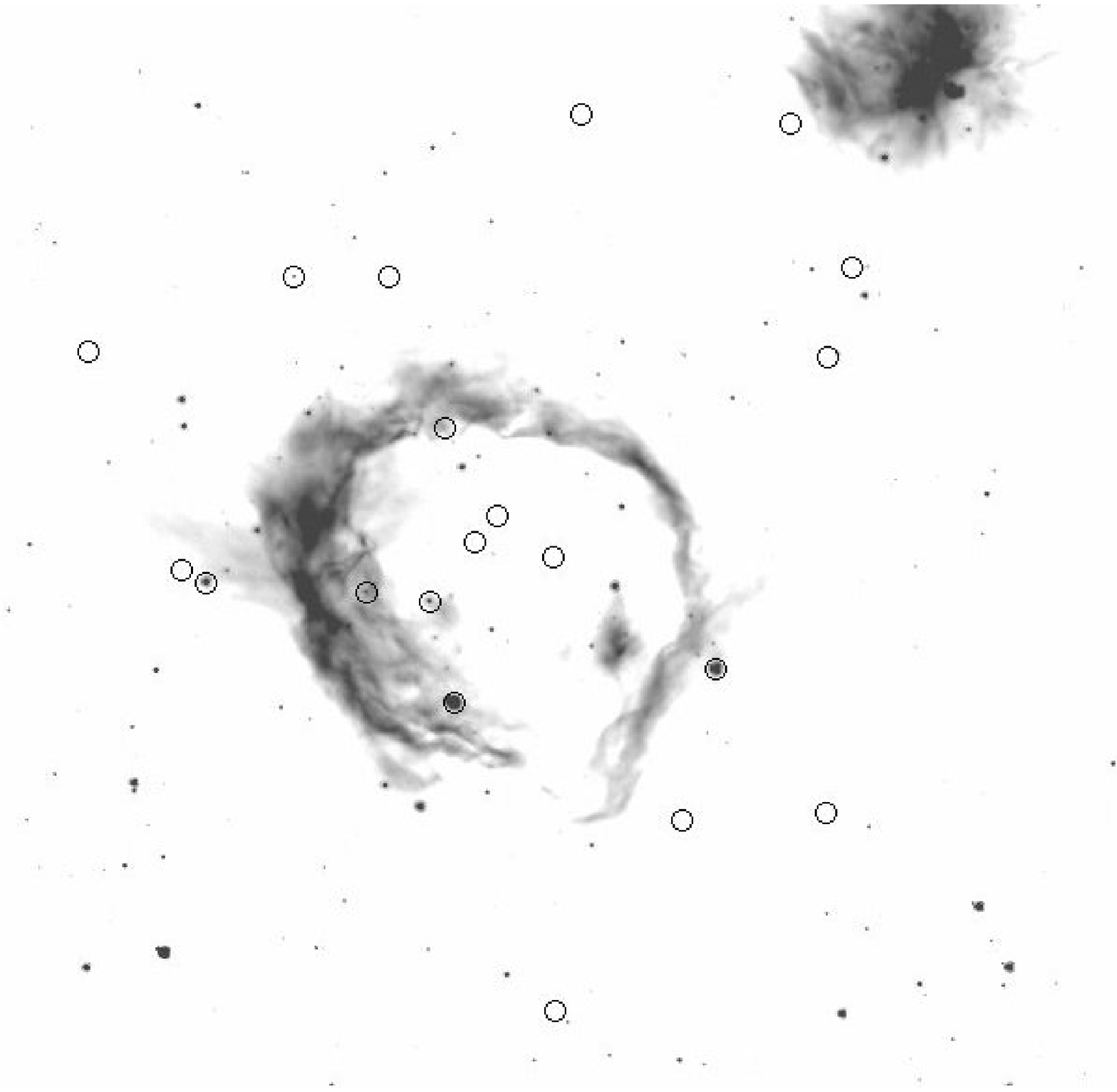}
\includegraphics[width=3in]{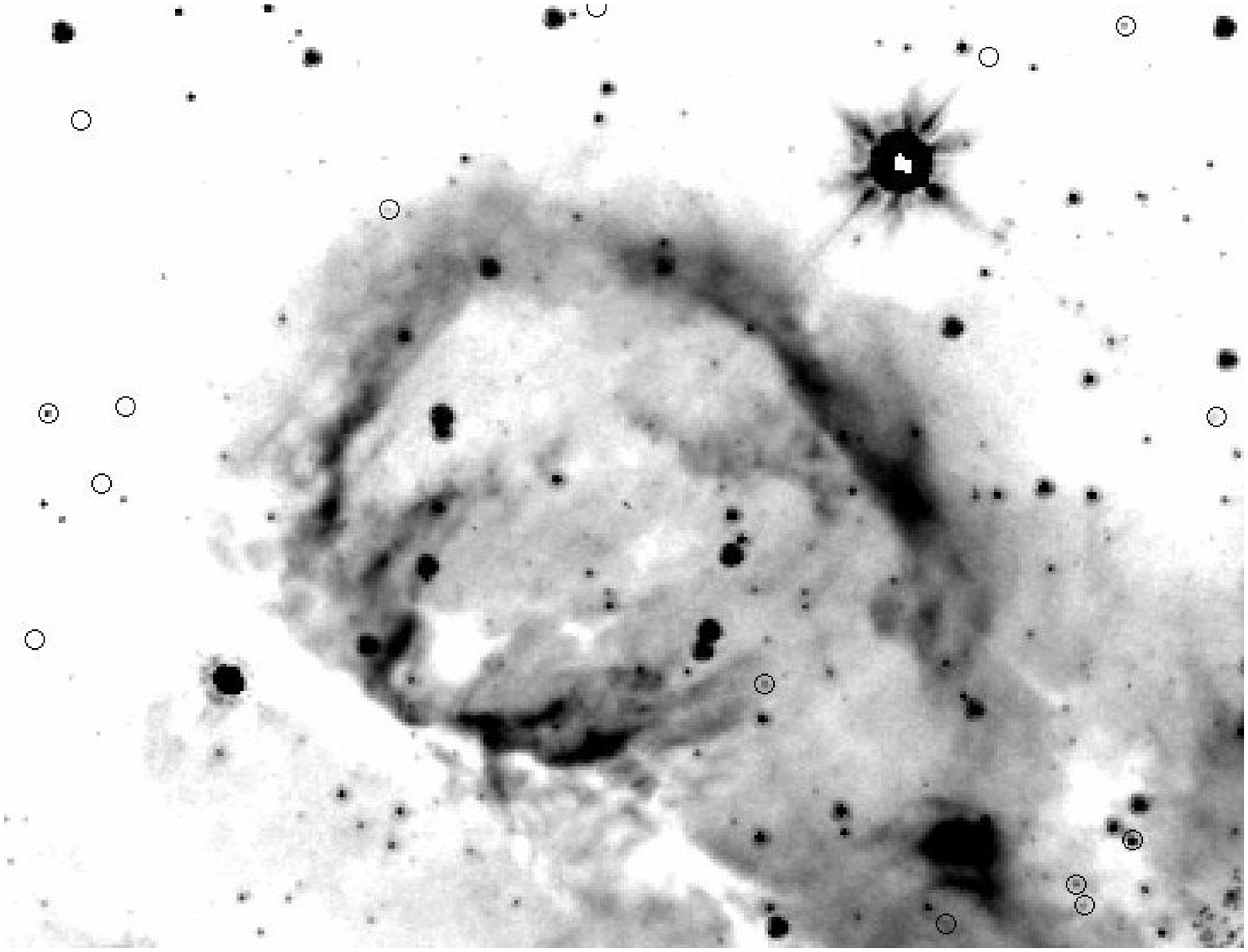}
\caption{Top (N90): 8 $\mu$m emission (greyscale) and candidate YSOs (squares), as identified using J-band to 8 $\mu$m photometry and the SED-fitting-method of Robitaille et al. (2007) (see $\S$3 for details); Middle (N4): Same; Bottom (N62): Same}
\label{yso}
\end{figure}

\begin{figure}
\psfig{file=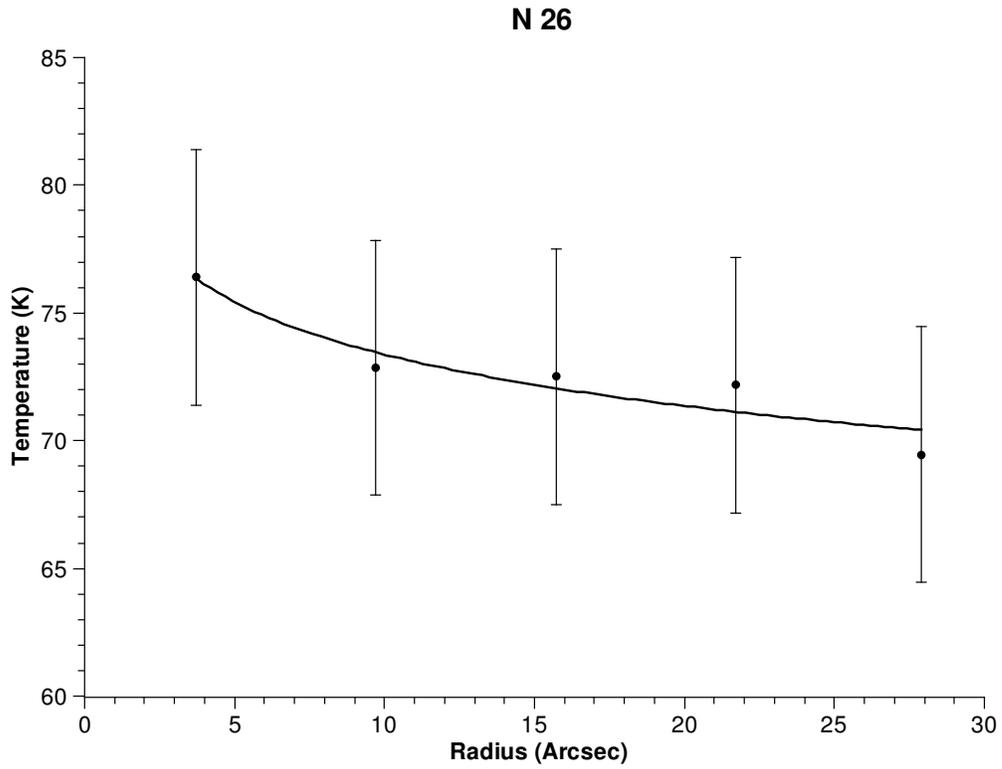,angle=-90,width=6in}
\caption{Temperature profile derived for N26. Temperatures are calculated by fitting modified blackbody curves to 24 and 70 $\mu$m emission integrated over concentric annuli. An error of 5 K is applied based on the uncertain background subtraction. See text for model fit parameters.}
\label{n26}
\end{figure}

The second group shows an 8 $\mu$m emission peak and YSO distribution peak that are coincident or nearly so. These members, listed in \S 4.2, are the best examples in this study of demonstrating potential triggered star formation. N90 is a typical member. In Figure \ref{yso}(top), the 8 $\mu$m emission and potential YSOs toward N90 are shown. The 8 $\mu$m emission forms a nearly symmetric circle, projected on the sky, with several YSOs coincident with this emission. The 8 $\mu$m emission is somewhat stronger on the upper half and, interestingly, the YSOs are concentrated on this part of the bubble, too. As is common for these bubbles, however, there are active regions surrounding the bubble, including N89 to the upper right of Figure \ref{yso}(top). If the collect-and-collapse mechanism is operating here, we expect molecular line emission would be coincident with the identified YSOs and would present an elongated morphology along the shell, as has been observed in other sources (see Zavagno et al., 2007 study of RCW 120).

The third group shows no clear YSO distribution peak and includes N4, N14, N54, N72, N74, N79, N80, N84, N89, N115, N124, N126, N127, N129, N130, N131, S11, S13, S14 and S29. A typical member is N4, whose 8 $\mu$m emission and coincident YSOs are shown in Figure \ref{yso}(middle). Several YSOs are identified, inside the shell, along the shell and outside the shell. As with N90, N4 is in a larger active region, in projection, and so associating any individual YSO with the expanding bubble is difficult. There does not appear to be an overdensity of YSOs along the shell or more than 2 or 3 YSOs associated with the brightest 8 $\mu$m emission. It should be stressed that these observations do not rule out the presence of triggered star formation toward N4. To do so would require a complete sample of the YSO population, a map of the molecular gas in several tracers and a map of the ionized gas. The interpretation offered here is only that we have not yet detected evidence of triggered star formation. This third group comprises 63\% of the bubbles with a distinct 8 $\mu$m emission peak, however. These results indicate that triggered star formation may be difficult to identify in most bubbles, either because of contamination in identifying YSOs or because the triggering mechanism does not dominate.

The fourth group show a clear 8 $\mu$m emission and YSO peak, but the two peaks are not coincident. This group includes only four sources: N62, N123, N133 and S17. We show the 8 $\mu$m emission and identified YSOs toward N62 as a typical example (see Fig \ref{yso},bottom). The YSOs present are $\sim$1' away from the shell (which has a radius of 1.4'). The closest YSOs are concentrated on the left side of the bubble. In an interesting contrast to N90, this side of the bubble shows weaker 8 $\mu$m emission. Although it may be possible that UV emission is leaking past the PDR, as traced by the 8 $\mu$m emission, it is also possible that the YSOs on the left side are not associated with N62 at all. Because of the small number of YSOs typically detected, it is also possible that the YSO distribution peak is a coincidental arrangement in projection.

To more precisely determine the relationship between the YSOs identified with the expanding H\,{\sc{ii}} region requires, above all, a census of the molecular gas toward each region. Collect-and-collapse makes specific predictions about the presence and morphology of molecular gas along the ionization front. If the third group discussed above is not triggering star formation, the molecular gas distribution around each source (or lack thereof) may help explain their different histories.

\section{Conclusions}
We conclude the following:\\
$\bullet$ Based on 24 and 70 $\mu$m emission detected in proximity to 15 bubbles, we calculate the average temperature of dust interior to the PDR to be between 49 K and 113 K.\\
$\bullet$ In eight morphologically simple sources, we calculated the dust temperature as a function of projected radius. As expected, the dust temperature decreases, but with a shallow power law between $\alpha$=0.05 and 0.19. A simple model of dust heating and cooling predicts a power law of $\alpha$=0.4.\\
$\bullet$ We find no evidence of a dust-free cavity, although several predictions of cavity radii are likely smaller than our resolution. These sources also may not be powered by late-O or early-B stars and not wind-dominated.\\
$\bullet$ Using the method of Robitaille et al. (2007), we have identified candidate YSOs located near 46 bubbles. In 20\% of the sources with a clear 8 $\mu$m emission peak, we observe a concentration of YSOs along the shell of the bubbles, one signpost of triggered star formation.\\

\acknowledgements
We acknowledge the helpful comments by the referees, which helped improve the paper significantly. C.W. and U.H. would like to acknowledge support through NASA/JPL Contract 1289406. A.M. would like to acknowledge support from the Office of Academic Affairs of Manchester College. This publication makes use of data products from the Two Micron All Sky Survey, which is a joint project of the University of Massachusetts and the Infrared Processing and Analysis Center/California Institute of Technology, funded by the National Aeronautics and Space Administration and the National Science Foundation.

\begin{deluxetable}{lrrlll}
\tablecaption{Bubbles analyzed for associated YSOs}
\tablehead{
\colhead{Name} &\colhead{l} &\colhead{b} &\colhead{citation} &\colhead{vel$_{LSR}$} &\colhead{distance}\\
&\colhead{($^\circ$)} &\colhead{($^\circ$)} &&\colhead{ (km s$^{-1}$) } &\colhead{(kpc)}}
\startdata
N4	&11.892	&0.748	&Blitz et al.(1982)	&28.5	&3.4\\
N72	&38.352	&-0.133	&Lockman et al.(1996)	&58.9	&1.5\\
N11	&13.218	&0.082	&Lockman et al.(1996)	&36.9	&3.8\\
N14	&14.002	&-0.135	&Lockman et al.(1996)	&38	&3.1\\
N37	&25.292	&0.293	&Blitz et al.(1982)	&43.8	&3.3\\
N39	&25.364	&-0.160	&Churchwell et al.(1990)	&95.4	&5.8\\
N47	&28.025	&-0.160	&Szymczak et al.(2000)	&104	&7.3\\
N48	&28.322	&0.154	&Szymczak et al.(2000)	&104	&7.3\\
N51	&29.158	&-0.262	&Szymczak et al.(2002)	&48.7	&3.4\\
N52	&30.749	&-0.019	&Szymczak et al.(2000)	&87.6	&5.7\\
N54	&31.164	&0.292	&Szymczak et al.(2000)	&42.4	&3.0\\
N62	&34.334	&0.216	&Szymczak et al.(2000)	&55.8	&3.9\\
N65	&35.000	&0.332	&Szymczak et al.(2002)	&44.2	&3.1\\
N74	&38.909	&-0.437	&Szymczak et al.(2000)	&32.3	&2.3\\
N77	&40.421	&-0.056	&Kuchar \& Bania(1994)	&68.5	&5.0\\
N79	&41.514	&0.030	&Kuchar \& Bania(1994)	&56.7	&2.8\\
N80	&41.930	&0.031	&Lockman et al.(1996)	&18.1	&1.4\\
N82	&42.102	&-0.623	&Kuchar \& Bania(1994)	&66	&5.2\\
N84	&42.831	&-0.161	&Szymczak et al.(2000)	&9.1	&1.1\\
N88	&43.265	&-0.186	&Kuchar \& Bania(1994)	&67.5	&2.8\\
N89	&43.739	&0.114	&Kuchar \& Bania(1994)	&71.6	&5.5\\
N90	&43.774	&0.059	&Bronfman et al.(1996)	&44.1	&3.1\\
N92	&44.333	&-0.839	&Szymczak et al.(2000)	&47.4	&3.7\\
N95	&45.393	&-0.717	&David et al.(1993)	&50.21	&4.4\\
N98	&47.027	&0.219	&Szymczak et al.(2000)	&56.1	&4.6\\
N101	&49.197	&-0.358	&Koo(1999)	&66.	&5.1\\
N115	&53.556	&-0.014	&Blitz et al.(1982)	&24.0	&2.7\\
N117	&54.112	&-0.064	&Blitz et al.(1982)	&18.7	&5.1\\
N123	&57.539	&-0.284	&Kuchar \& Bania(1994)	&25	&2.6\\
N124	&58.605	&0.638	&Kuchar \& Bania(1994)	&29.4	&3.2\\
N126	&59.606	&0.330	&Kuchar \& Bania(1994)	&42.8	&6.3\\
N127	&60.648	&-0.057	&Szymczak et al.(2000)	&3.6	&0.9\\
N128	&61.673	&0.946	&Szymczak et al.(2000)	&17.3	&2.8\\
N129	&61.755	&0.839	&Szymczak et al.(2000)	&10.45	&2.8\\
N130	&62.370	&-0.540	&de Gregorio-Monsalvo et al.(2004)	&20.5	&3.0\\
N131	&63.084	&-0.395	&Watson et al.(2003)	&22.6	&2.4\\
N133	&63.159	&0.451	&Watson et al.(2003)	&21.4	&2.1\\
S6	&348.263	&-0.976	&Bronfman et al.(1996)	&-13.7	&2.6\\
S8	&347.401	&0.265	&Caswell(1999)	&-96.6	&6.3\\
S11	&345.480	&0.399	&Bronfman et al.(1996)	&-16.7	&2.0\\
S13	&345.041	&-0.737	&Caswell(1999)	&-27	&1.8\\
S14	&344.756	&-0.554	&Zinchenko et al.(2000)	&-27.58	&2.9\\
S17	&343.482	&-0.044	&Vilas-Boas \& Abraham(2000)	&-30.0	&2.9\\
S23	&341.281	&-0.349	&Georgelin et al.(1996)	&-38	&3.3\\
S29	&338.901	&0.609	&Vilas-Boas \& Abraham(2000)	&-66	&4.4\\
S41	&336.483	&-0.214	&Bronfman et al.(1996)	&-81.1	&5.1\\
\enddata
\label{ysobubbles}
\end{deluxetable}

\begin{deluxetable}{lrrrr}
\tablecaption{Bubbles analyzed for 24 and 70 $\mu$m flux.}
\tablehead{
	\colhead{Name} &\colhead{l} &\colhead{b} &\colhead{S$_{24 \mu m}$} &\colhead{S$_{70 \mu m}$}\\
&\colhead{($^\circ$)} &\colhead{($^\circ$)} &\colhead{ (MJy) } &\colhead{(MJy)} }
\startdata
N26	&19.587	&-0.051	&120.1	&150.0\\
N41	&26.266	&0.282 	&210.0 	&193.4\\
N56	&32.583	&0.002	&248.8	&173.8\\
N57	&32.763	&-0.150	&31.7	&27.6\\
N72	&38.352	&-0.133	&265.6	&255.8\\
N78	&41.229	&0.170	&21.6	&13.1\\
N90	&43.774	&0.059	&505.2	&298.8\\
N93	&44.777	&-0.550	&82.6	&42.7\\
S2	&349.215	&0.142	&662.5	&119.9\\
S21	&341.357	&-0.288	&159.2	&121.6\\
S83	&323.975	&0.057	&46.6	&33.0\\
S87	&322.418	&0.207	&32.4	&33.6\\
S115	&315.979	&-0.182	&160.2	&71.3\\
S130	&312.151	&-0.313	&41.0	&36.1\\
S146	&307.795	&-0.483	&65.0	&28.0\\
\enddata
\label{tempbubbles}
\end{deluxetable}

\begin{deluxetable}{lr}
\tablecaption{Results of modified blackbody fits to 24 $\mu$m and 70 $\mu$m emission from entire bubble interior.}
\tablehead{
	\colhead{Name} &\colhead{Temperature (K)}\\}
\startdata
N26	&71\\
N41	&76\\
N56	&80\\
N57	&76\\
N72	&75\\
N78	&82\\
N90	&83\\
N93	&85\\
S2	&113\\
S21	&78\\
S83	&80\\
S87	&49\\
S115	&89\\
S130	&76\\
S146	&89\\
\enddata
\label{tempresults}
\end{deluxetable}

\begin{figure}
\psfig{file=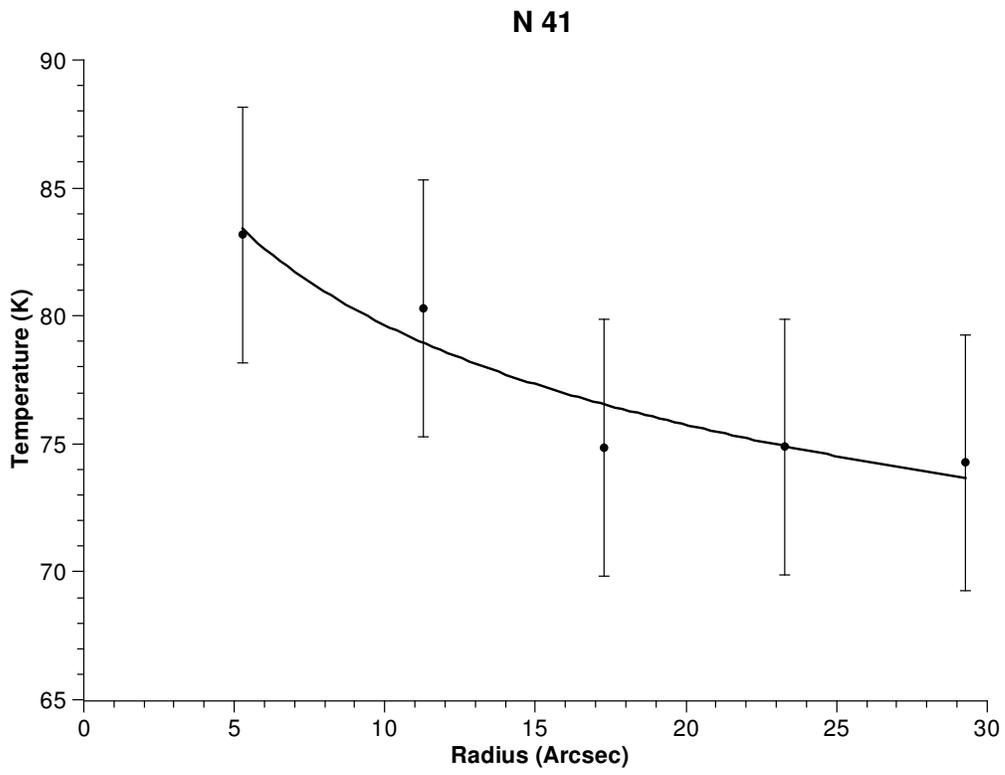,angle=-90,width=6in}
\caption{Same as Fig \ref{n26} for N41.}
\label{n41}
\end{figure}

\begin{figure}
\psfig{file=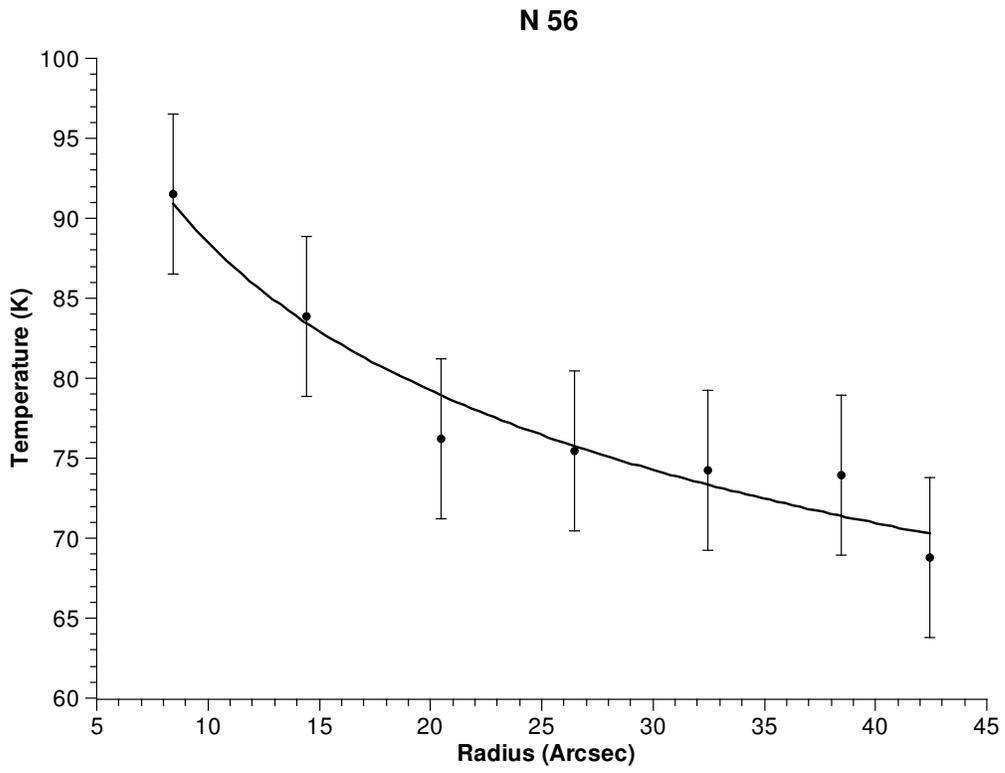,angle=-90,width=6in}
\caption{Same as Fig \ref{n26} for N56.}
\label{n41}
\end{figure}

\begin{figure}
\psfig{file=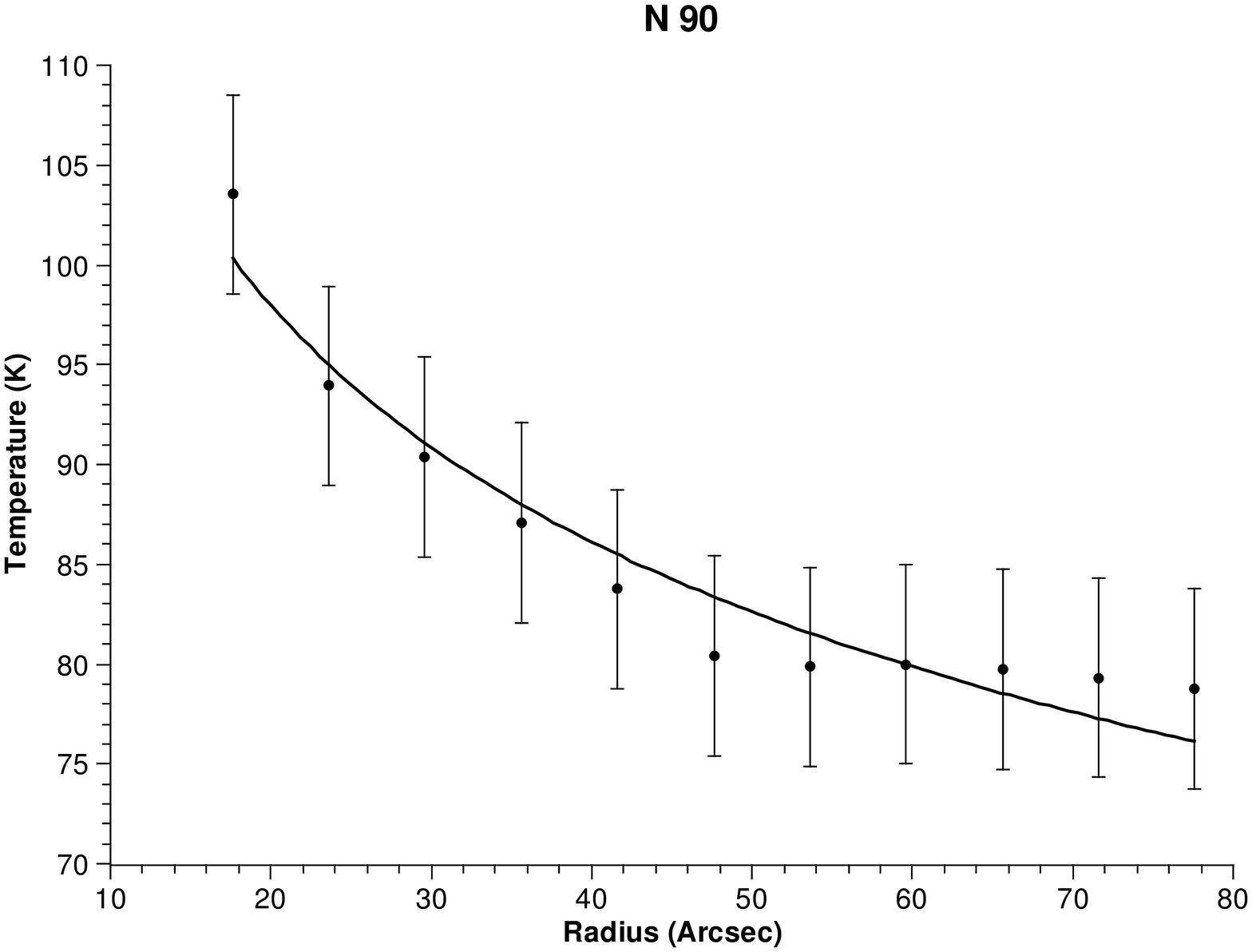,angle=-90,width=6in}
\caption{Same as Fig \ref{n26} for N90.}
\label{n90}
\end{figure}

\begin{figure}
\includegraphics[angle=-90,width=6in]{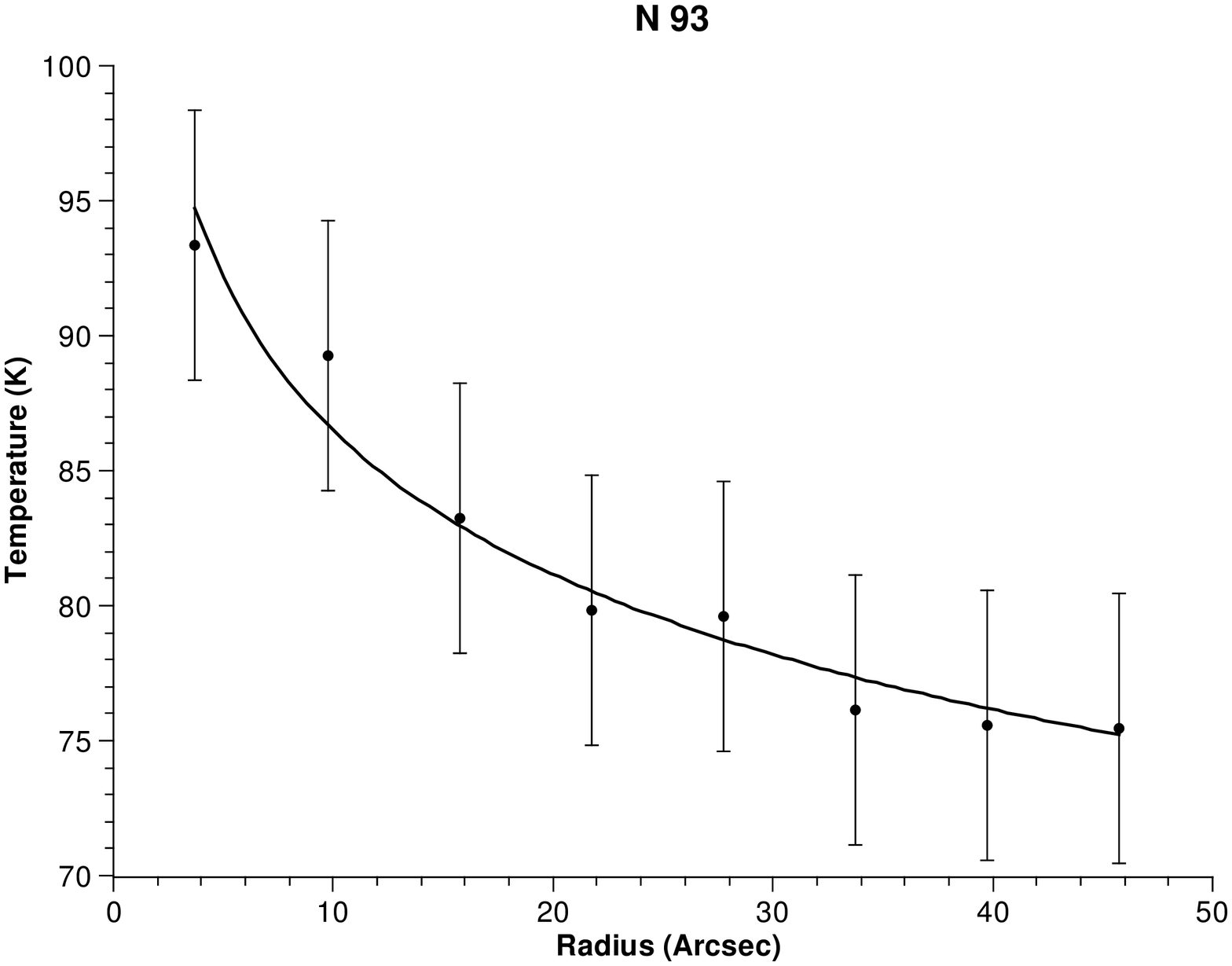}
\caption{Same as Fig \ref{n26} for N93.}
\label{n93}
\end{figure}

\begin{figure}
\includegraphics[angle=-90,width=6in]{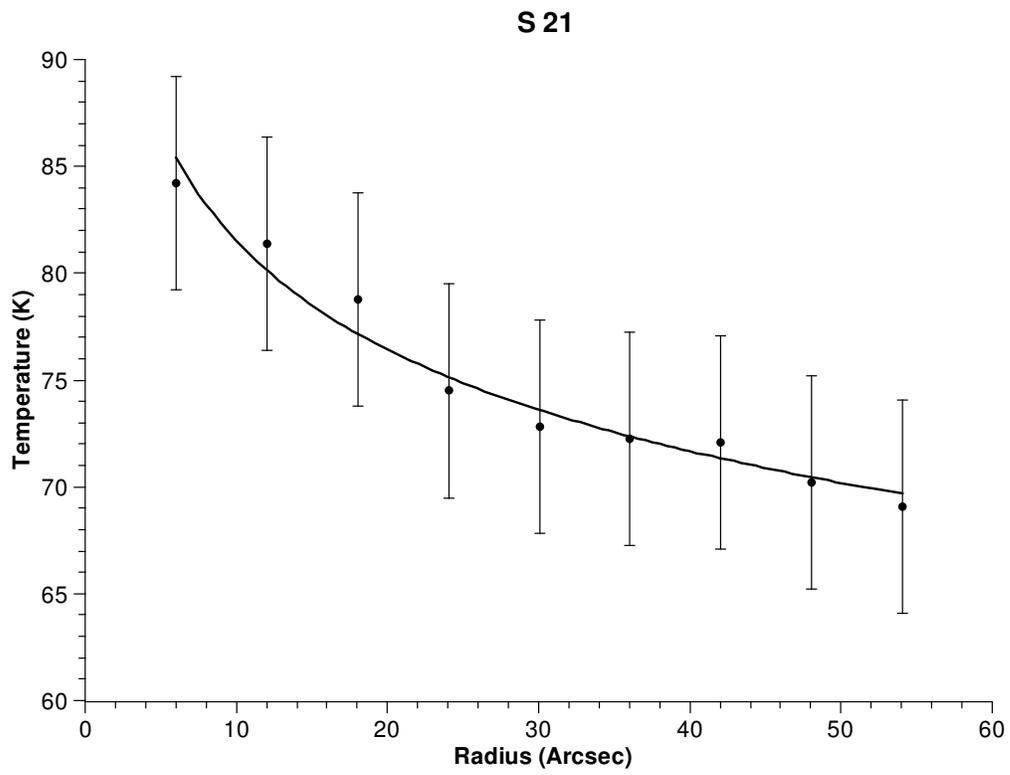}
\caption{Same as Fig \ref{n26} for S21.}
\label{s83}
\end{figure}

\begin{figure}
\includegraphics[angle=-90,width=6in]{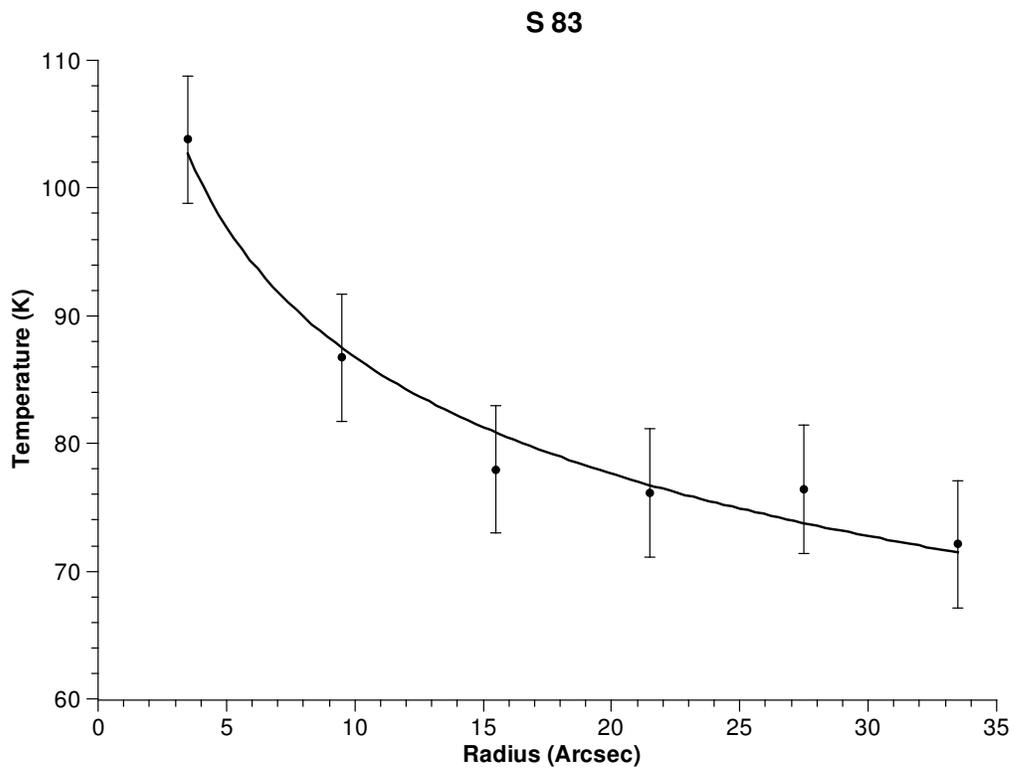}
\caption{Same as Fig \ref{n26} for S83.}
\label{s83}
\end{figure}

\begin{figure}
\includegraphics[angle=-90,width=6in]{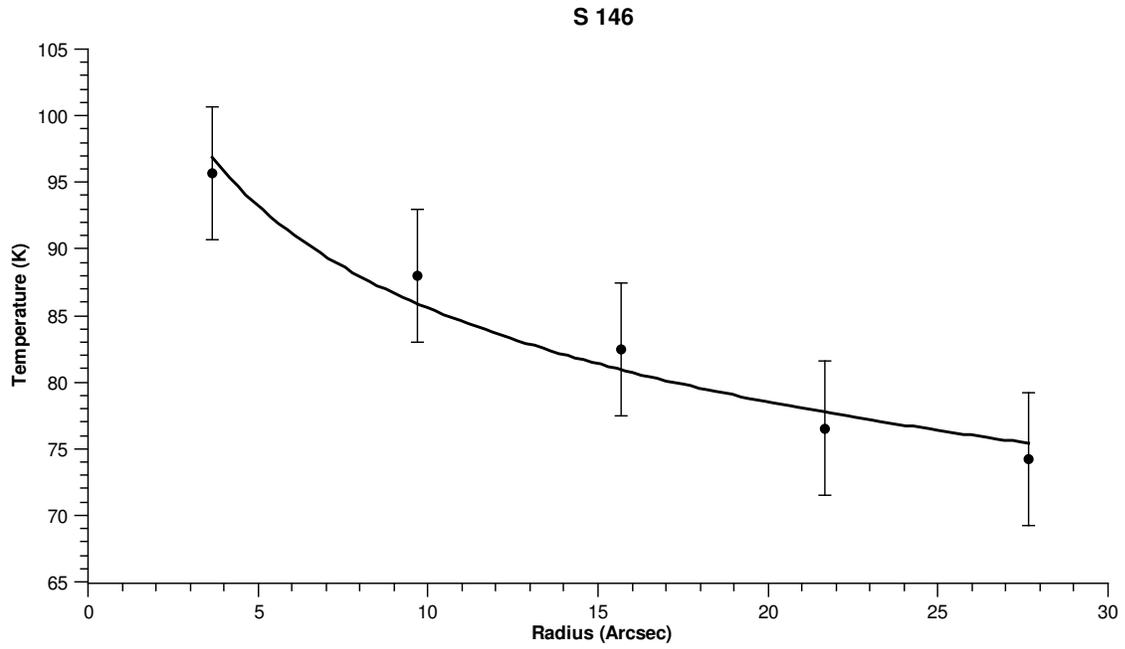}
\caption{Same as Fig \ref{n26} for S146.}
\label{s146}
\end{figure}

\begin{deluxetable}{llllll}
\tablecaption{Results of power-law fit to temperature profiles derived from 24 $\mu$m and 70 $\mu$m emission from bubble interior.}
\tablehead{
	\colhead{Name} &\colhead{Power-law Exponent}\\}
\startdata
N26 &0.04 $\pm$ 0.01\\
N41 &0.07 $\pm$ 0.01\\
N56 &0.16 $\pm$ 0.02\\
N90 &0.19 $\pm$ 0.02\\
N93 &0.09 $\pm$ 0.01\\
S21 &0.09 $\pm$ 0.01\\
S82 &0.16 $\pm$ 0.01\\
S146 &0.12 $\pm$ 0.01\\
\enddata
\label{tempprofiles}
\end{deluxetable}

\clearpage
\begin{figure}
\plottwo{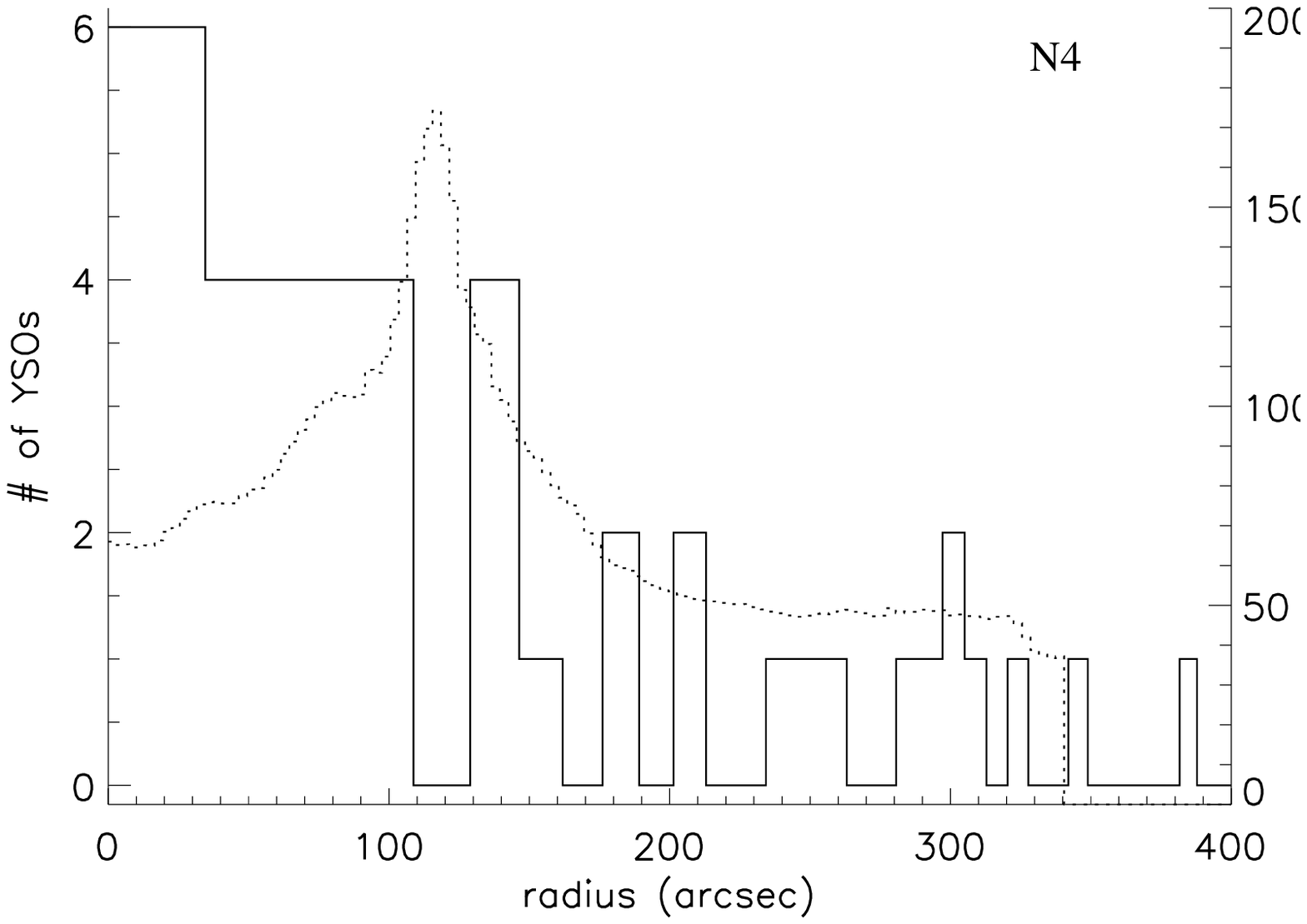}{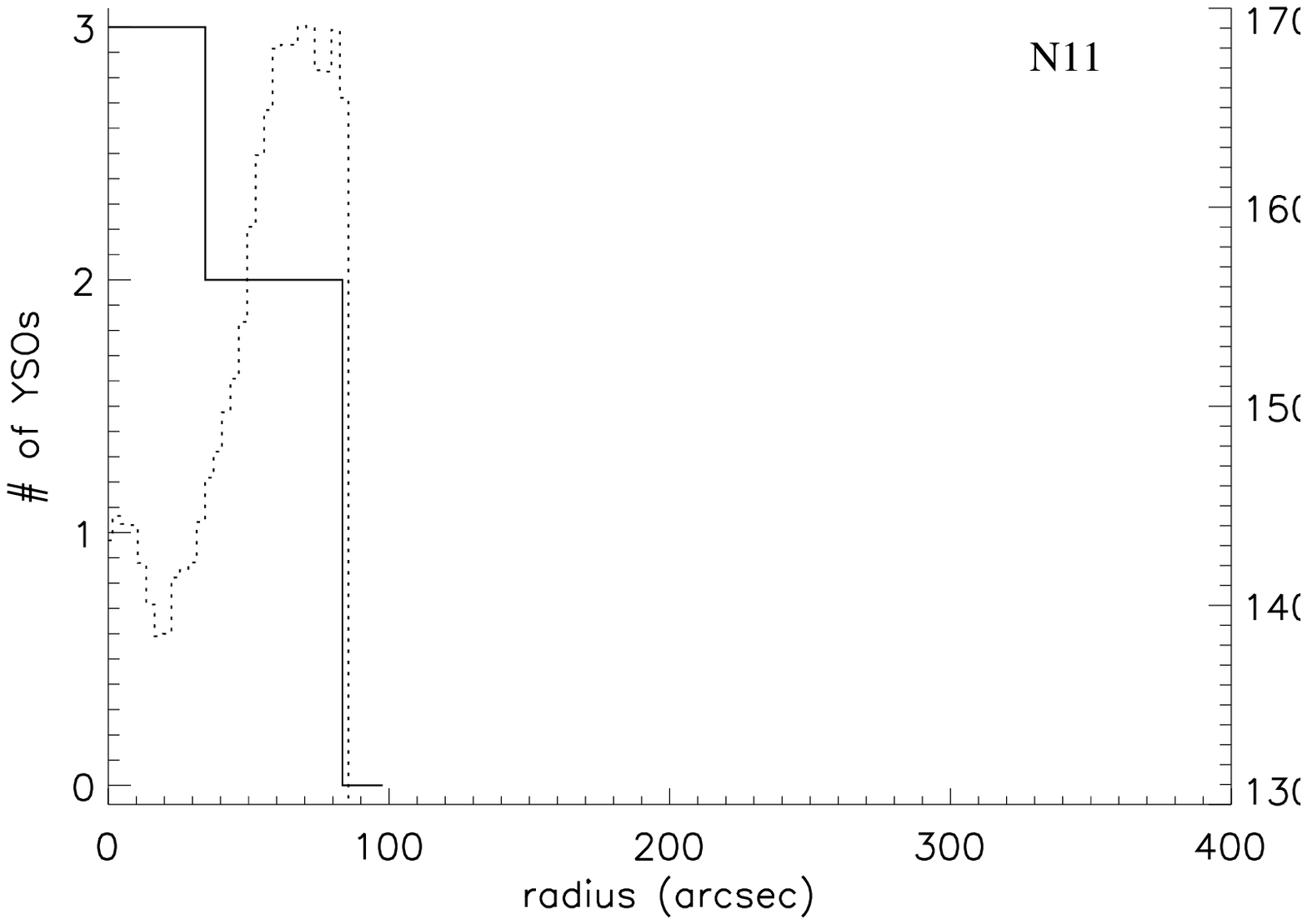}
\plottwo{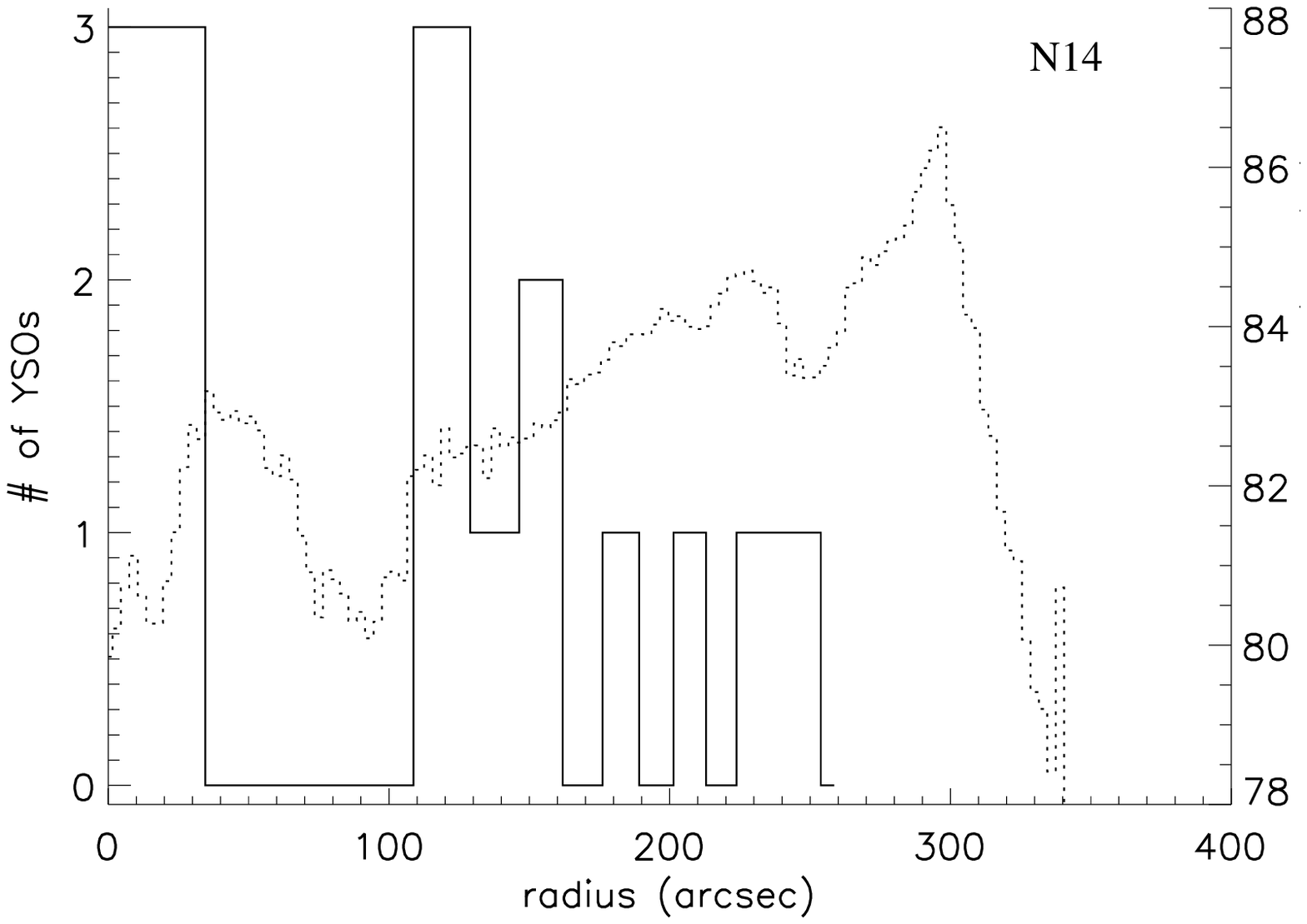}{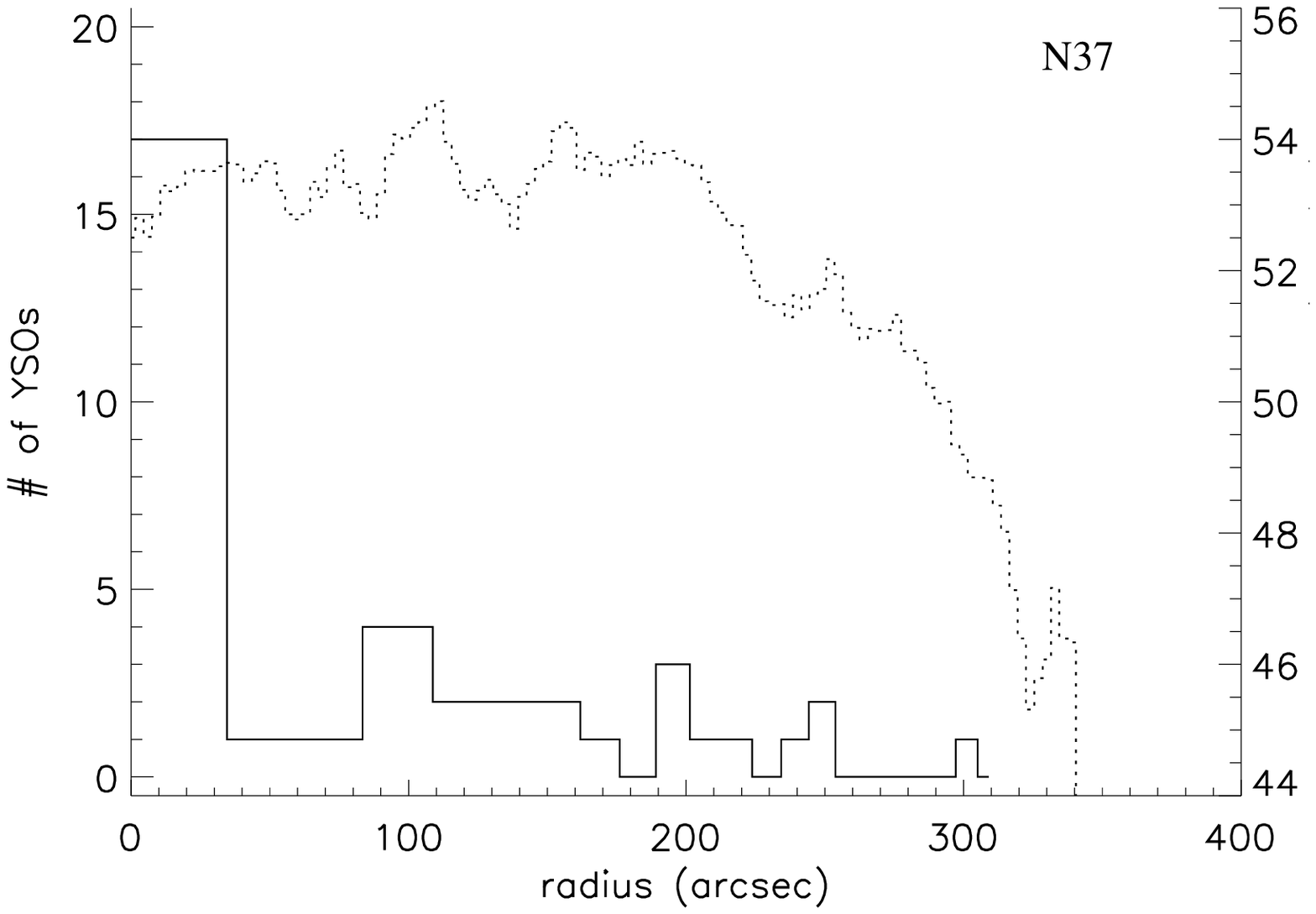}
\plottwo{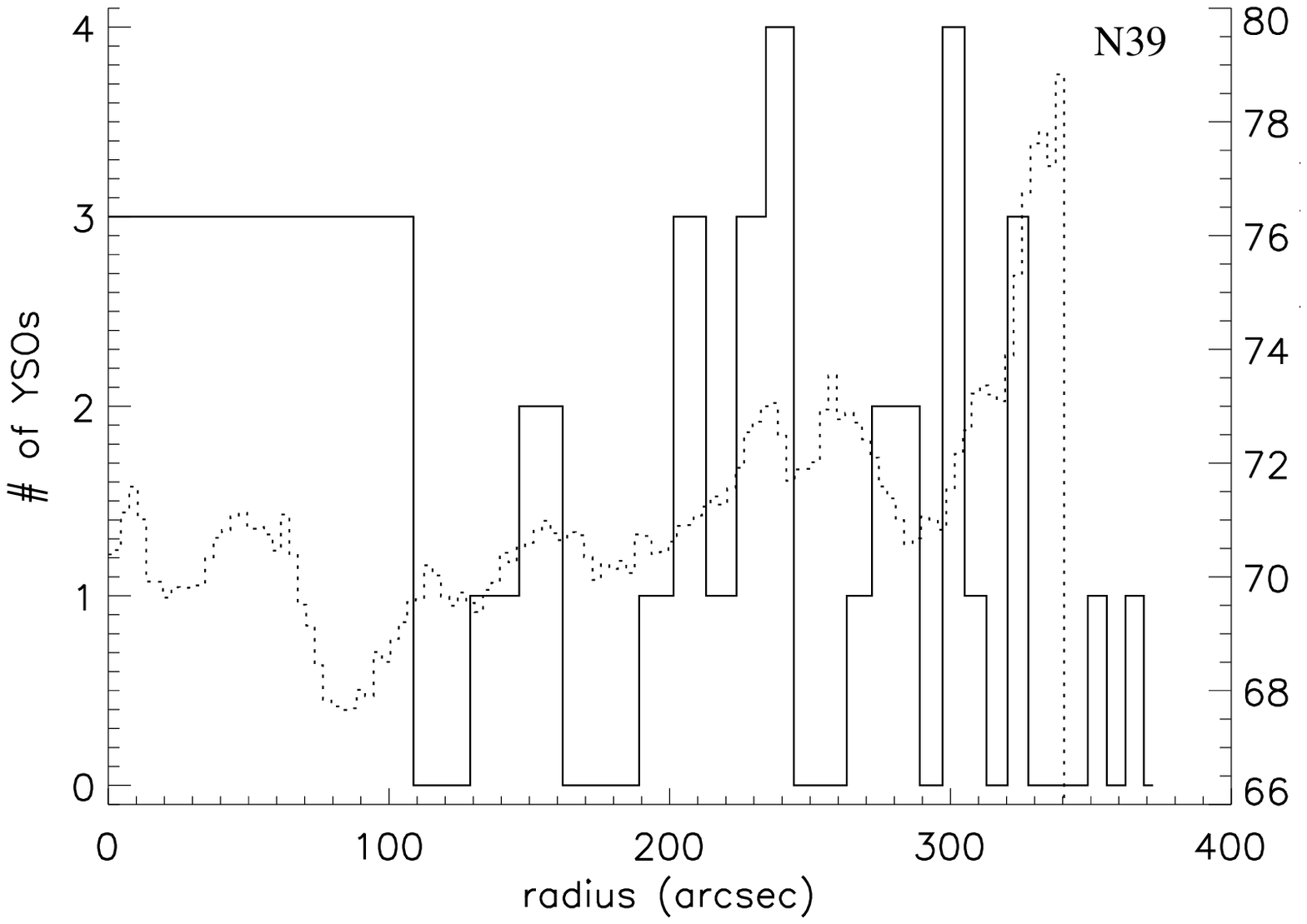}{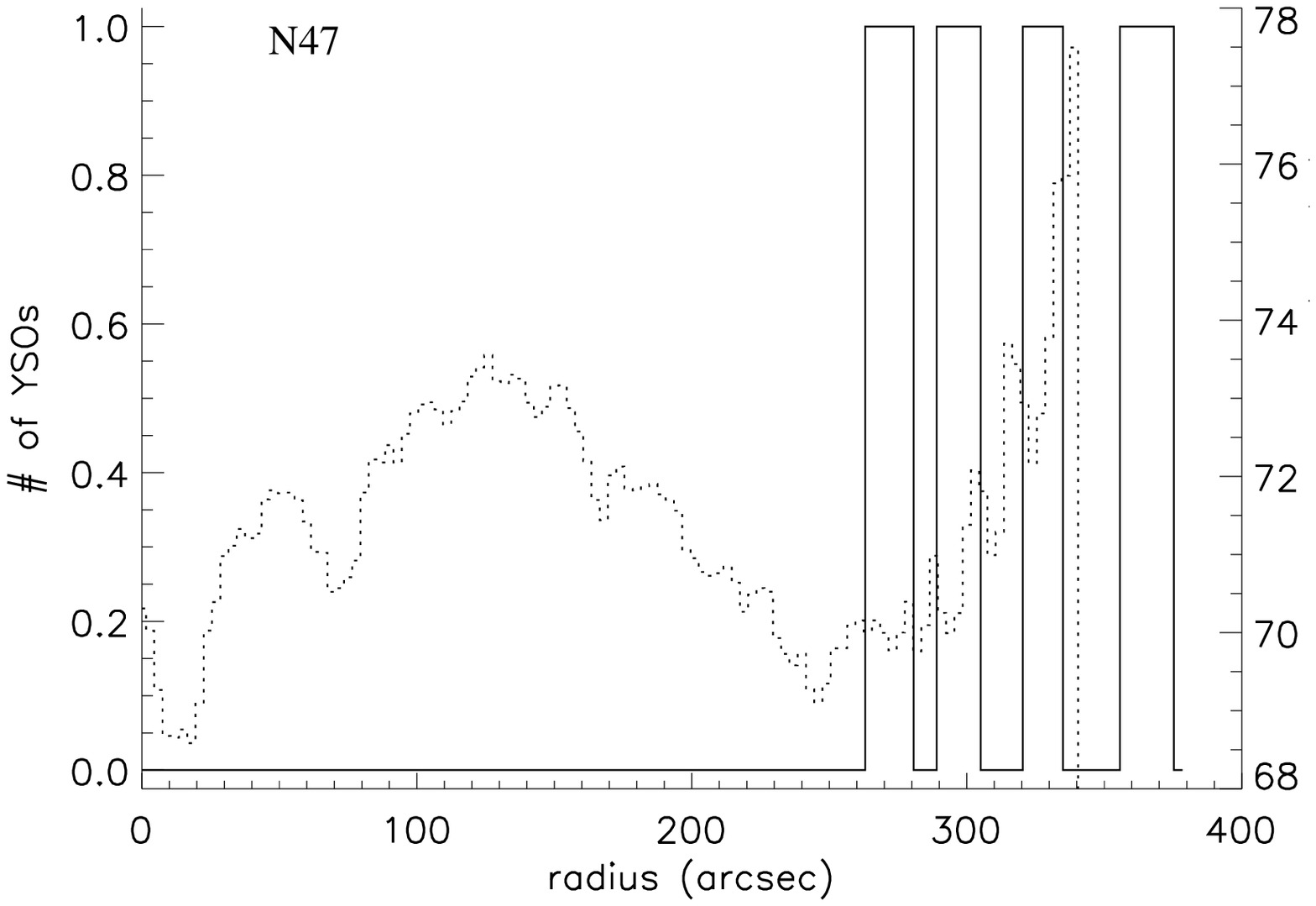}
\plottwo{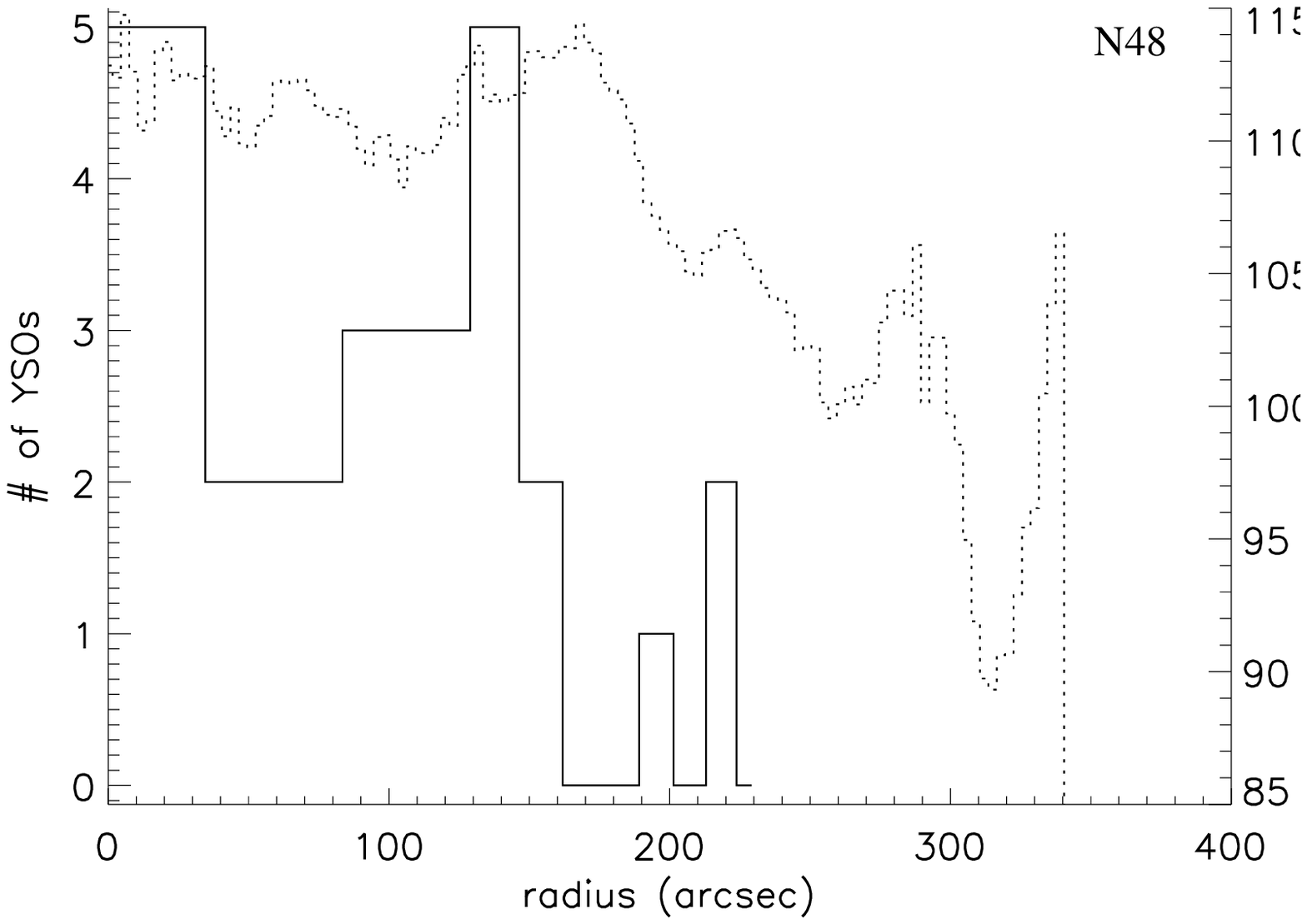}{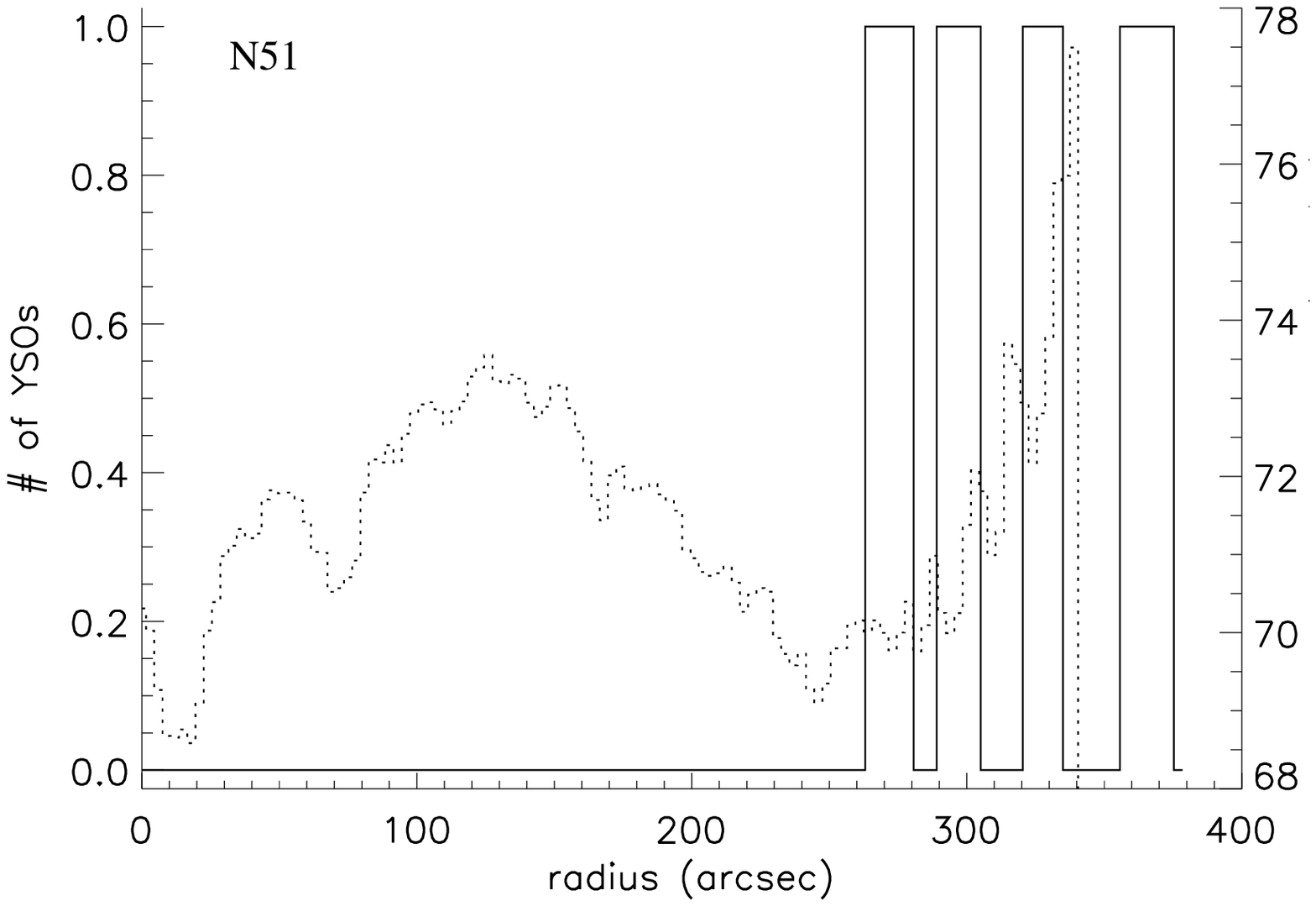}
\caption{Number of YSOs in equal-area annuli (solid) and 8 $\mu$m brightness azimuthally-averaged (dashed) for (upper-left by rows): N4, N11, N14, N37, N39, N47, N48, N51}
\label{profile1}
\end{figure}
\clearpage
\begin{figure}
\plottwo{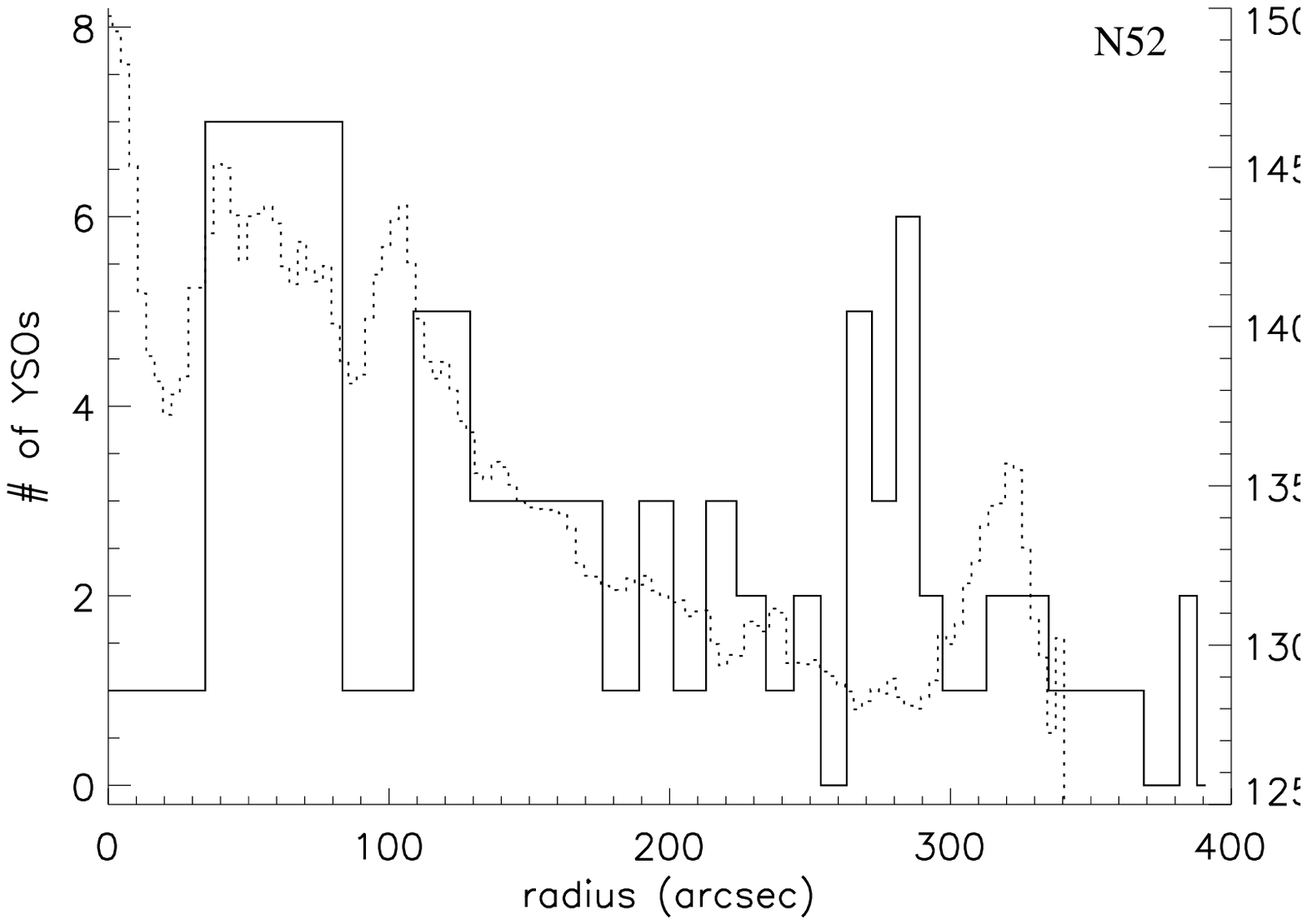}{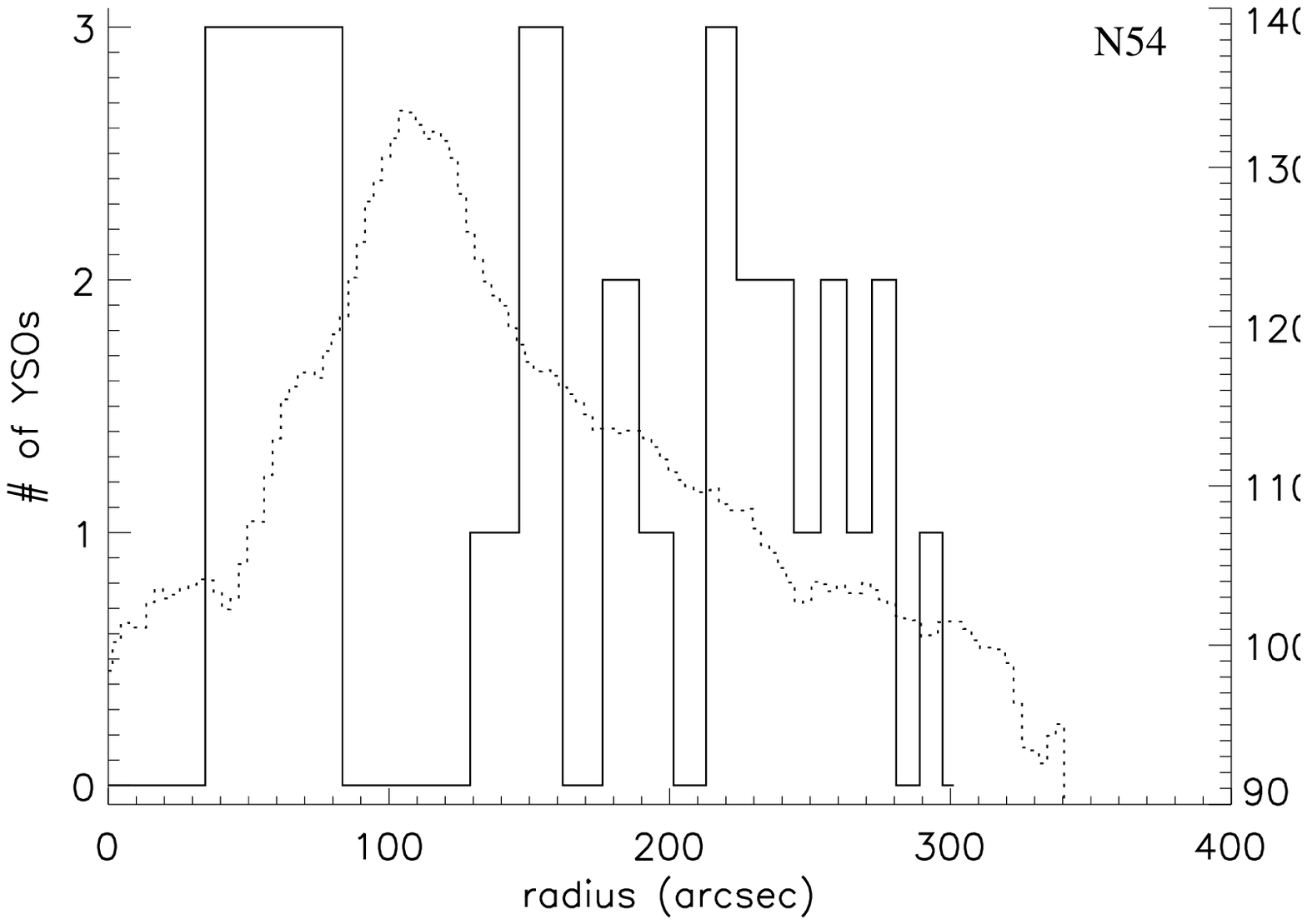}
\plottwo{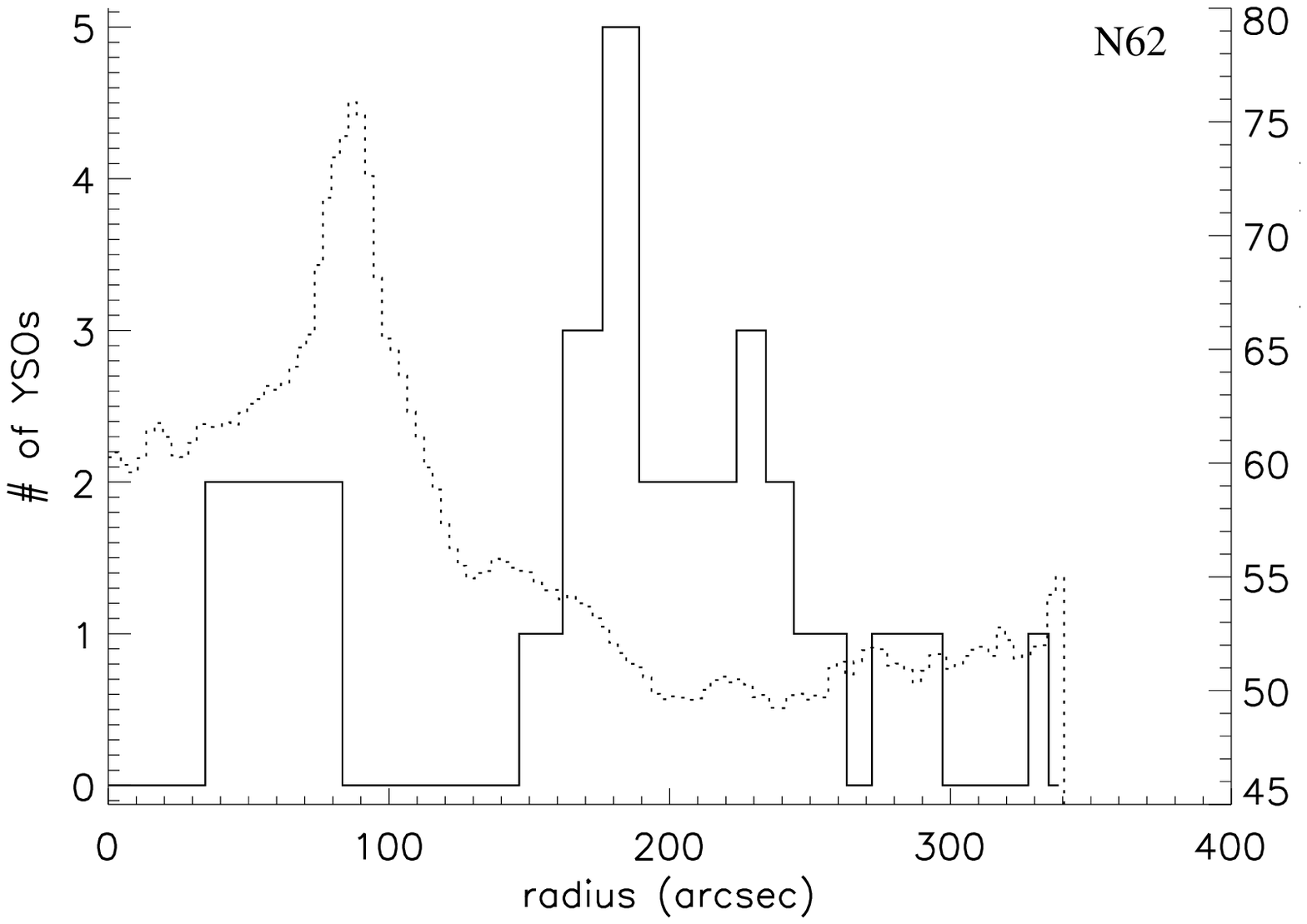}{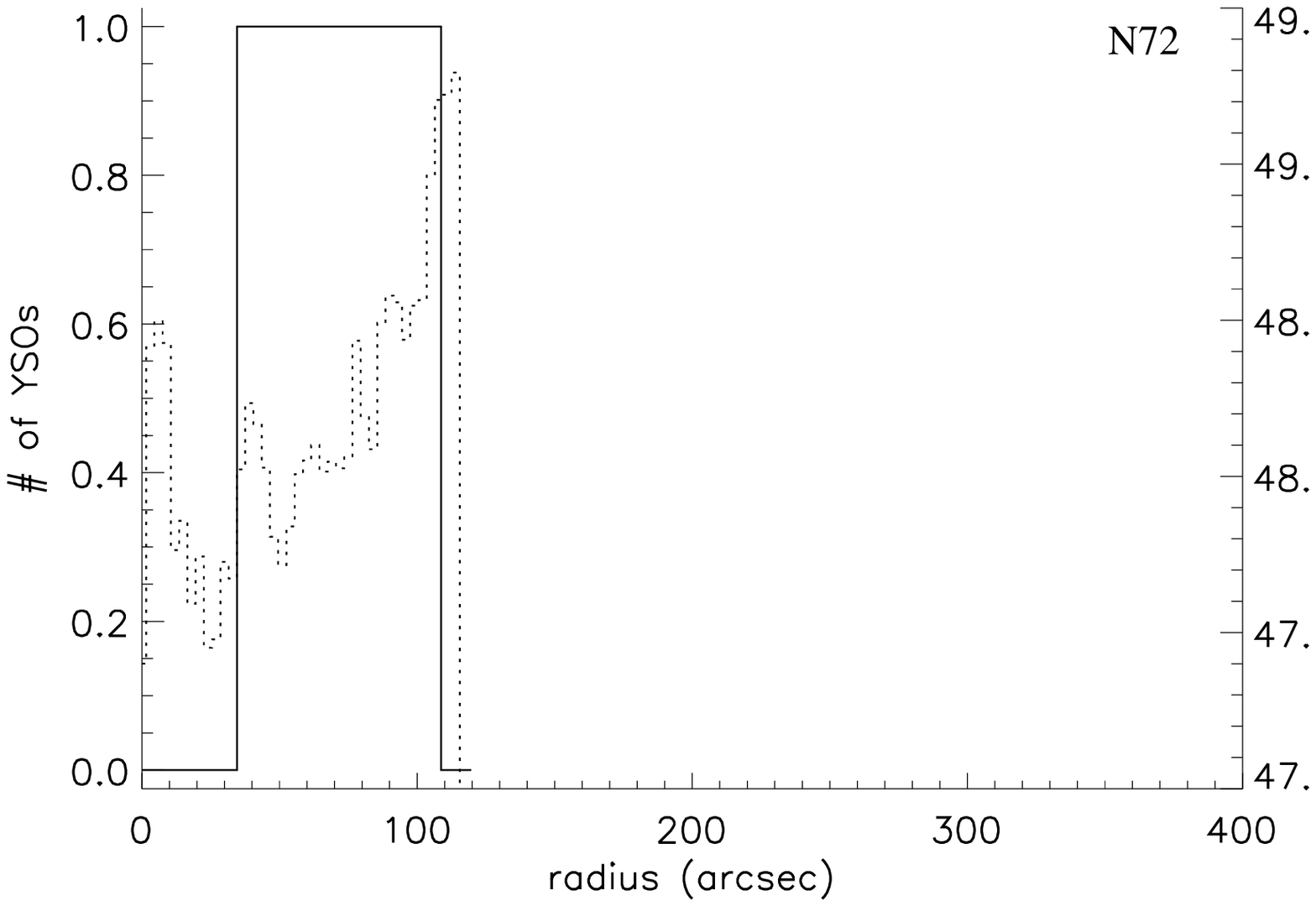}
\plottwo{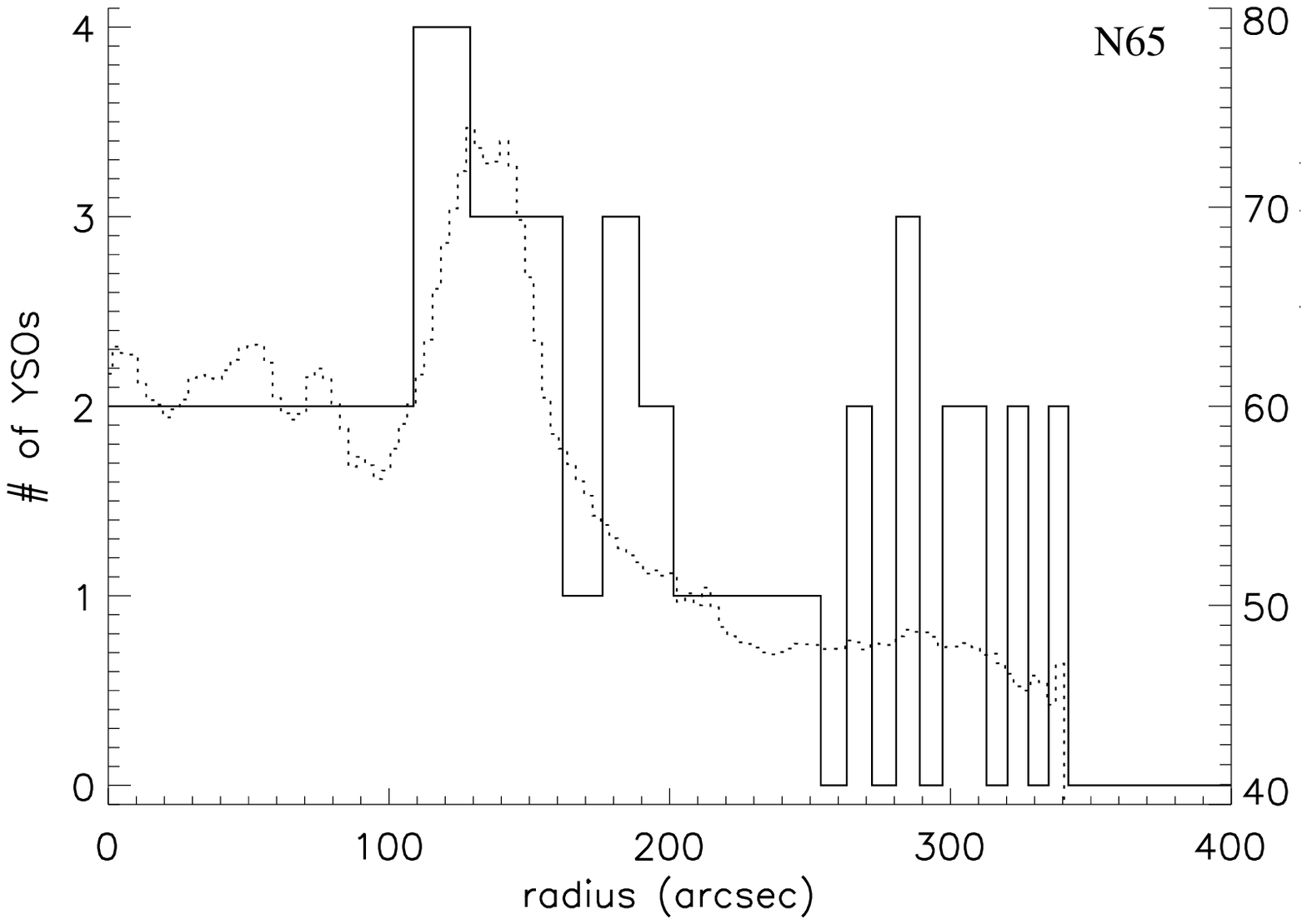}{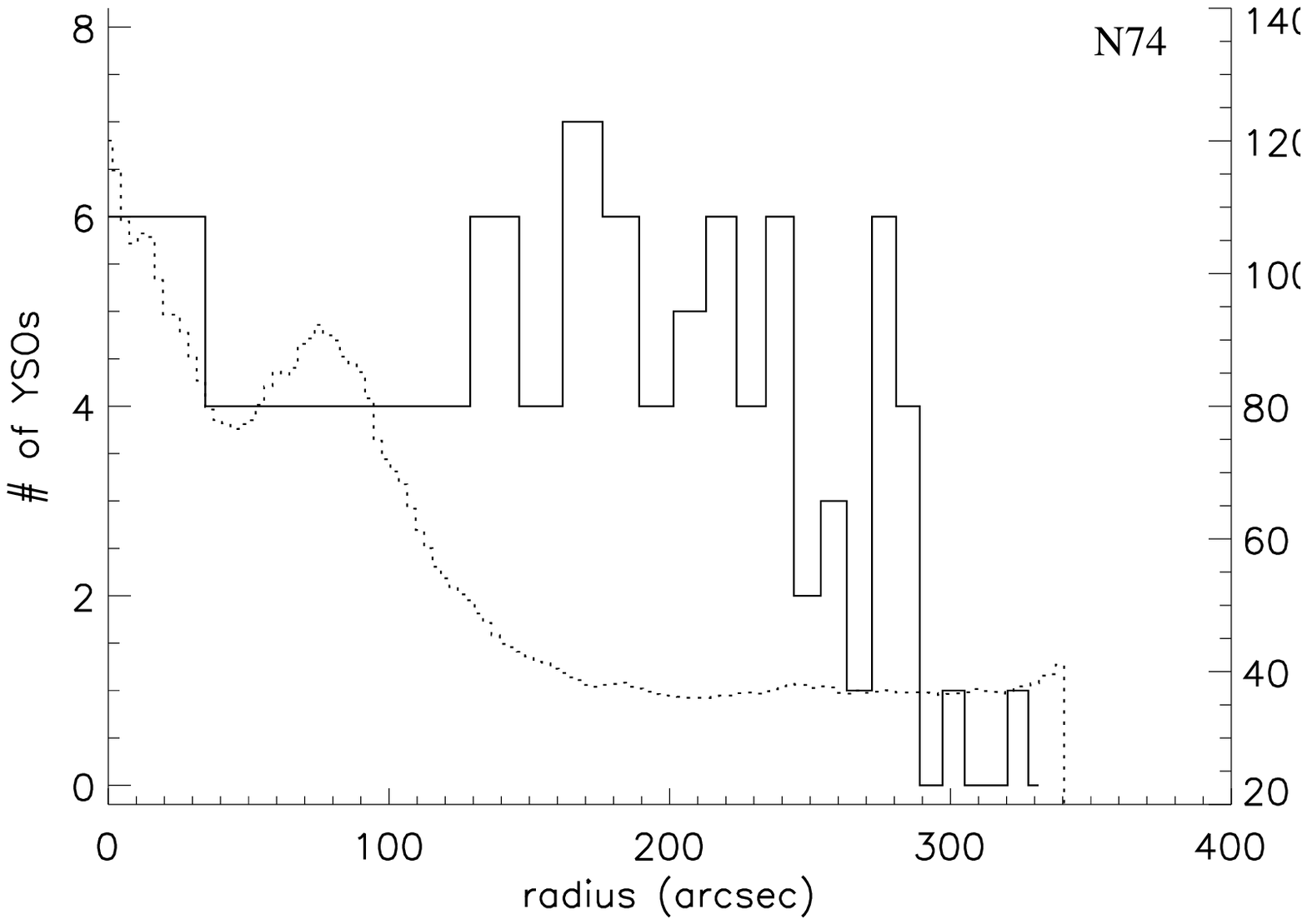}
\plottwo{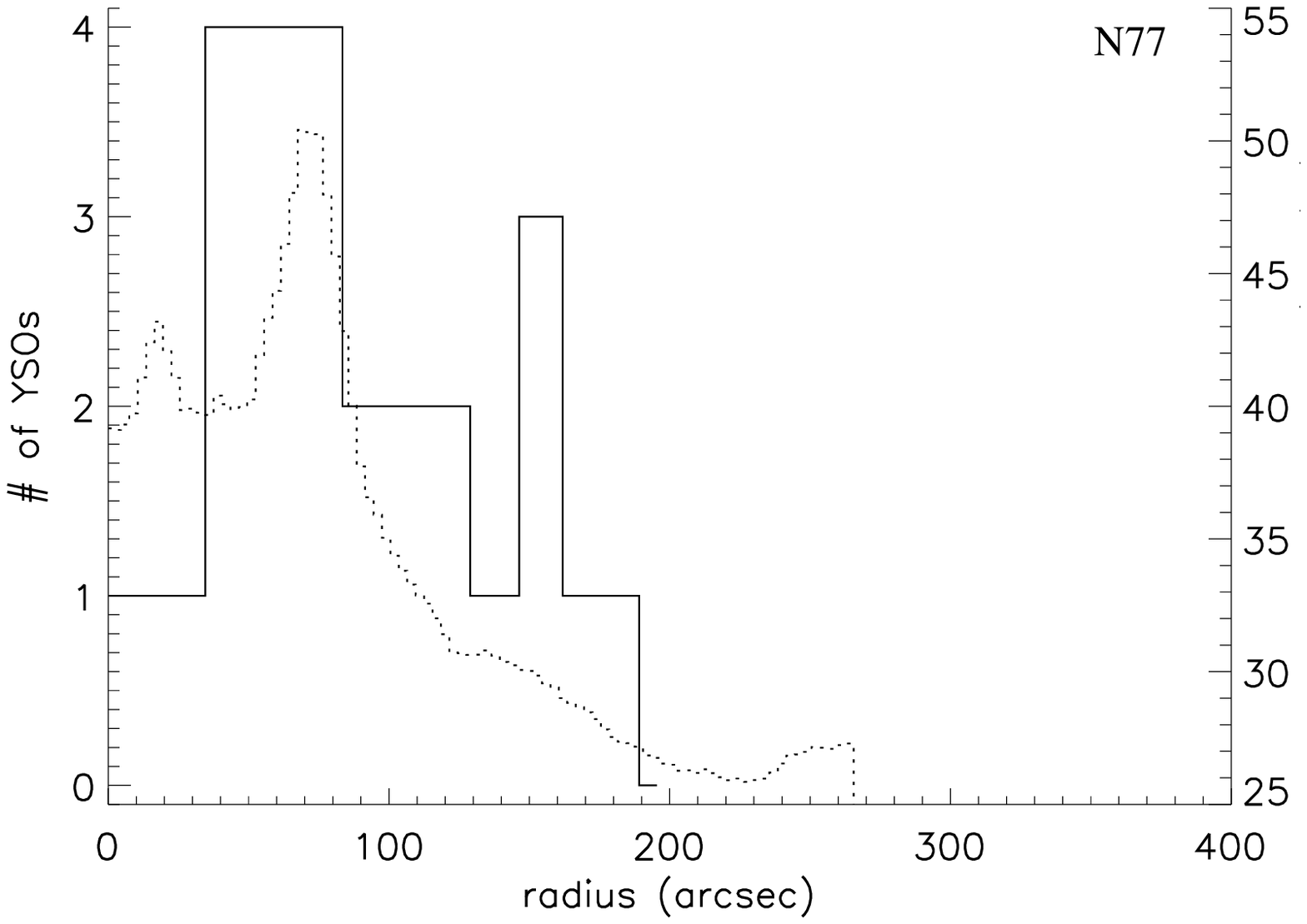}{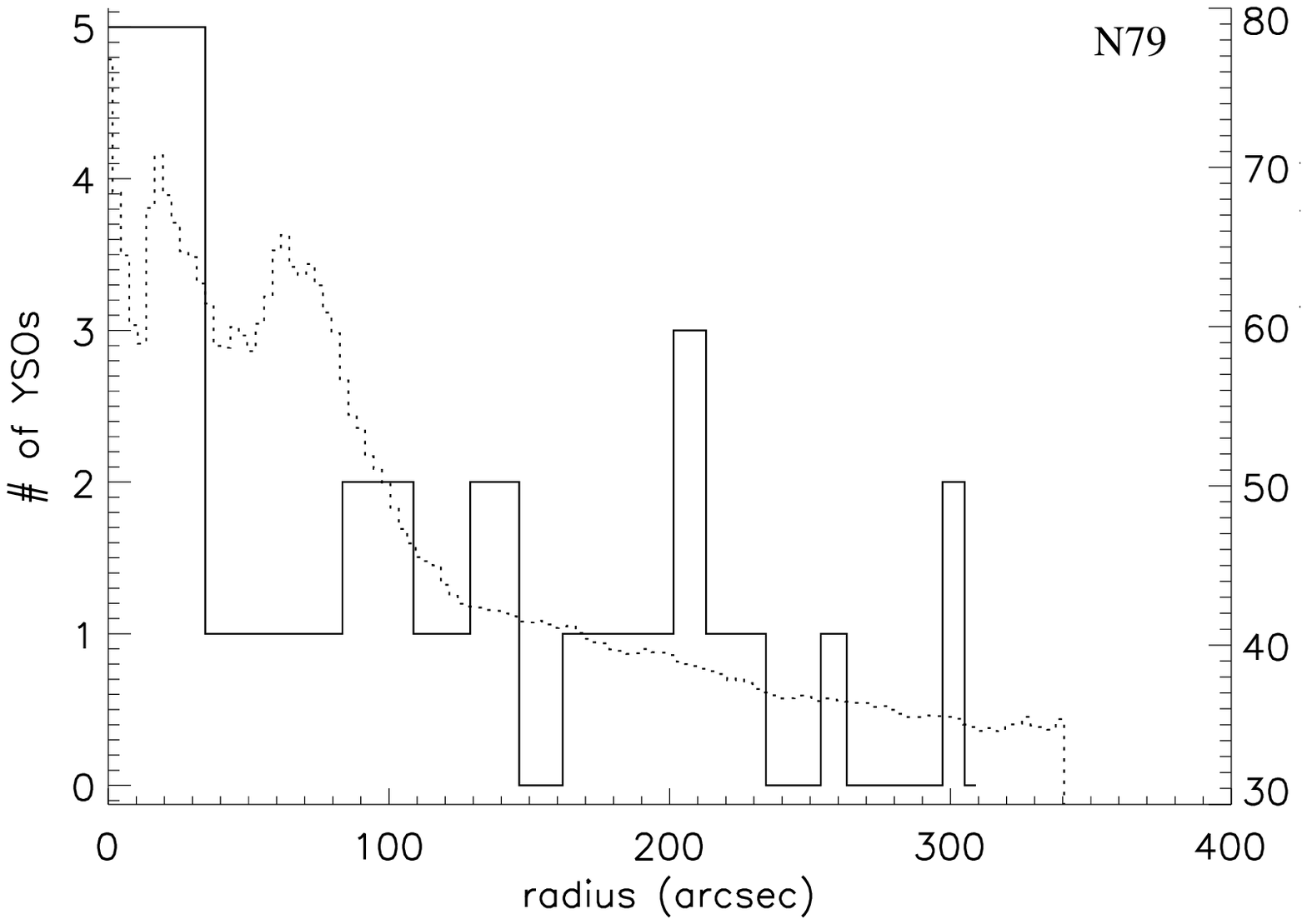}
\caption{Number of YSOs in equal-area annuli (solid) and 8 $\mu$m brightness azimuthally-averaged (dashed) for (upper-left by rows): N52, N54, N62, N72, N65, N74, N77, N79}
\label{profile2}
\end{figure}
\clearpage
\begin{figure}
\plottwo{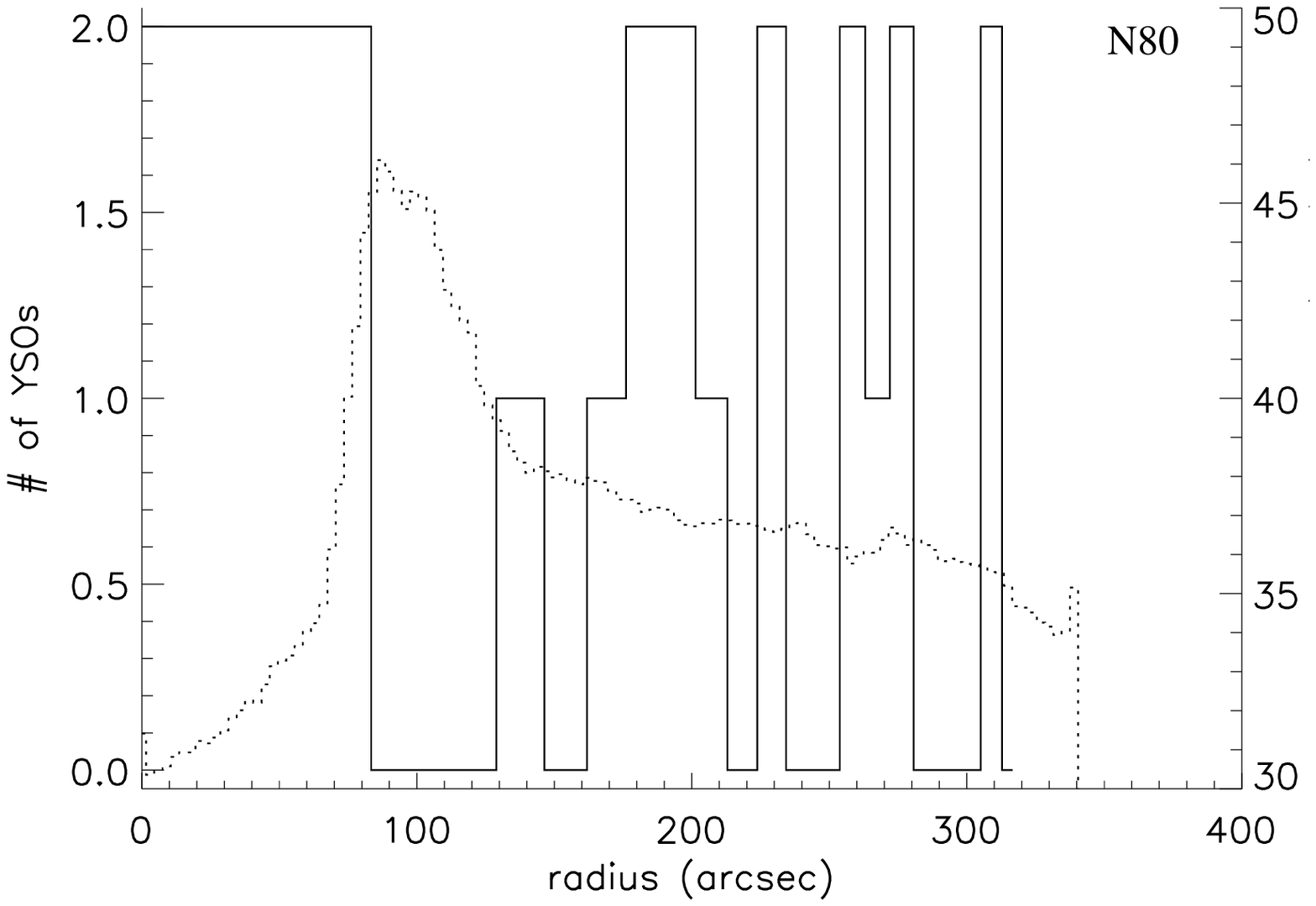}{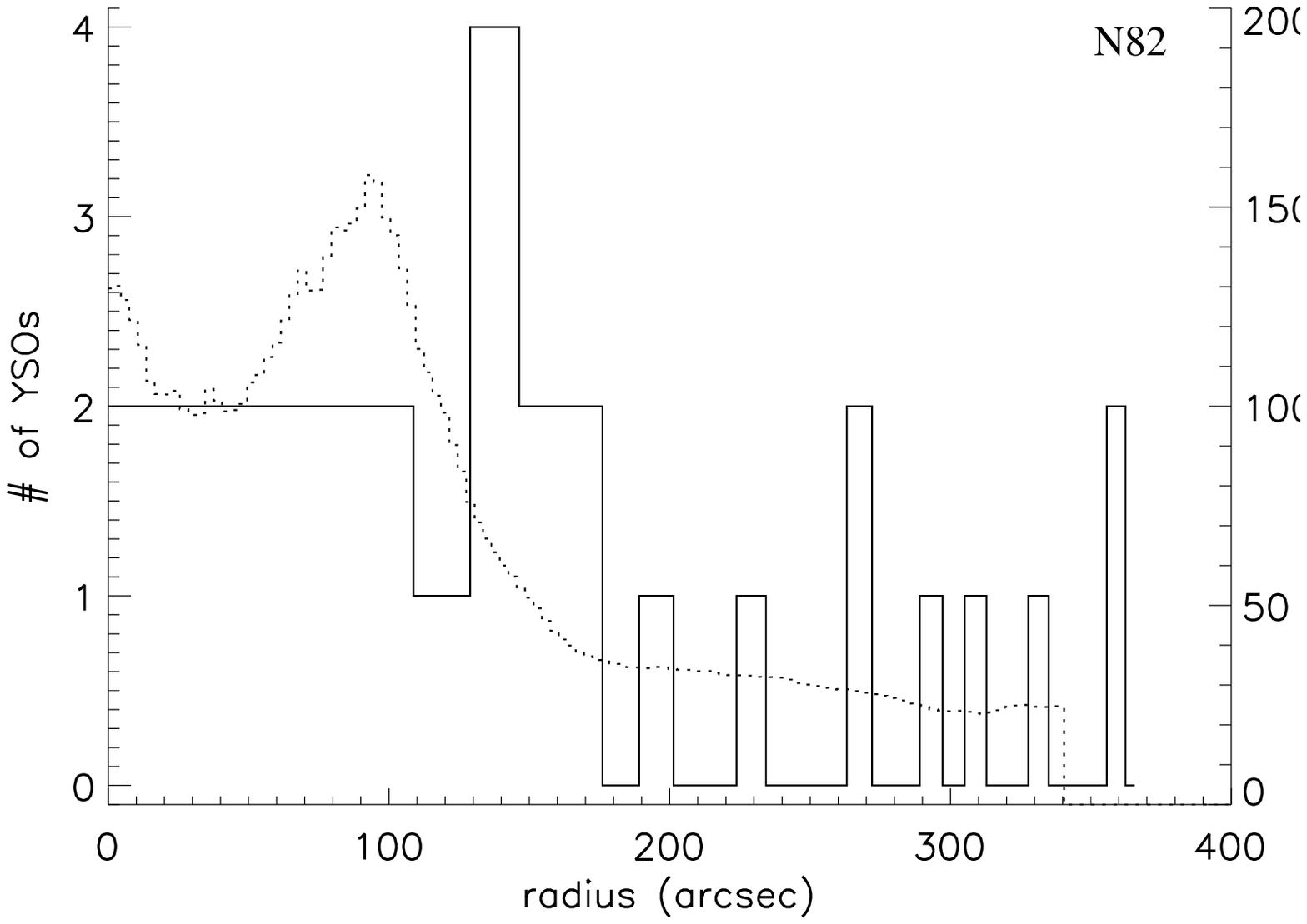}
\plottwo{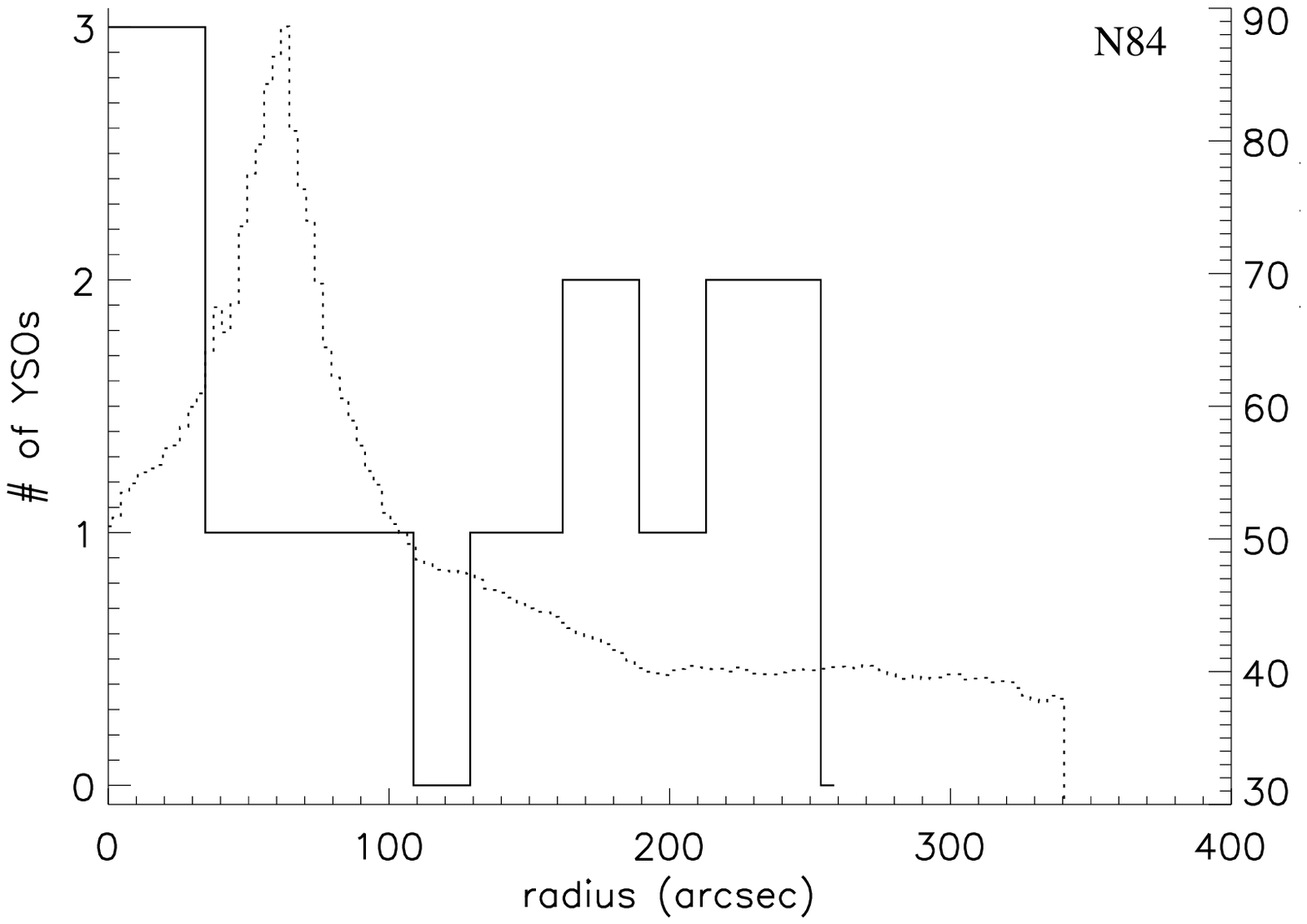}{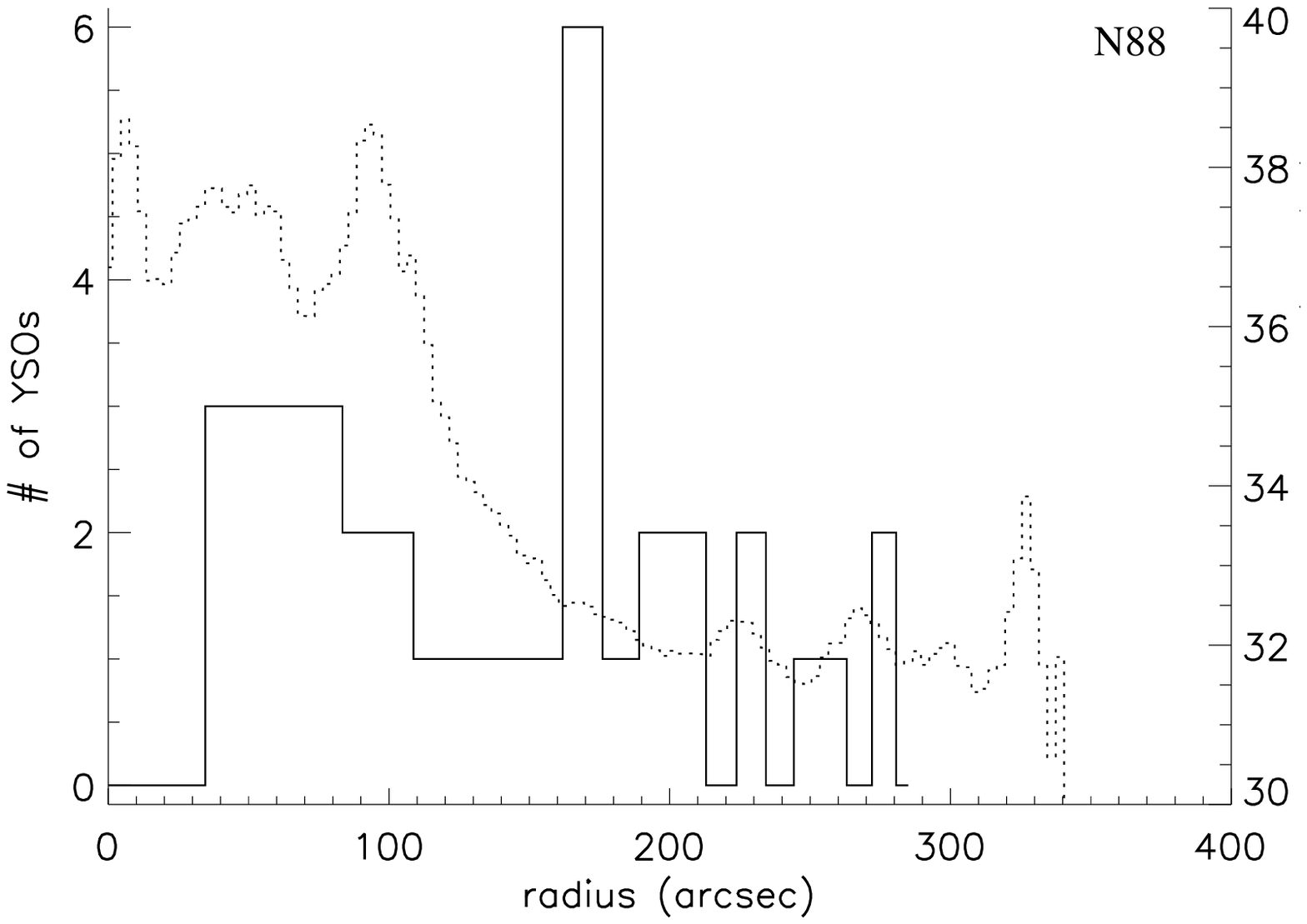}
\plottwo{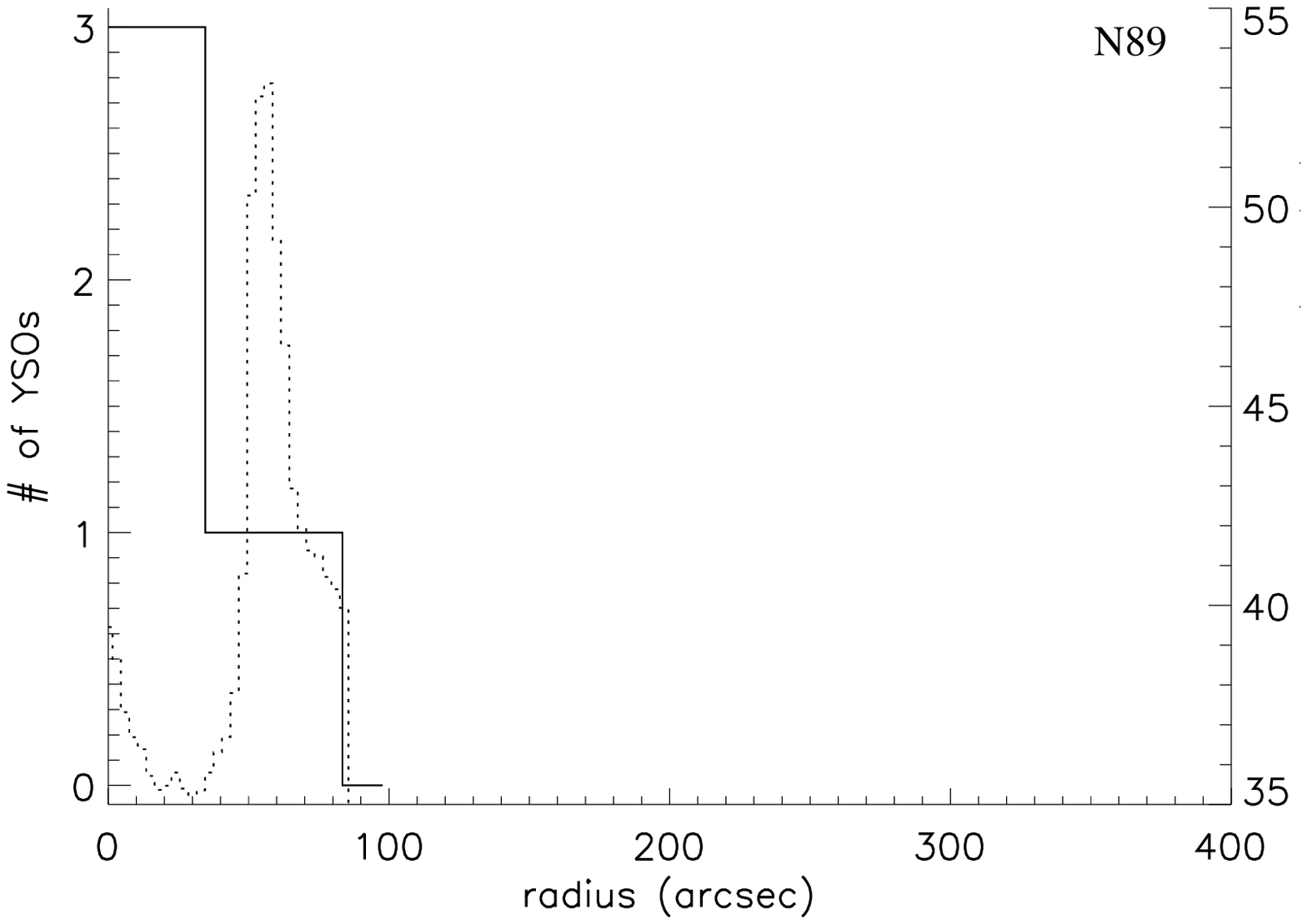}{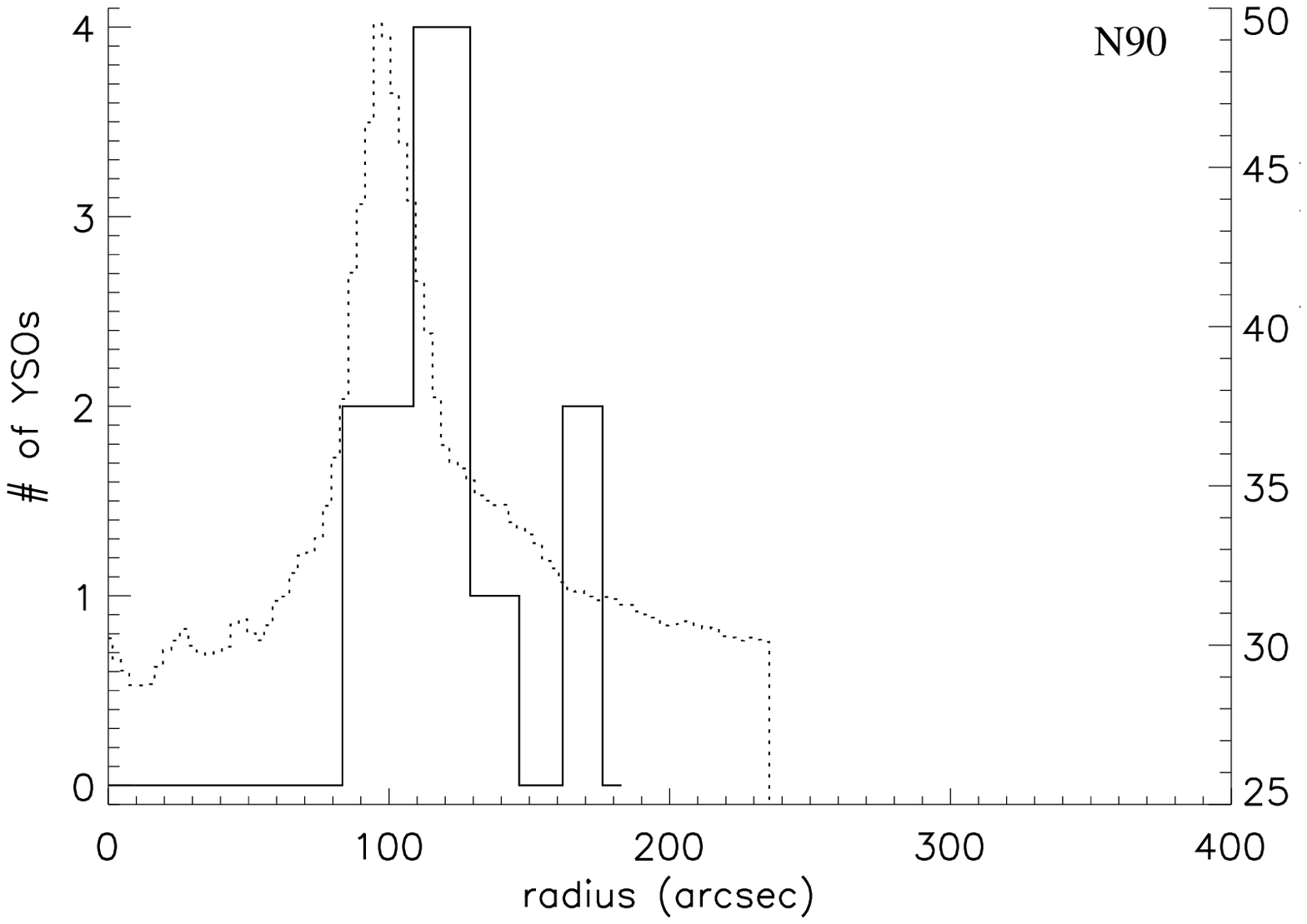}
\plottwo{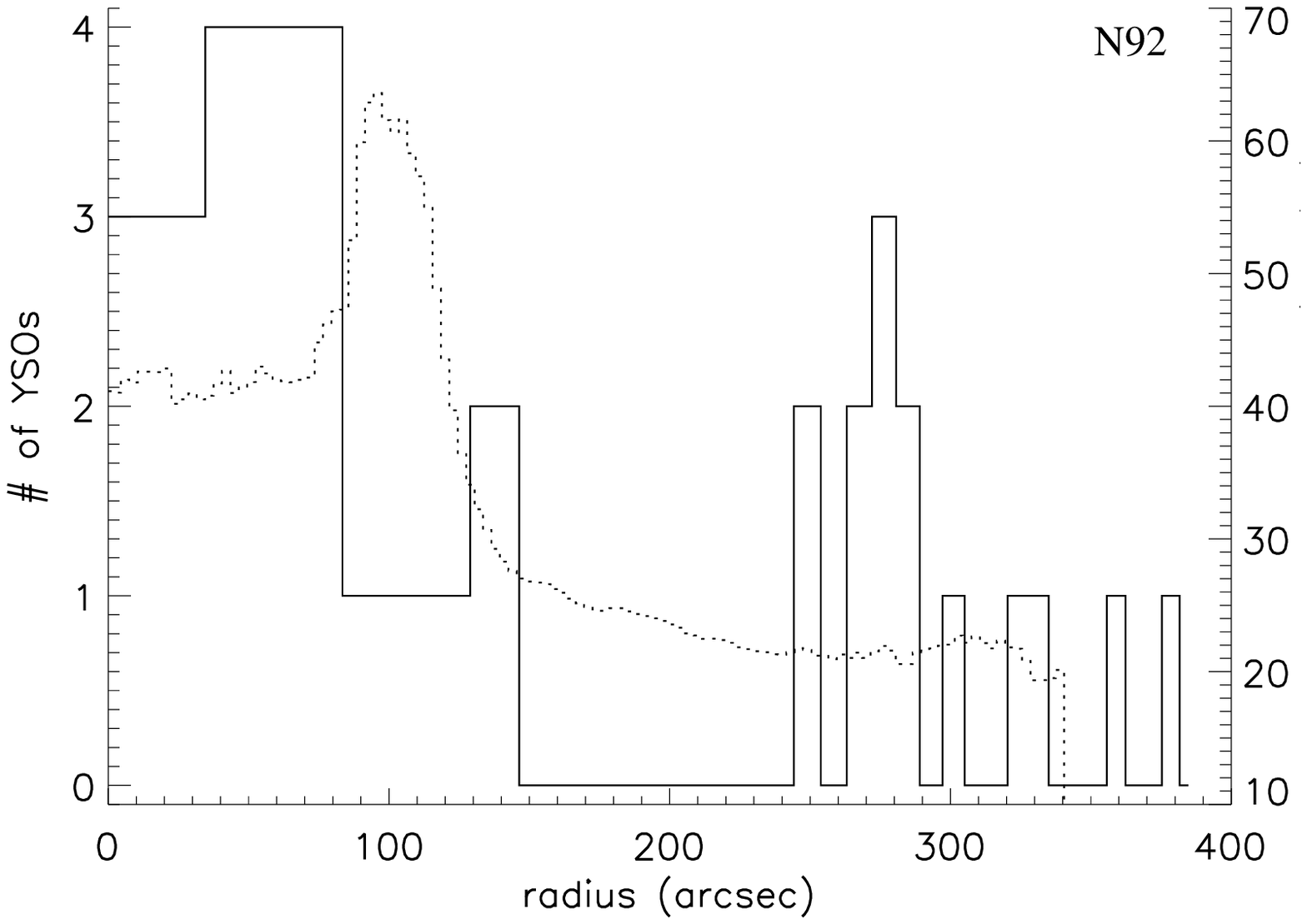}{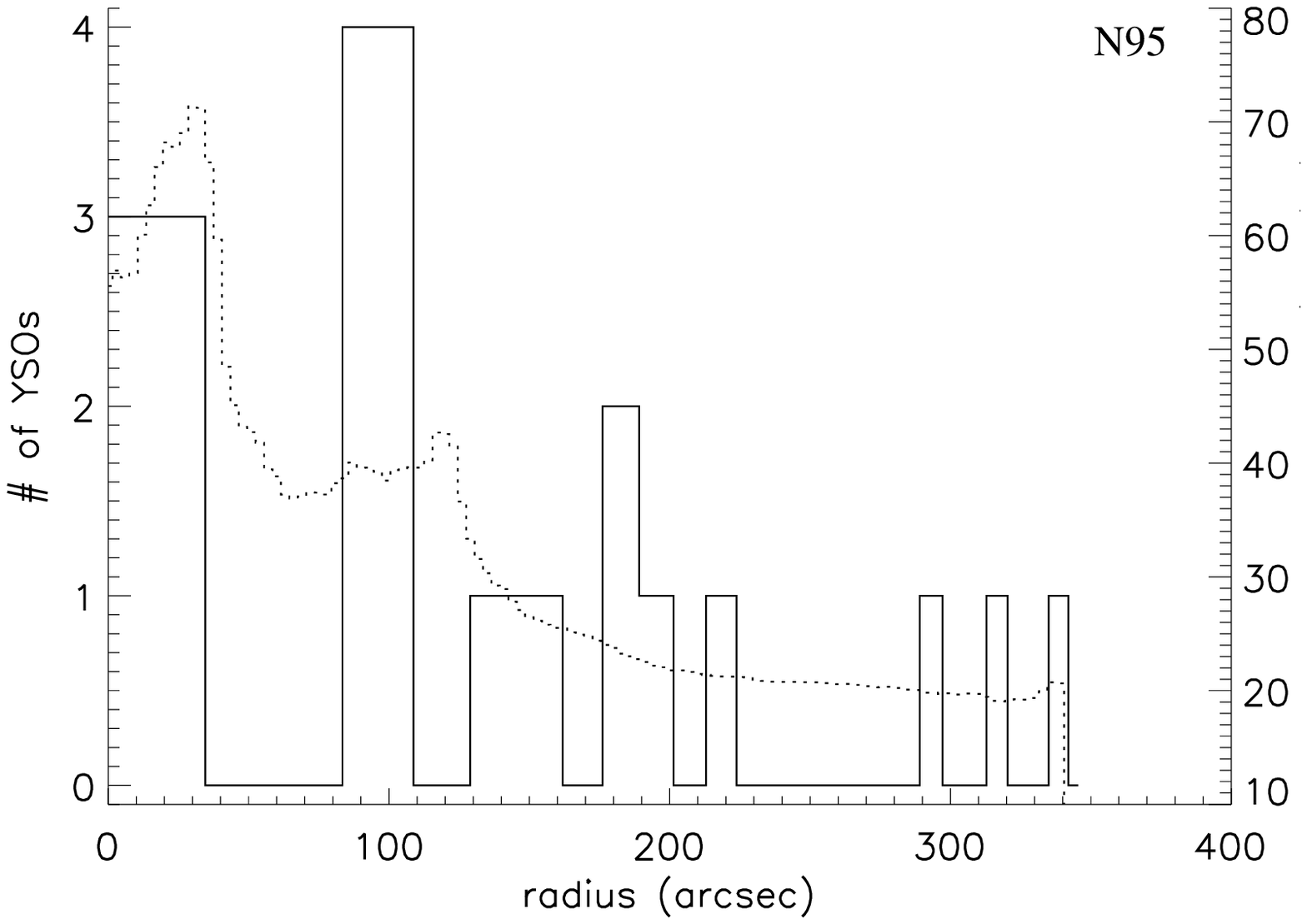}
\caption{Number of YSOs in equal-area annuli (solid) and 8 $\mu$m brightness azimuthally-averaged (dashed) for (upper-left by rows): N80, N82, N84, N88, N89, N90, N92, N95}
\label{profile3}
\end{figure}
\clearpage
\begin{figure}
\plottwo{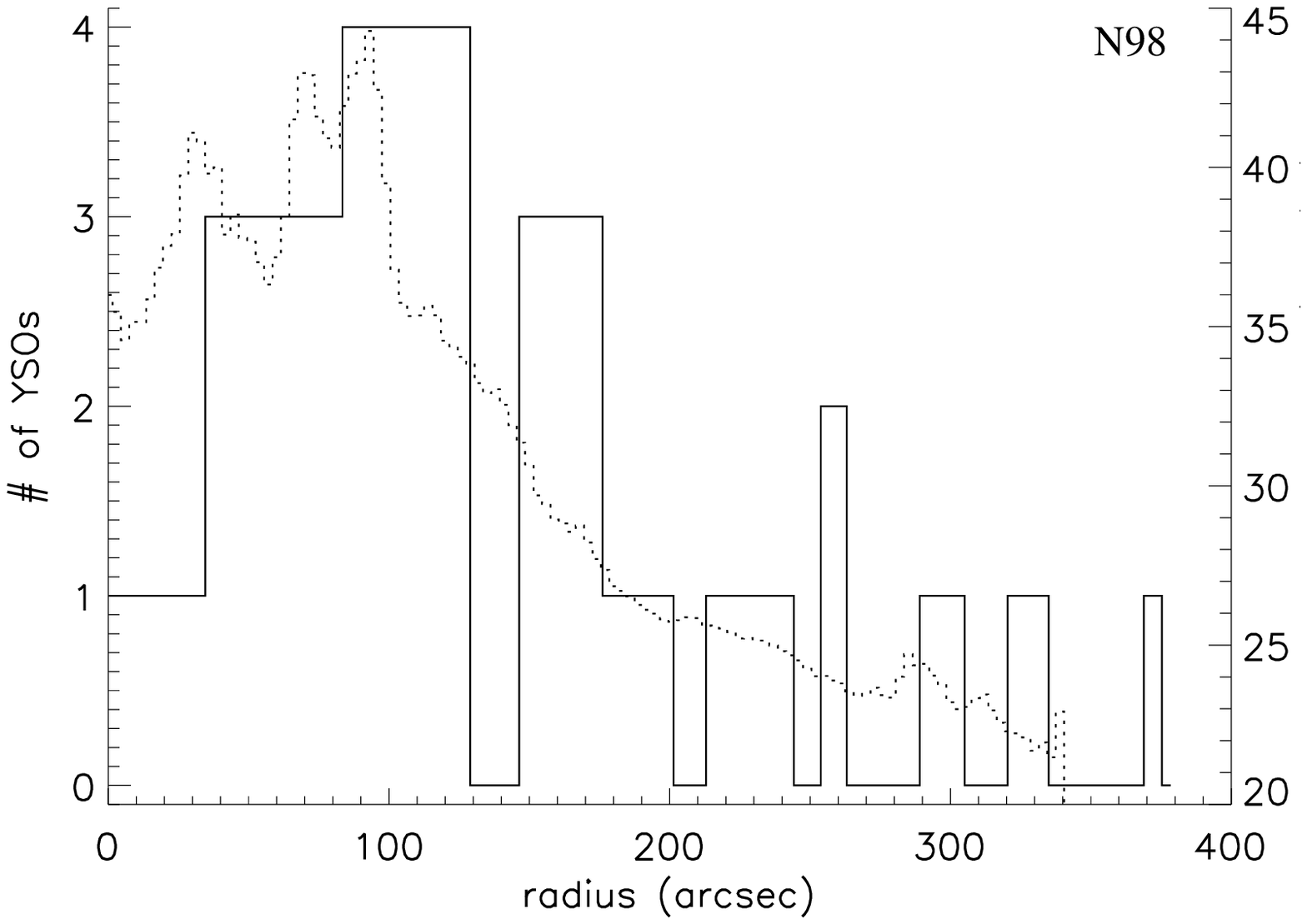}{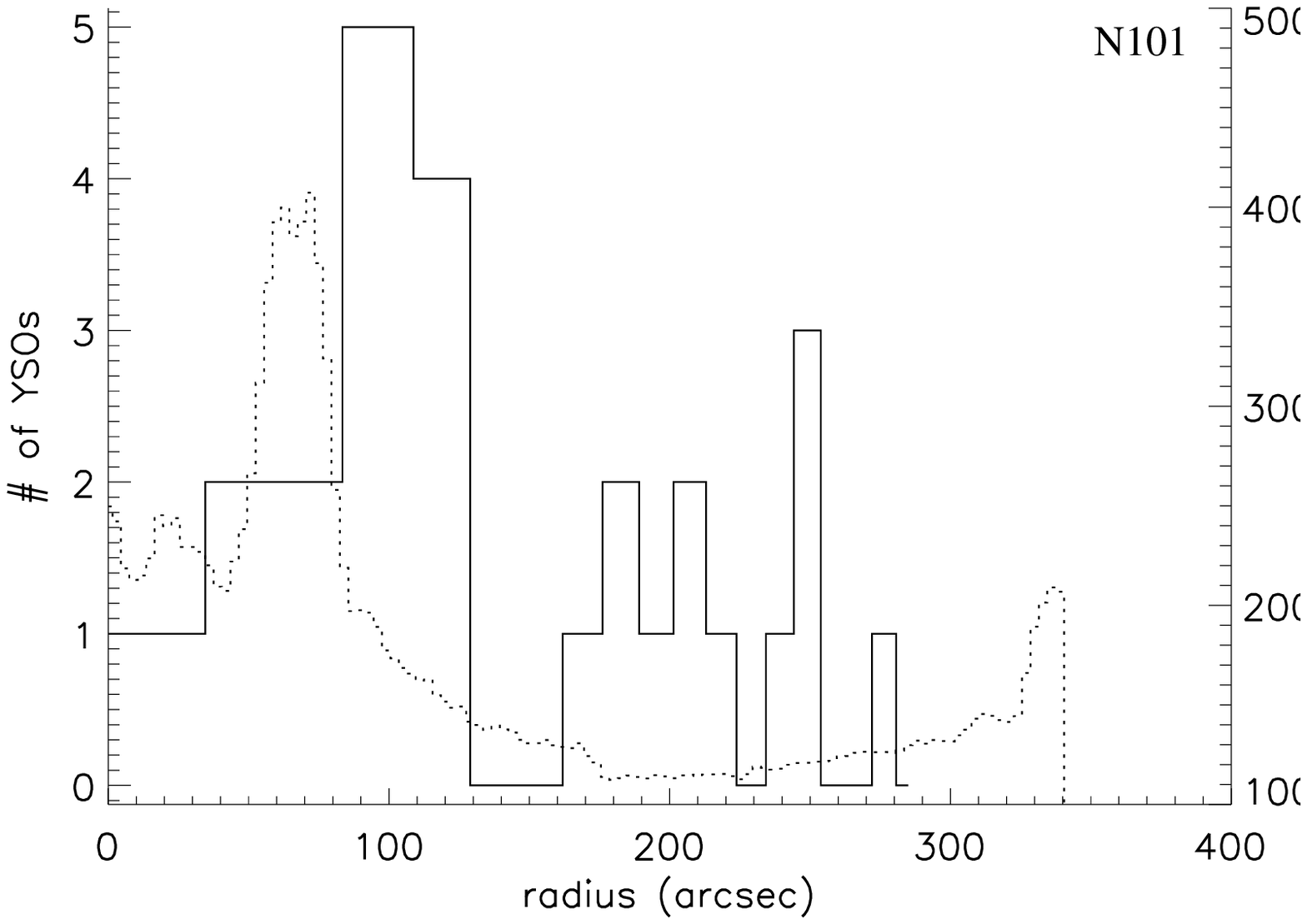}
\plottwo{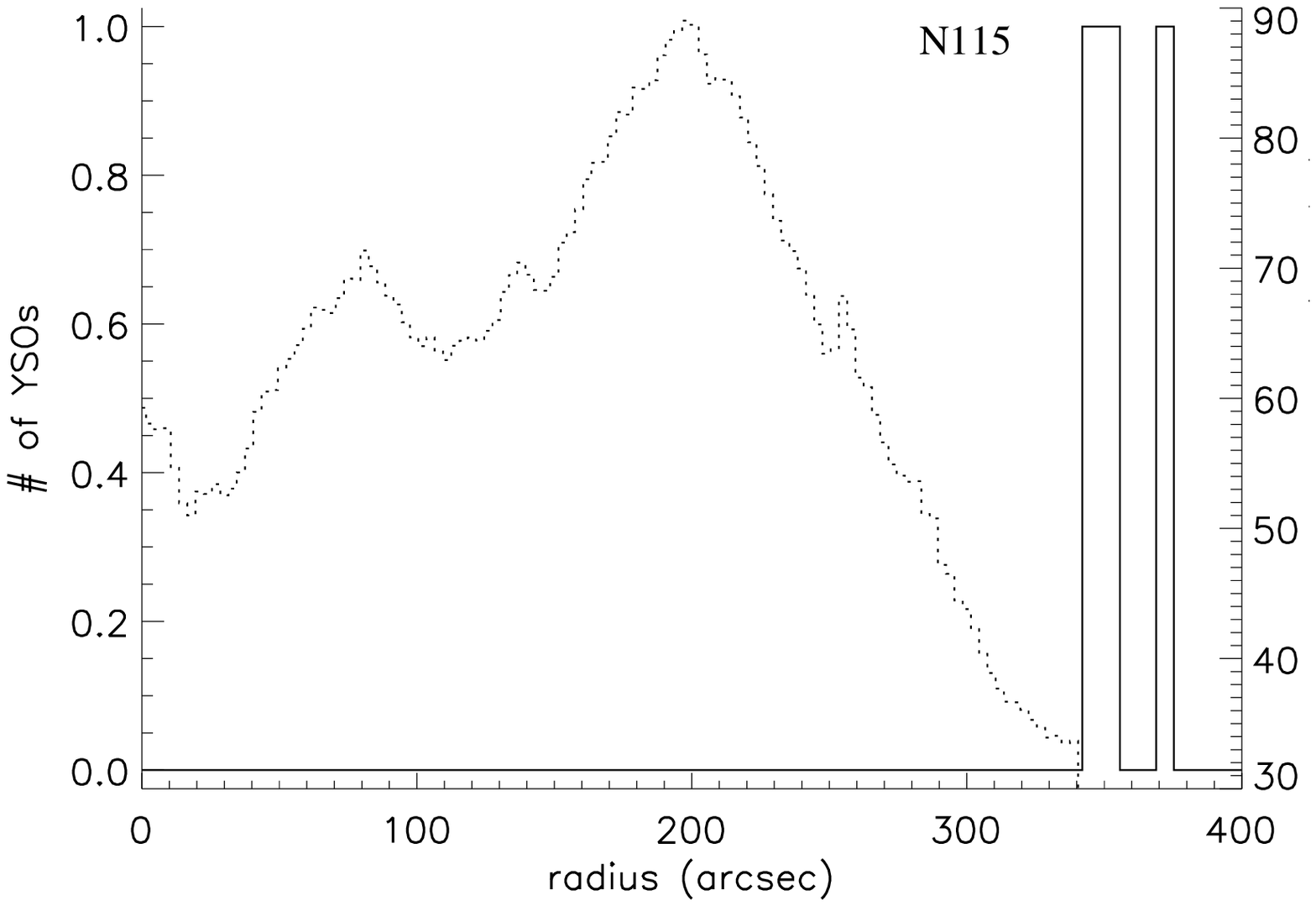}{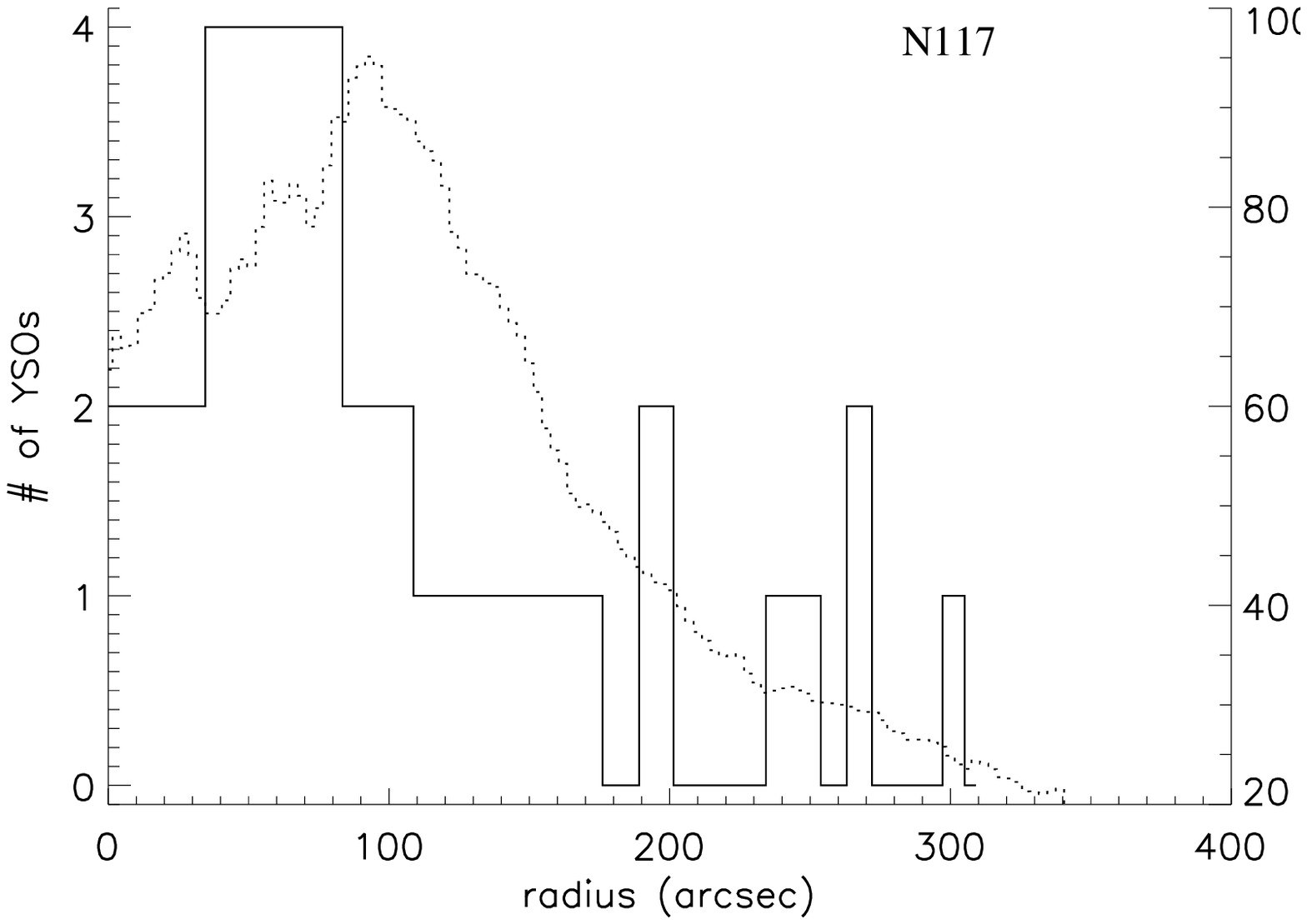}
\plottwo{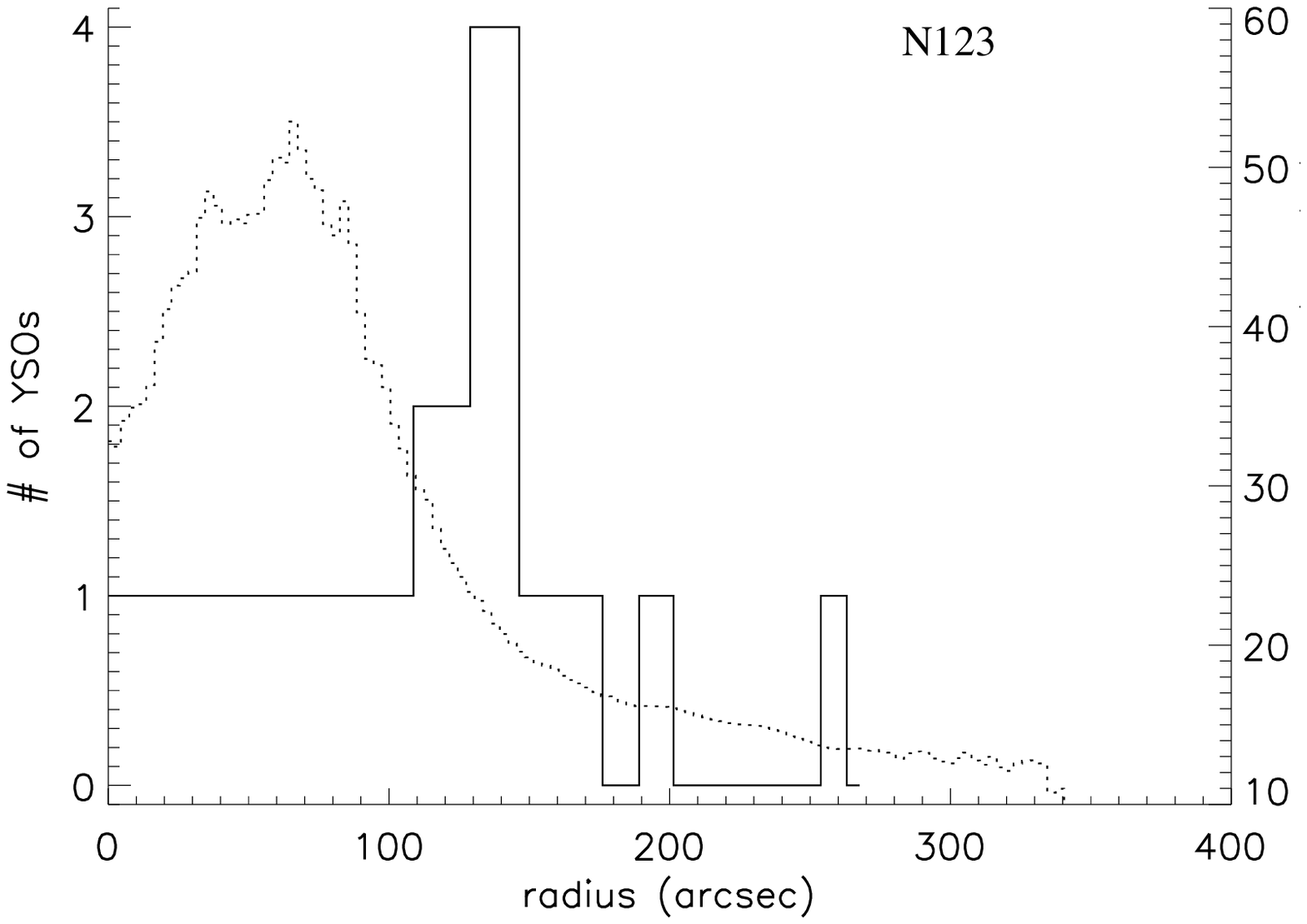}{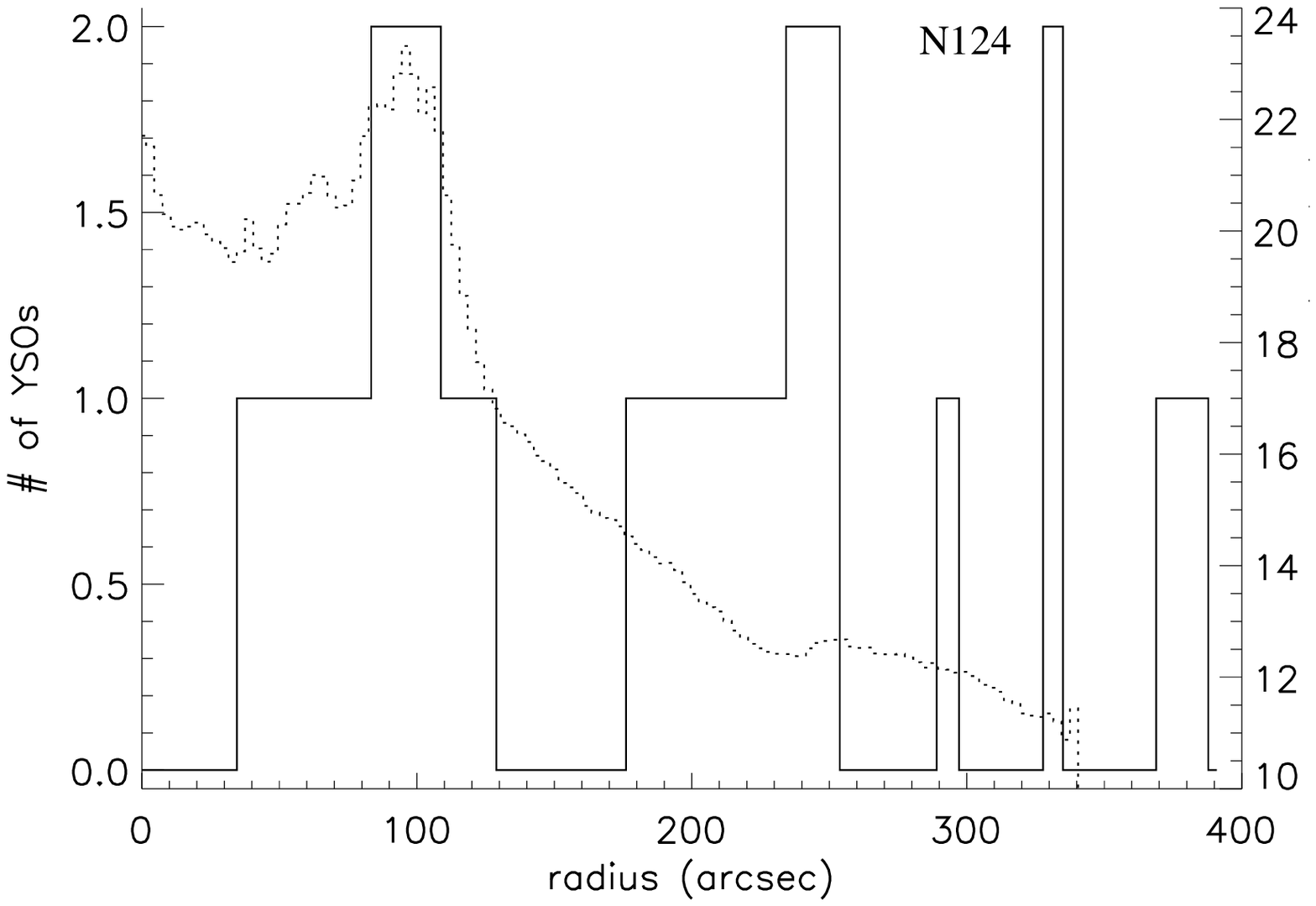}
\plottwo{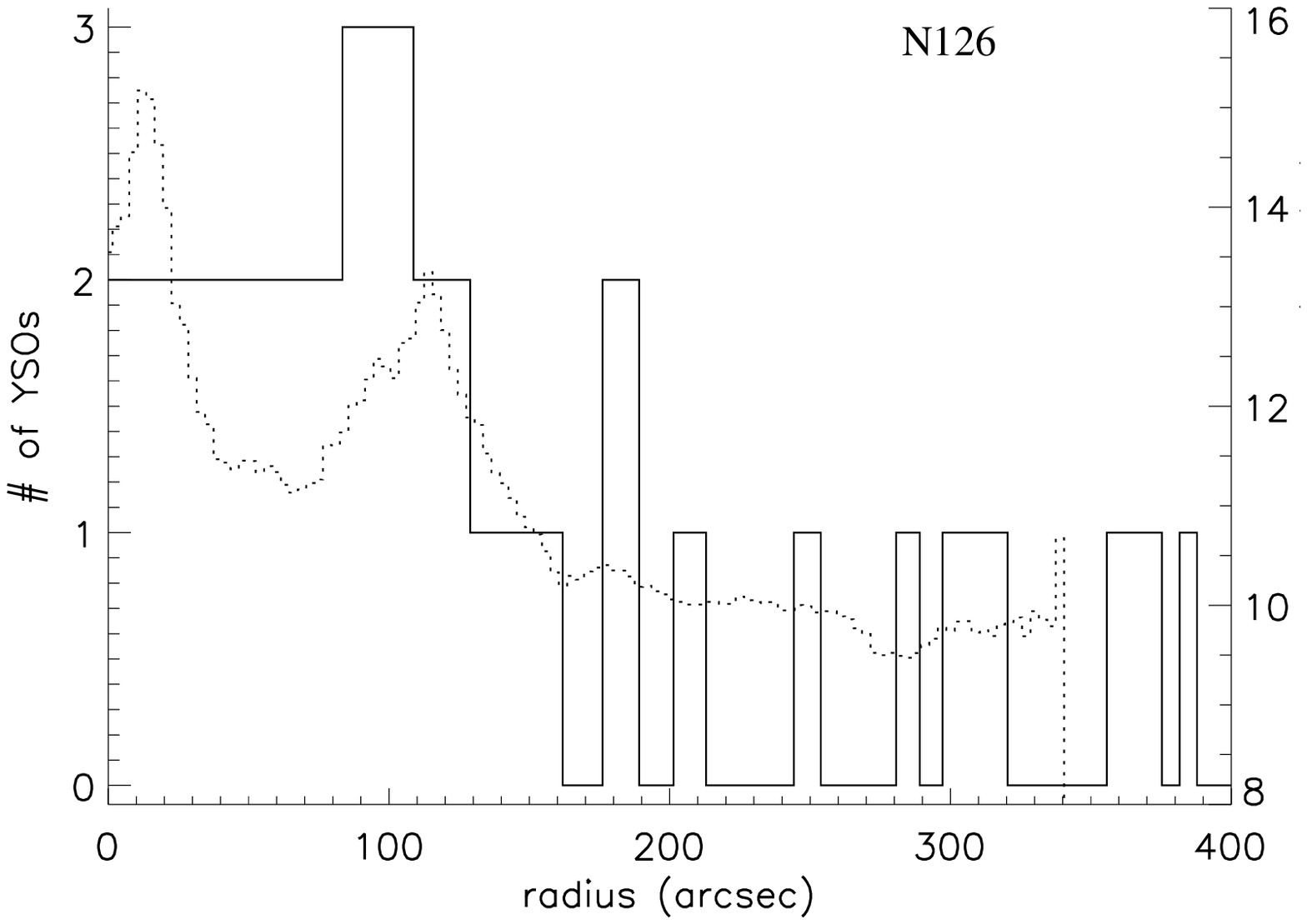}{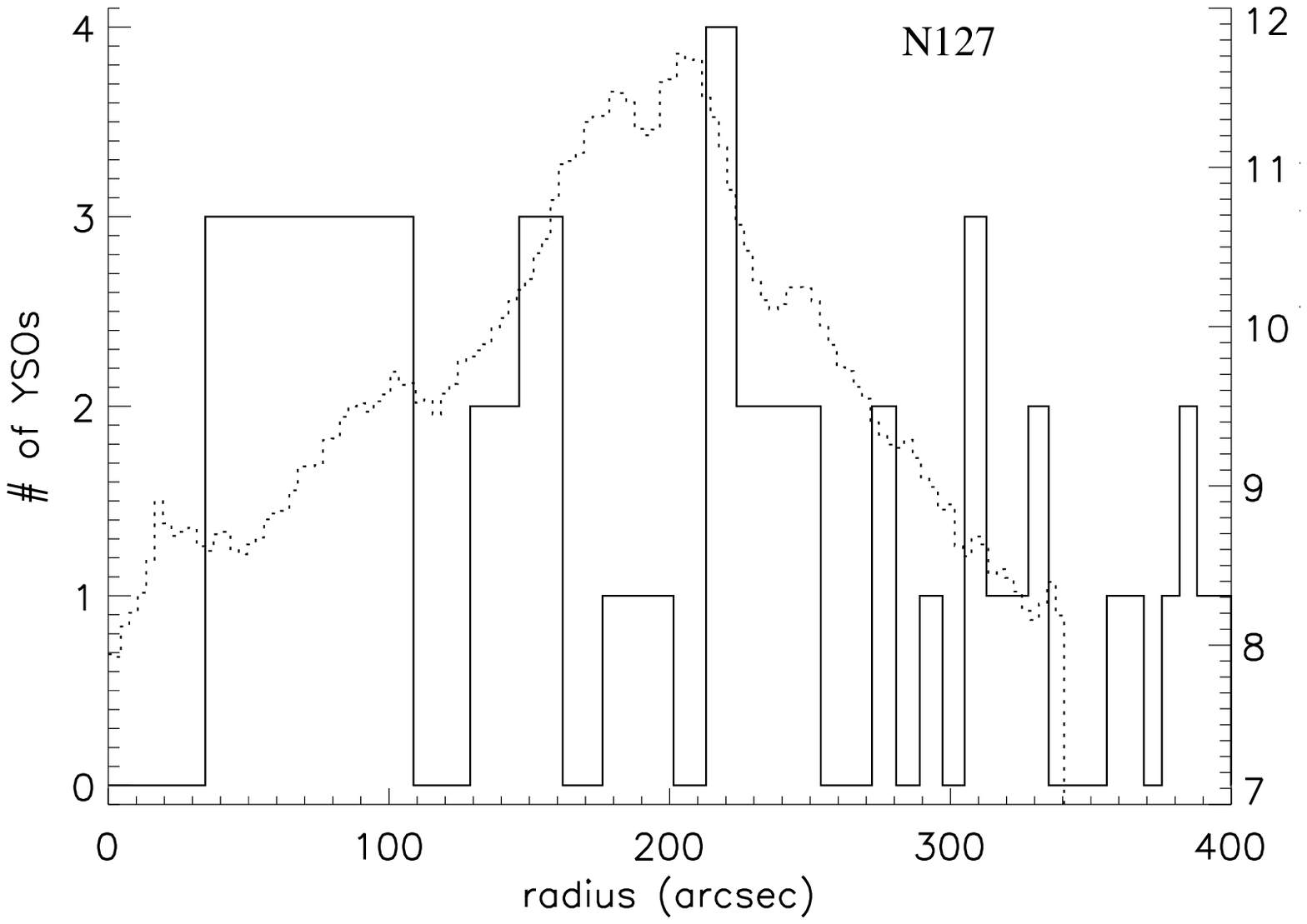}
\caption{Number of YSOs in equal-area annuli (solid) and 8 $\mu$m brightness azimuthally-averaged (dashed) for (upper-left by rows): N98, N101, N115, N117, N123, N124, N126, N127}
\label{profile4}
\end{figure}

\begin{figure}
\plottwo{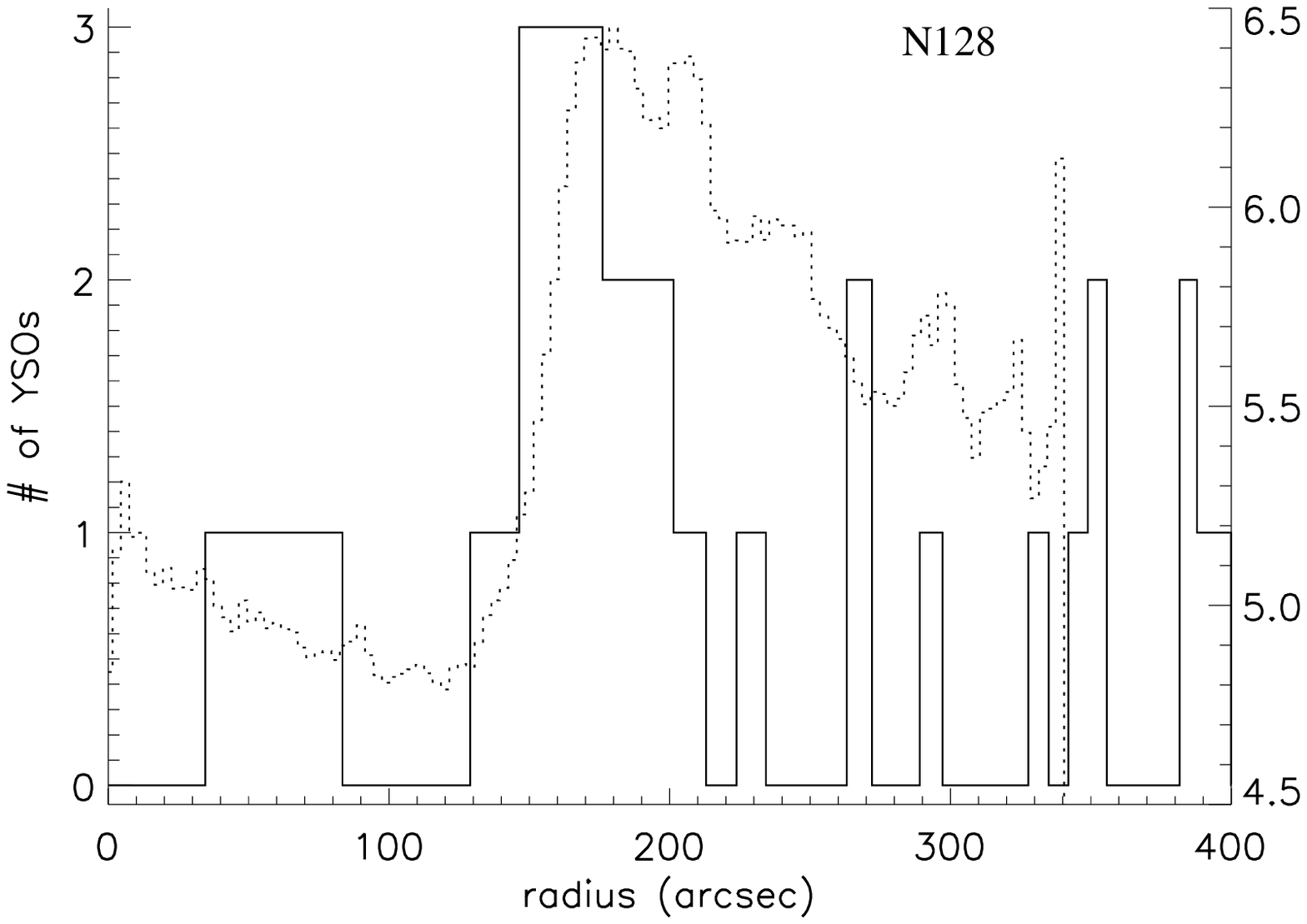}{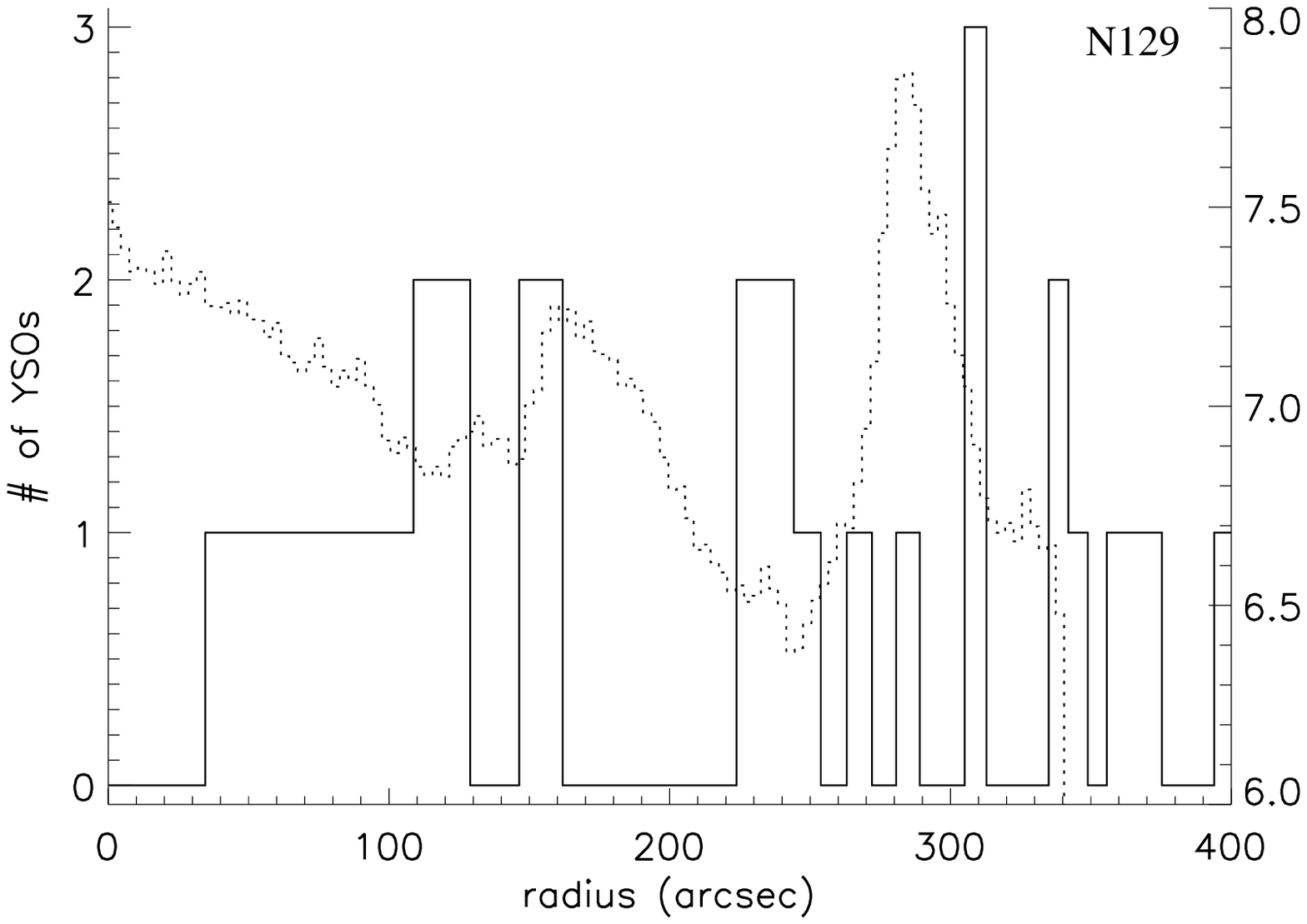}
\plottwo{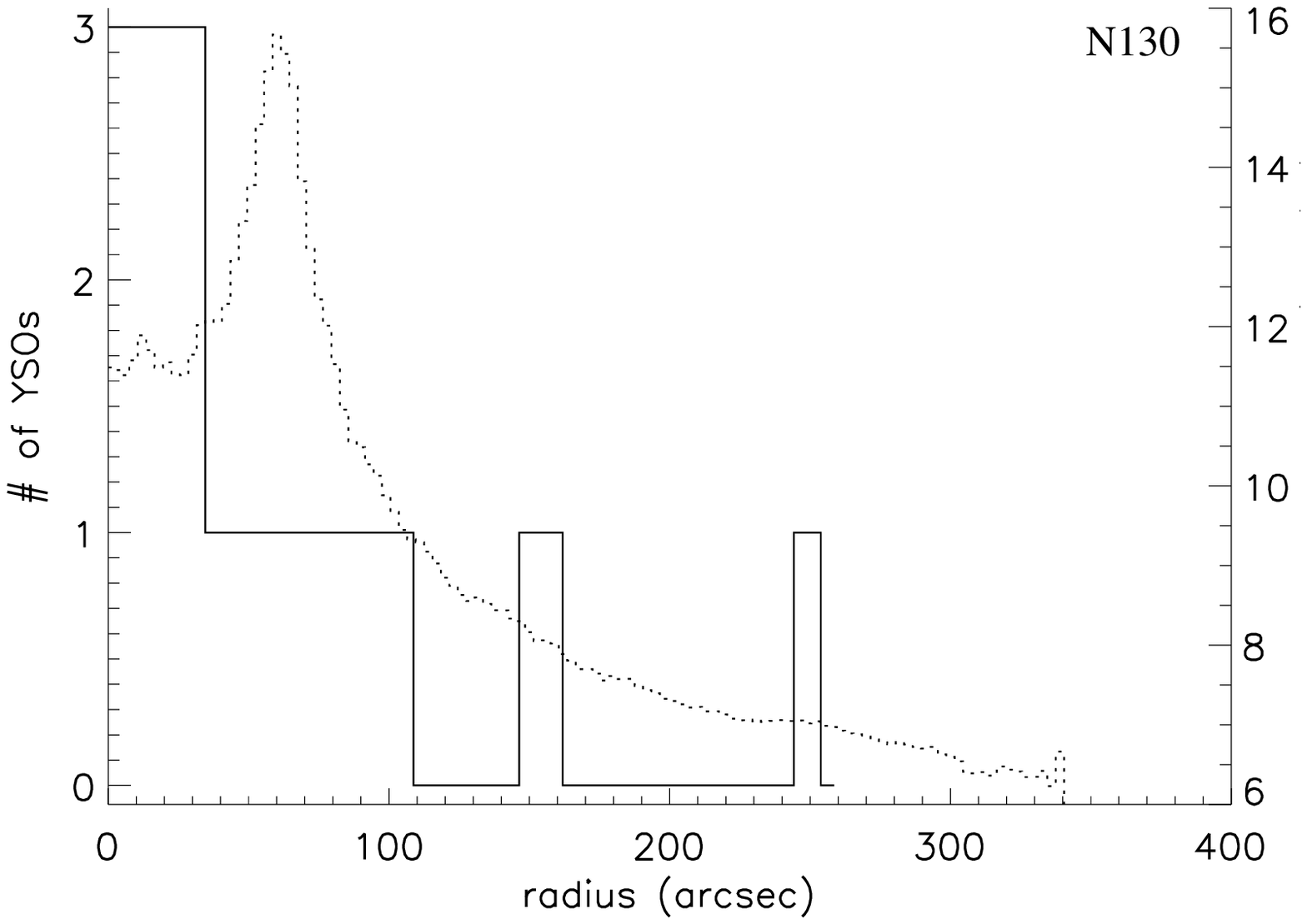}{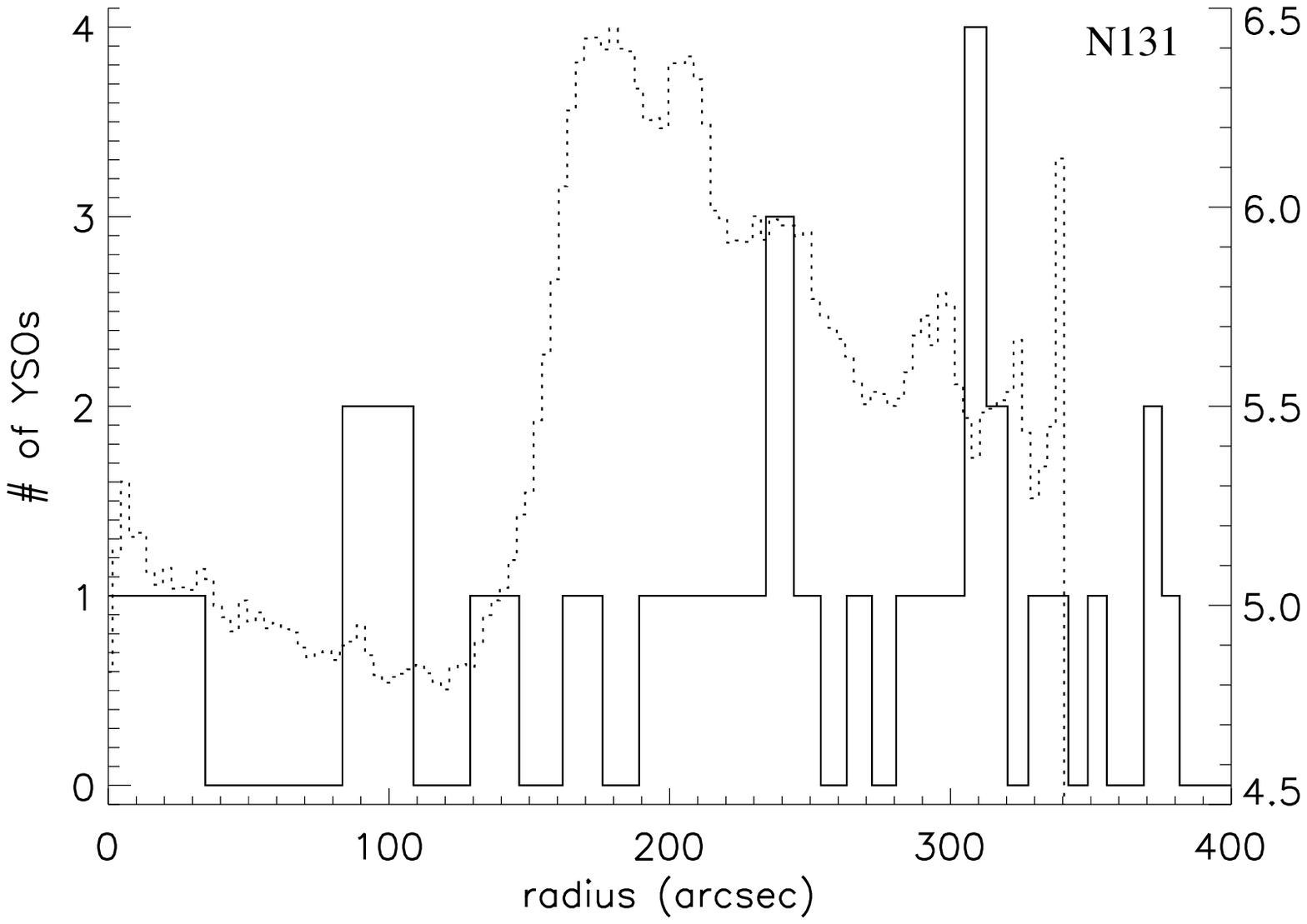}
\plottwo{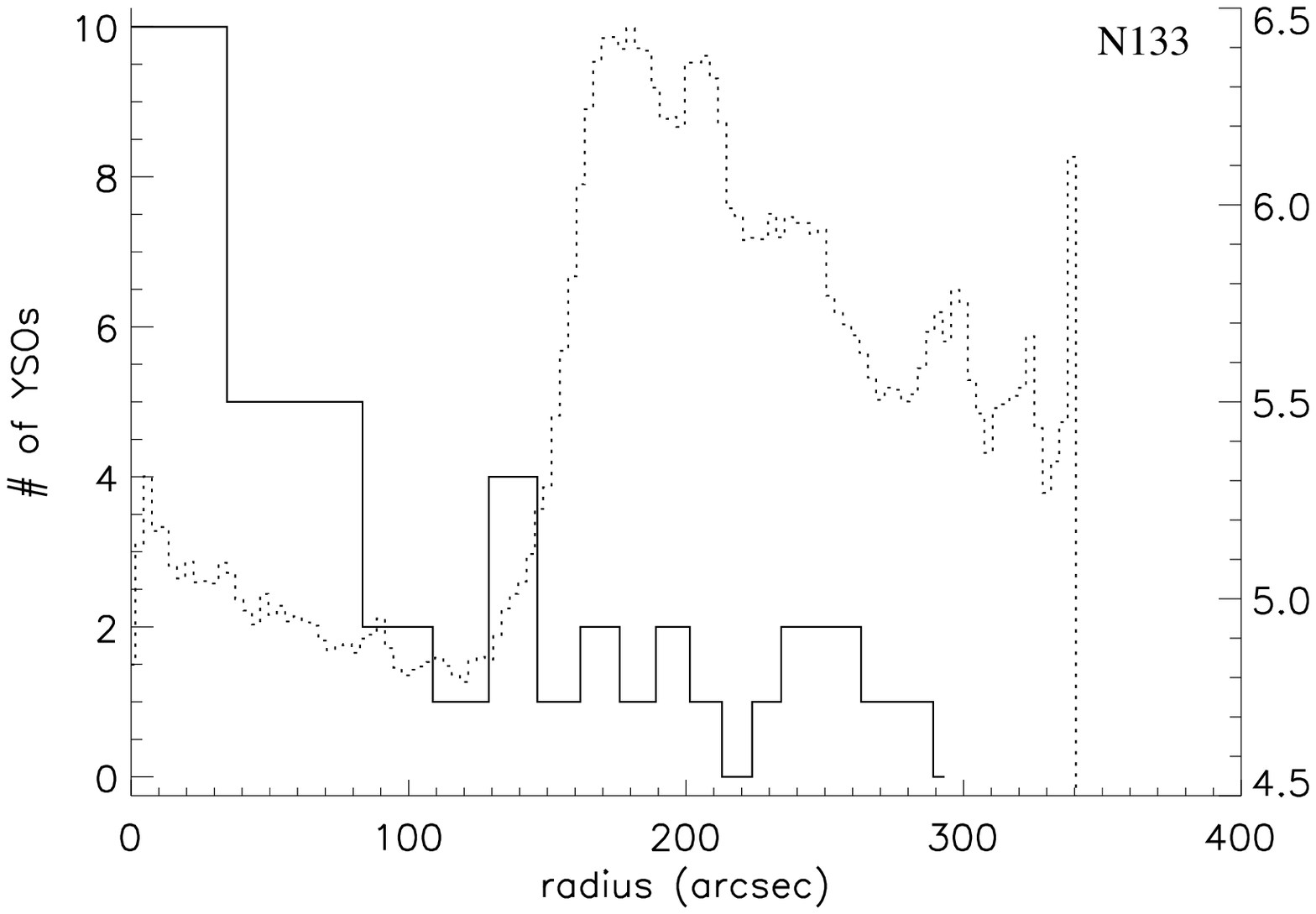}{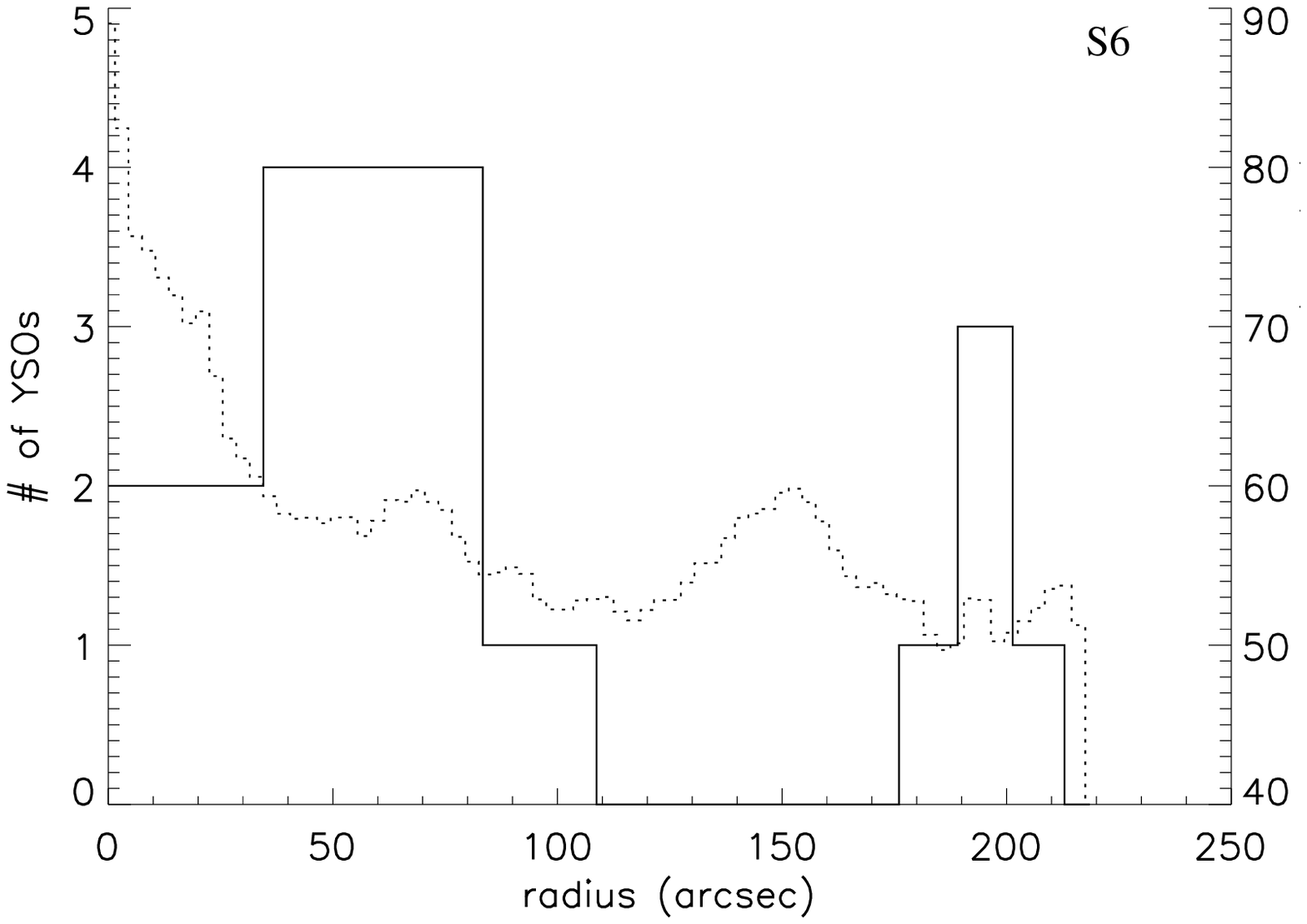}
\plottwo{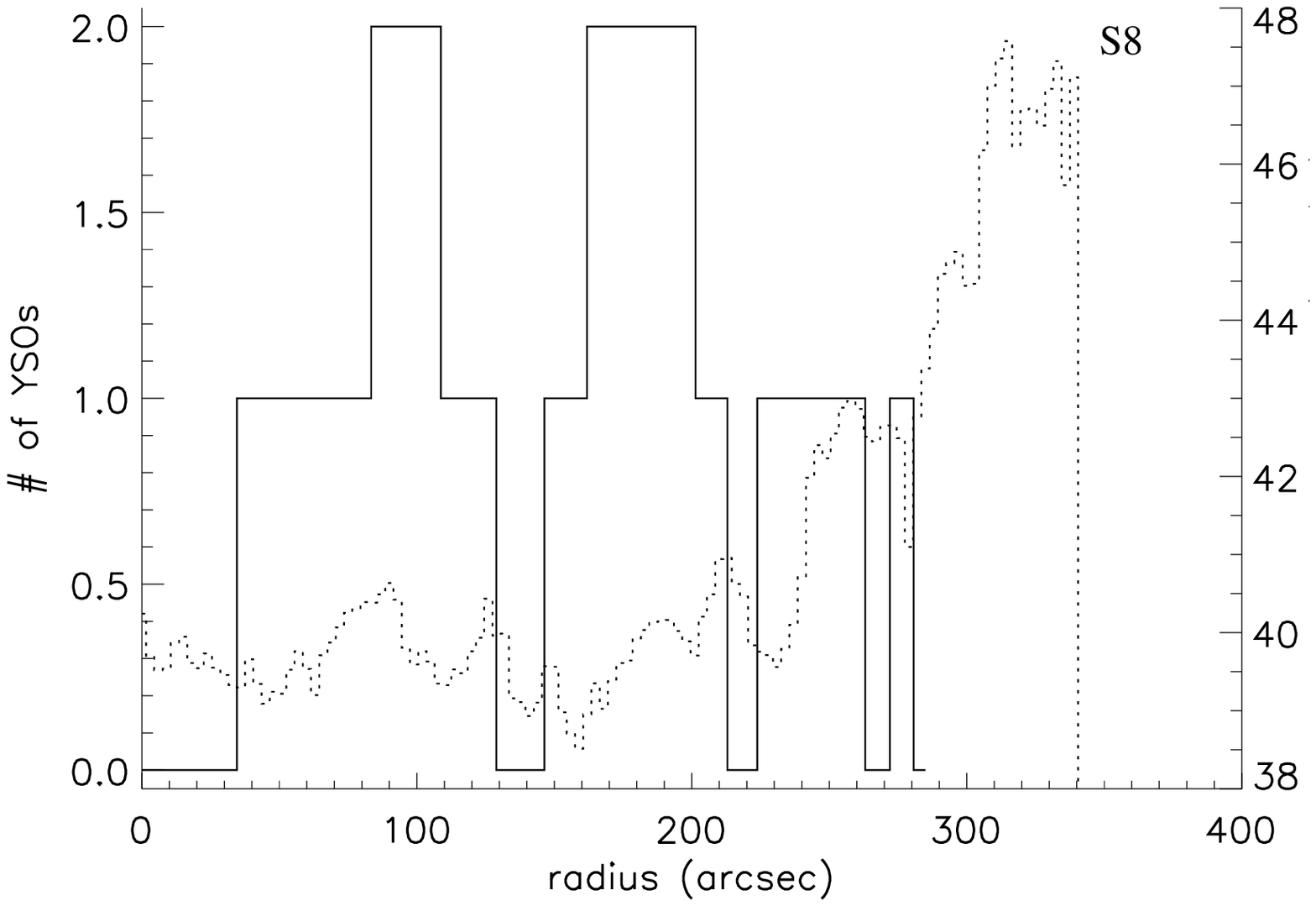}{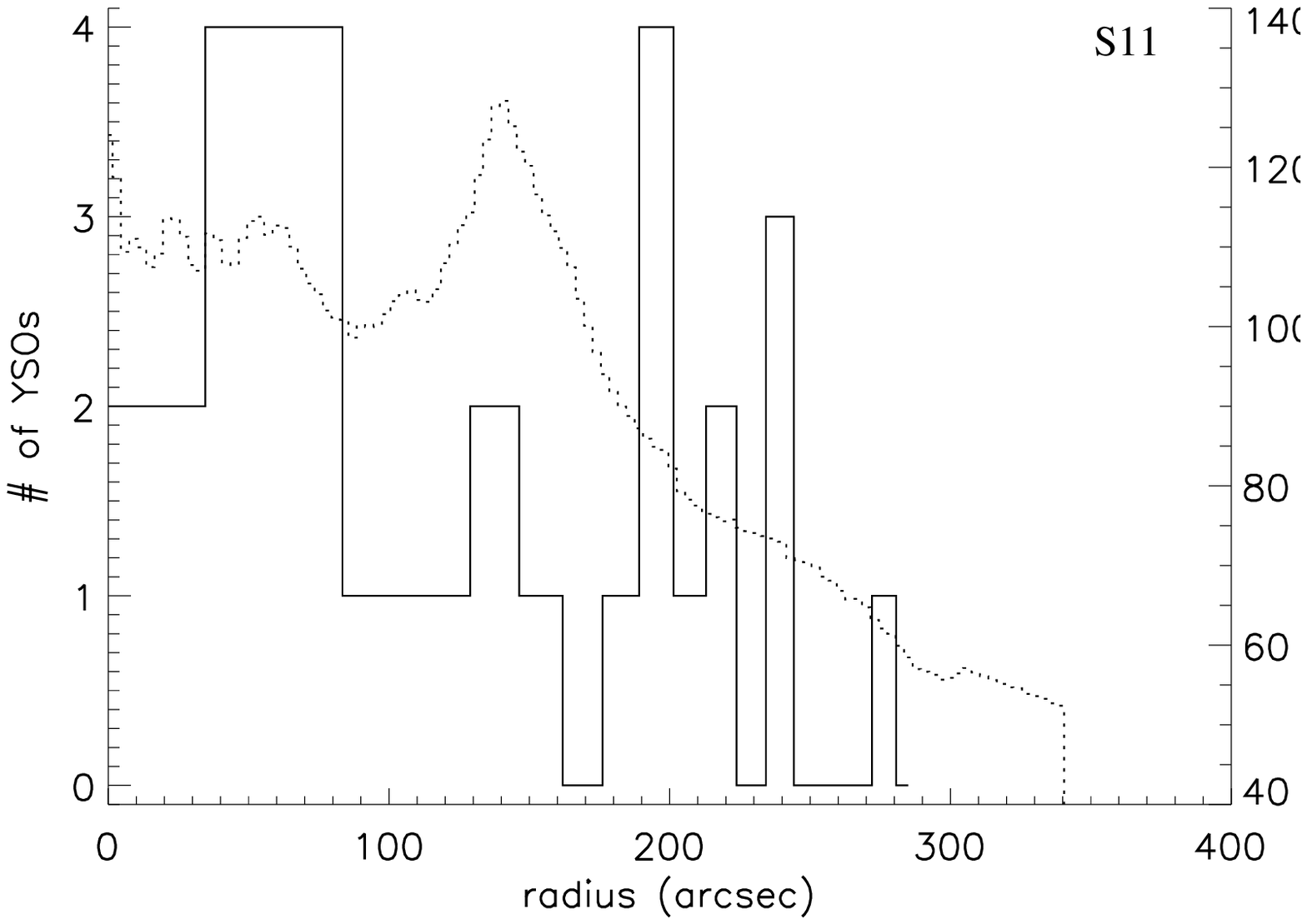}
\caption{Number of YSOs in equal-area annuli (solid) and 8 $\mu$m brightness azimuthally-averaged (dashed) for (upper-left by rows): N128, N129, N130, N131, N133, S6, S8, S11}
\label{profile5}
\end{figure}
\clearpage
\begin{figure}
\plottwo{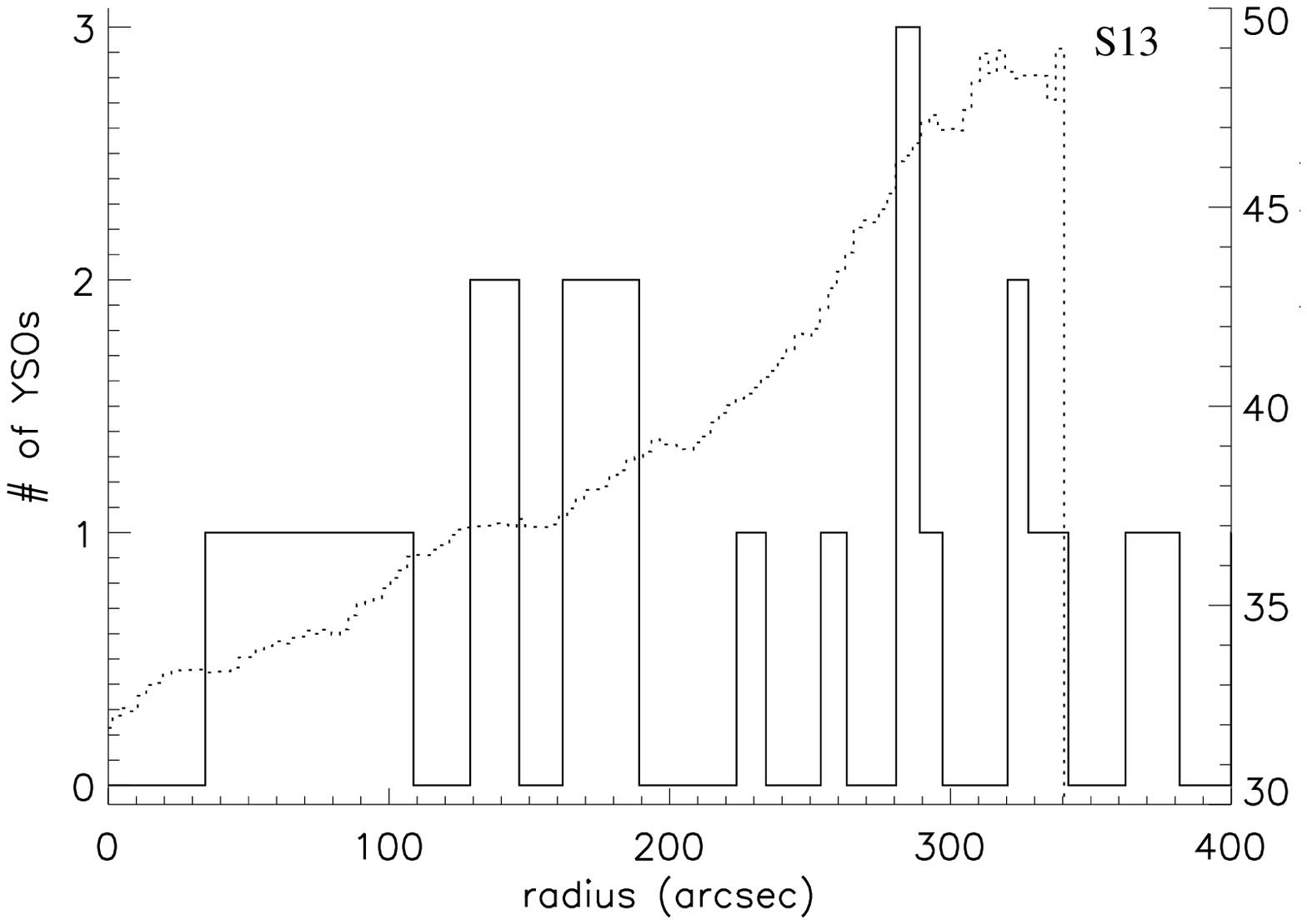}{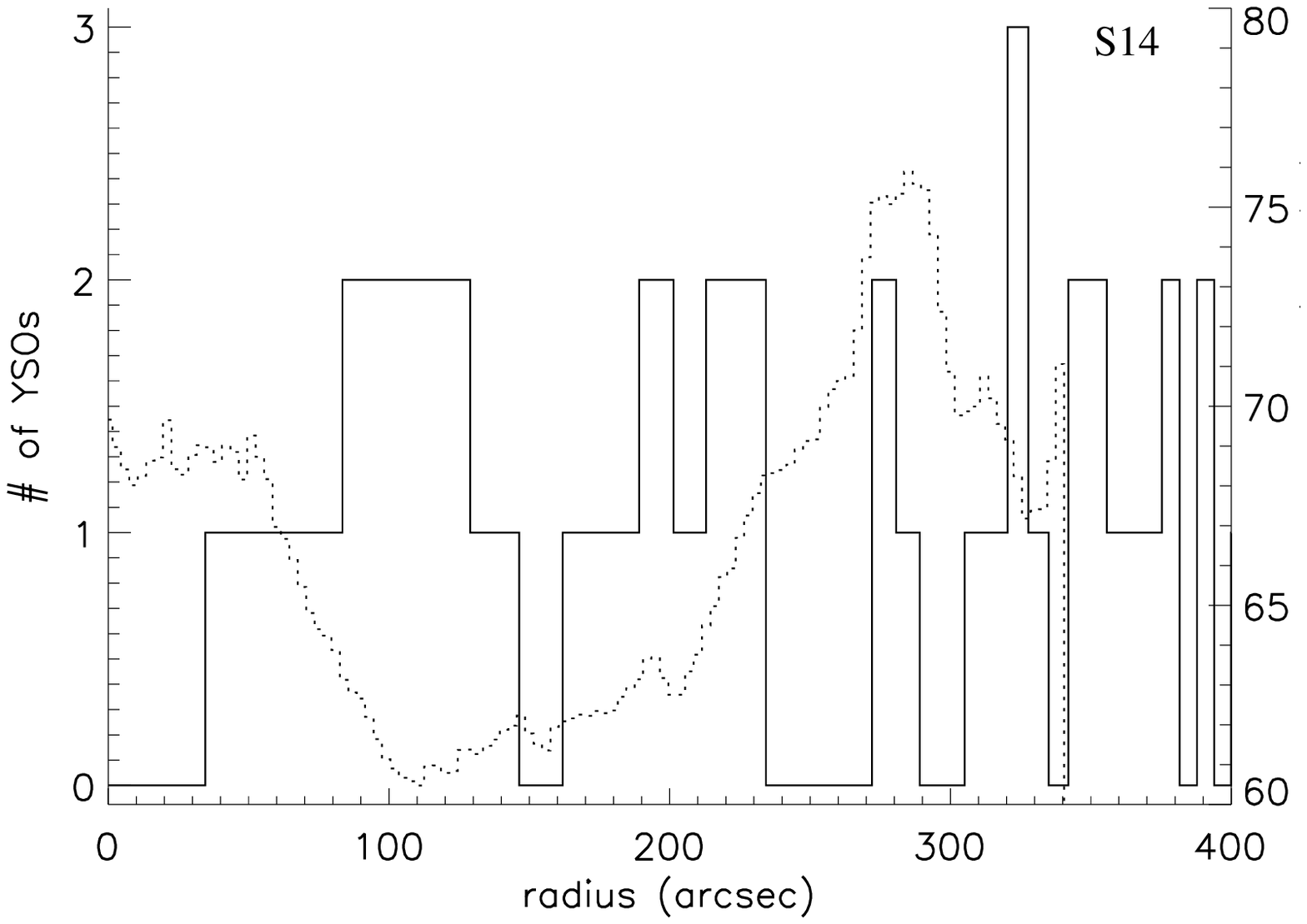}
\plottwo{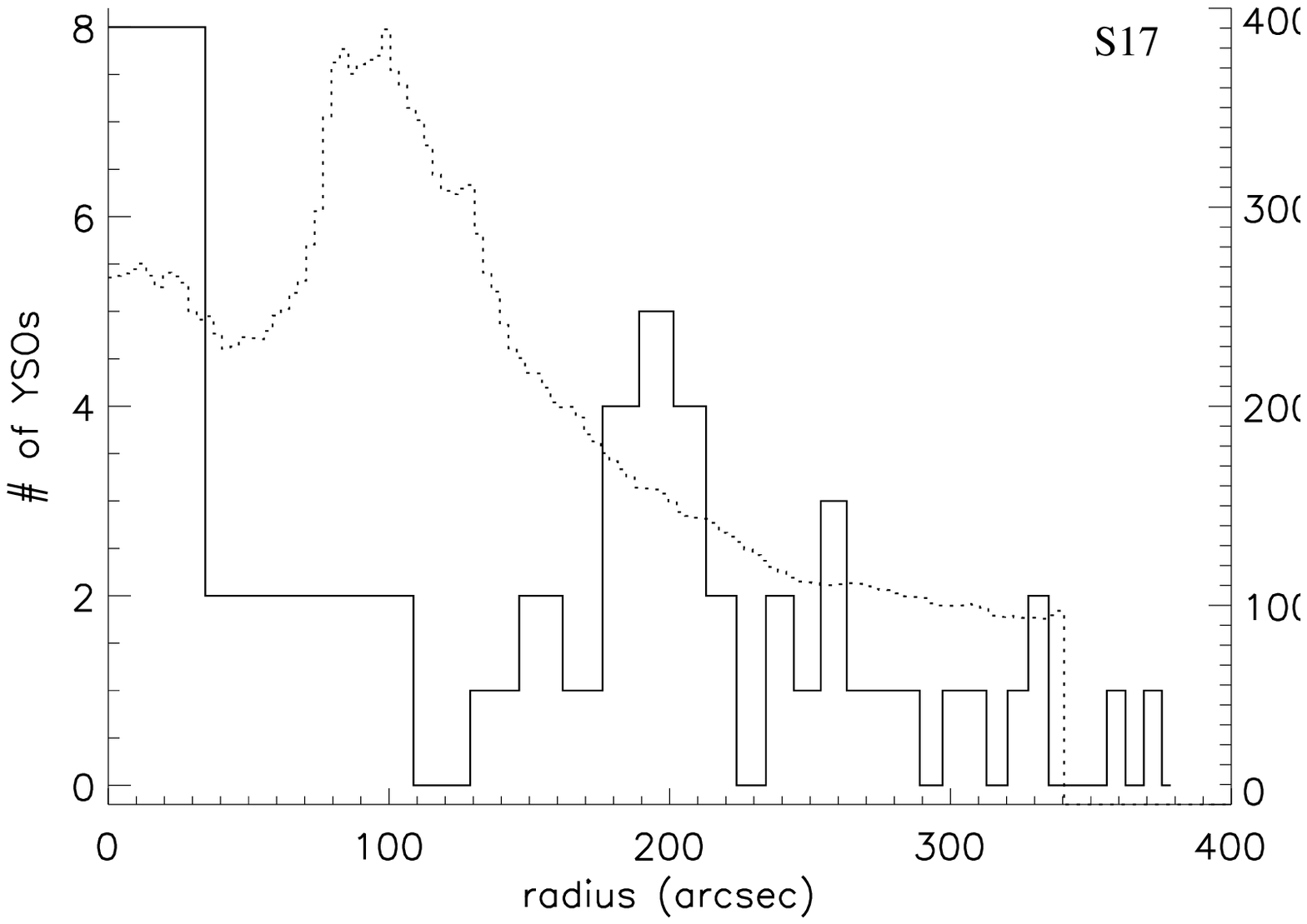}{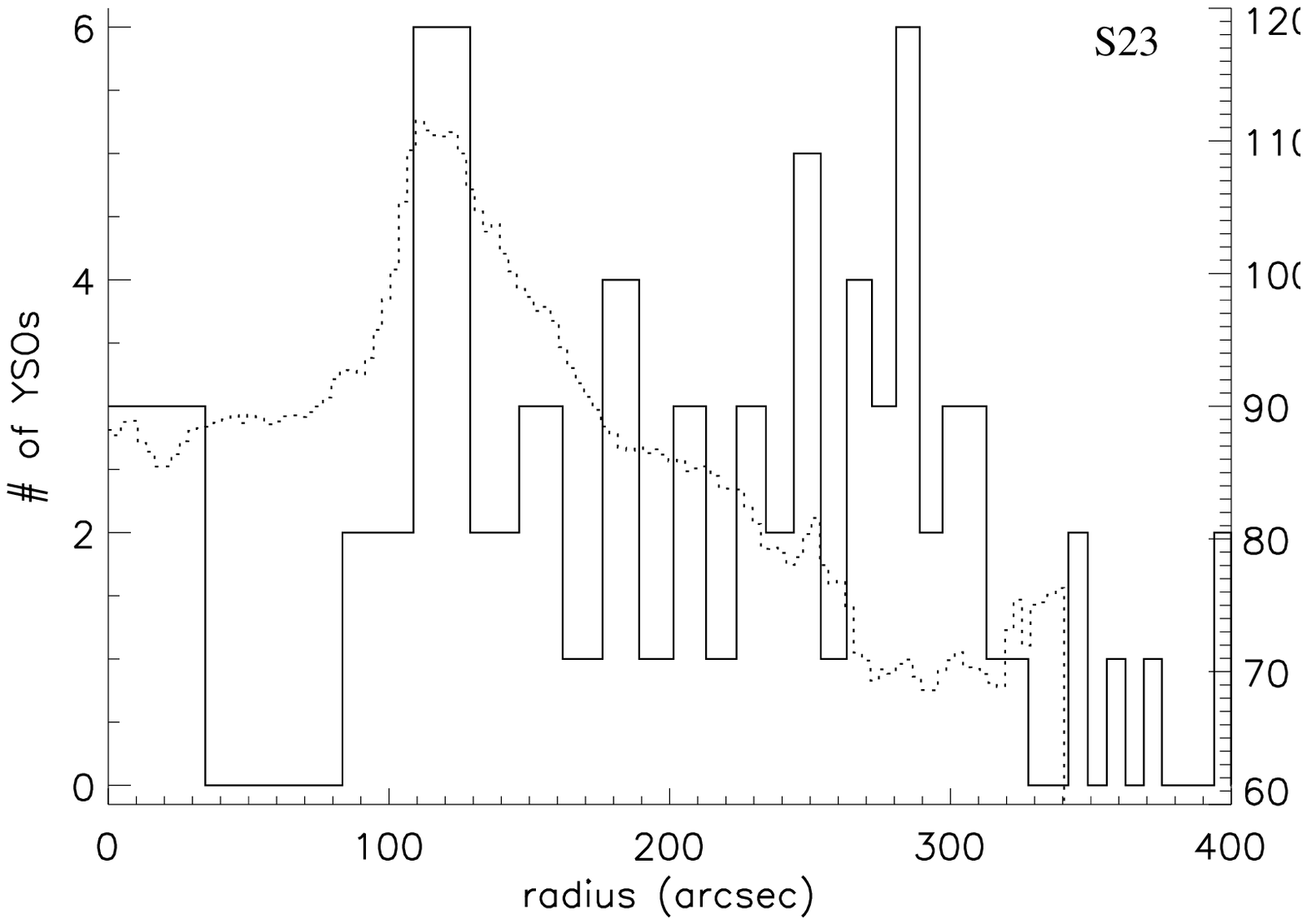}
\plottwo{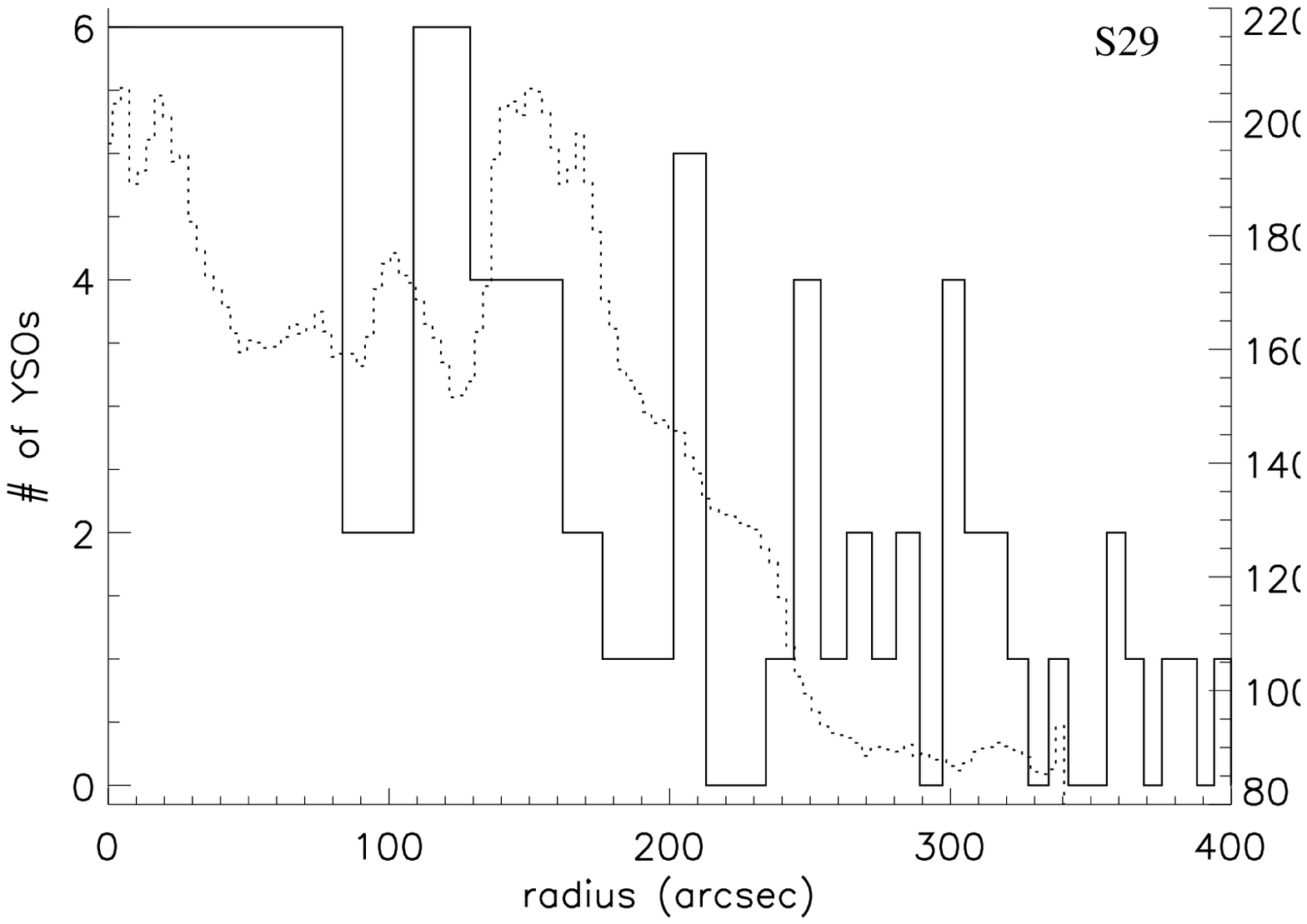}{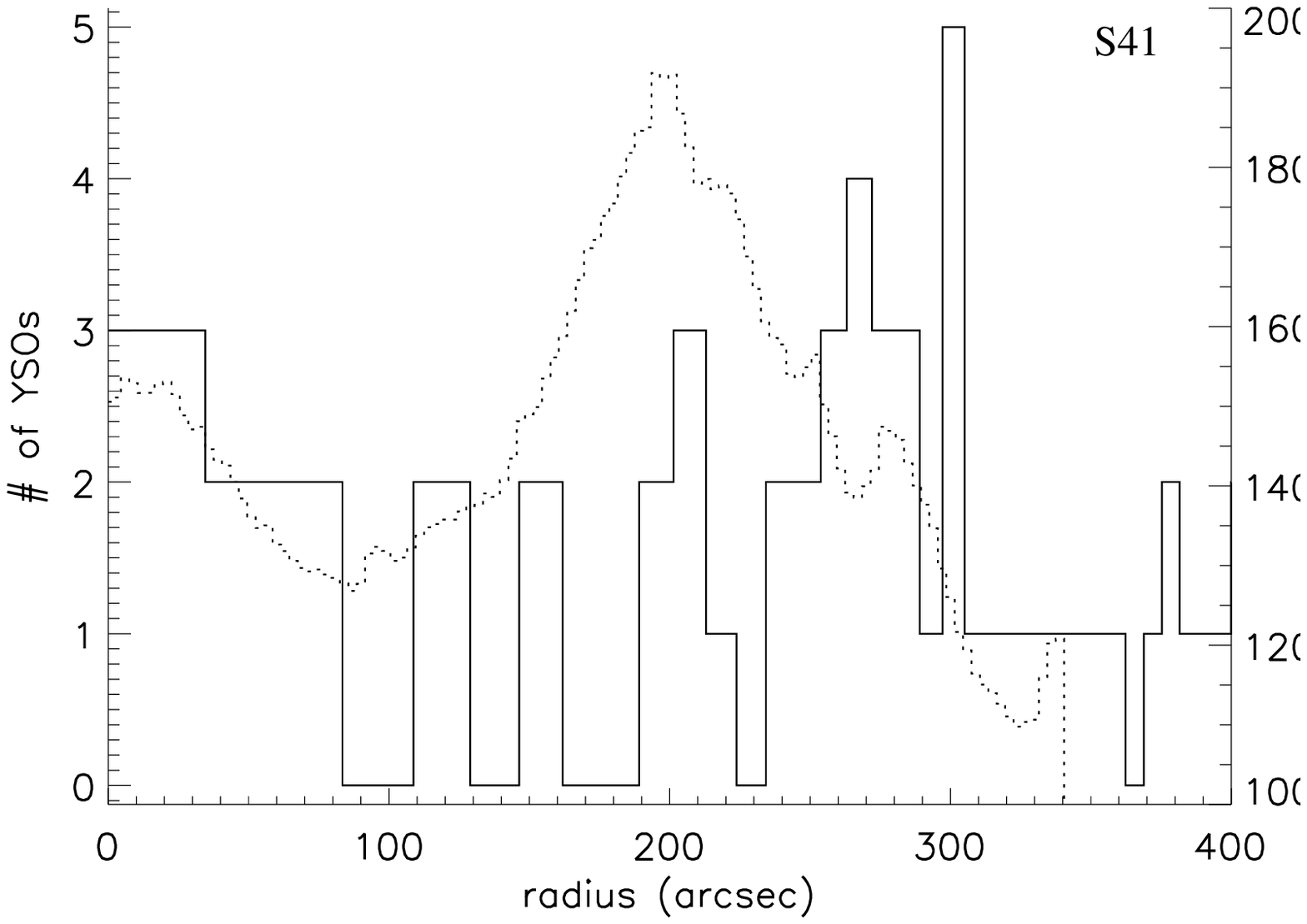}
\caption{Number of YSOs in equal-area annuli (solid) and 8 $\mu$m brightness azimuthally-averaged (dashed) for (upper-left by rows): S13, S14, S17, S23, S29, S41}
\label{profile6}
\end{figure}

\begin{deluxetable}{lrrrrrrrr}
\tablecaption{Simulated and observed YSO properties for bubbles we identify as potential sites for triggered star-formation.}
\tablehead{
	\colhead{Name} &\colhead{l} &\colhead{b} &\colhead{Radius} &\colhead{Distance} &\colhead{Radius} &\colhead{\# YSOs} &\colhead{Average} &\colhead{\% of Taurus}\\
&&&\colhead{(')} &\colhead{(kpc)} &\colhead{(pc)}&&\colhead{YSO A$_V$} &\colhead{YSOs detectable}}
\startdata

N62 & 34.334 & 0.216 & 1.5 & 3.9 & 1.7 & 2 & 10.1 & 76 \\
N65 &35.000  &0.332  & 2.0 & 3.1 & 1.8 &11 & 9.0 &83\\
N77 & 40.421 & -0.056 & 1.3 & 5.0 & 1.9 & 3 & 9.7 & 55 \\
N82 & 42.102 & -0.623 & 1.7 & 5.2 & 2.5 & 6 & 9.5 & 55 \\
N90 & 43.774 & 0.050 & 1.7 & 3.0 & 1.5 & 7 & 8.6 & 83 \\
N92 & 44.333 & -0.839 & 1.9 & 3.7 & 2.1 & 7 & 10.2 & 83 \\
N101 & 49.197 & -0.358 & 1.2 & 5.1 & 1.7 & 10 & 7.1 & 66 \\
N117 & 54.112 & -0.064 & 1.7 & 5.1 & 2.5  & 10 & 10.2 & 55 \\
N128 & 61.673 & 0.946 & 3.2 & 2.8 & 2.6 & 12 & 7.5 & 83 \\
N123 & 57.539 & -0.284 & 1.4 & 2.6 & 1.1 & 1 & 10.9 & 83 \\
N133 &63.159 & 0.451 &1.83 & 2.1 &1.1 &6 &9.0 &83\\
S17 & 343.482 & -0.044 & 1.9 & 3.0 & 1.6 & 12 & 9.5 & 83 \\
S23 &341.281 &-0.349 &1.8 &3.0 &1.6 &3 &9.2 &83\\
\enddata
\label{ysobubbleprop}
\end{deluxetable}

\end{document}